\documentclass[prl,twocolumn,superscriptaddress,nobibnotes,a4paper]{revtex4-2}
\usepackage[pdftex]{graphicx}
\usepackage[pdftex]{epsfig}
\usepackage{amsmath}
\usepackage{amssymb}
\usepackage{amsfonts}
\usepackage{color}
\usepackage{wrapfig}
\usepackage{eucal}
\usepackage{hhline}
\usepackage{threeparttable}
\usepackage{supertabular}
\usepackage{titlesec}

\usepackage{blindtext}
\usepackage{multirow}
\usepackage{tabularx}
\usepackage{upgreek}
\usepackage{float}
\usepackage{siunitx}
\usepackage[normalem]{ulem}

\DeclareMathOperator{\Tr}{Tr}

\DeclareRobustCommand{\orderof}{\ensuremath{\mathcal{O}}}

\usepackage[export]{adjustbox}

\usepackage{soul}

\usepackage[version=4]{mhchem} % for chemical formulas
\usepackage{array} % make column bigger (vertical)
\usepackage[colorlinks=true, allcolors=blue]{hyperref} %hyperrefs

\usepackage{comment}

\newcommand{\bra}[1]{\left\langle #1\right|}
\newcommand{\ket}[1]{\left|#1\right\rangle}

\setstcolor{red}

\graphicspath{ {./fig_pdf/} }
\usepackage{graphicx}
\begin{document}

\makeatletter
\renewcommand*{\@fnsymbol}[1]{\ensuremath{\ifcase#1\or \dagger \or* \or \ddagger\or
   \mathsection\or \mathparagraph\or \|\or **\or \dagger\dagger
   \or \ddagger\ddagger \else\@ctrerr\fi}}
\makeatother

\title{Low-noise Optomechanical Single Phonon-photon Conversion for Quantum Networks}
\thanks{This work was published in \href{https://doi.org/10.1038/s41467-025-67956-2}{Nat. Commun. \textbf{17}, 1187 (2026).}}

\author{Liu Chen}
\thanks{These authors contributed equally to this work}
\affiliation{Kavli Institute of Nanoscience, Department of Quantum Nanoscience, Delft University of Technology, 2628CJ Delft, The Netherlands}

\author{Alexander Rolf Korsch}
\thanks{These authors contributed equally to this work}

\affiliation{Kavli Institute of Nanoscience, Department of Quantum Nanoscience, Delft University of Technology, 2628CJ Delft, The Netherlands}
\affiliation{Department of Physics, Fudan University, Shanghai 200433, P.R.\ China}
\affiliation{Department of Physics, School of Science, Westlake University, Hangzhou 310030, P.R.\ China}

\author{Cau\^{e} Moreno Kersul}
\affiliation{Instituto de F\'isica Gleb Wataghin, Universidade Estadual de Campinas (UNICAMP), 13083-859 Campinas, SP, Brazil}

\author{Rodrigo Benevides}
\affiliation{Instituto de F\'isica Gleb Wataghin, Universidade Estadual de Campinas (UNICAMP), 13083-859 Campinas, SP, Brazil}

\author{Yong Yu}
\affiliation{Kavli Institute of Nanoscience, Department of Quantum Nanoscience, Delft University of Technology, 2628CJ Delft, The Netherlands}

\author{Thiago P. Mayer Alegre}
\affiliation{Instituto de F\'isica Gleb Wataghin, Universidade Estadual de Campinas (UNICAMP), 13083-859 Campinas, SP, Brazil}

\author{Simon Gr\"oblacher}
\email{s.groeblacher@tudelft.nl}
\affiliation{Kavli Institute of Nanoscience, Department of Quantum Nanoscience, Delft University of Technology, 2628CJ Delft, The Netherlands}

%\date{\today}

\begin{abstract}
Nano-structured optomechanical crystals (OMC) form an interface between mechanical modes with long coherence times and telecom optical photons, ideal for long-distance distribution of quantum information. However, the implementation of scalable quantum networks based on OMCs has been inhibited by thermal mechanical noise. Here, we overcome this limitation using a quasi-two-dimensional OMC and generate single photons via single phonon-photon conversion. In this work, we verify the low thermal noise and high purity of the generated single photons through a Hanbury Brown-Twiss experiment with $g^{(2)}(0)=0.35^{+0.10}_{-0.08}$. We perform Hong-Ou-Mandel interference of the emitted photons showcasing the indistinguishability and coherence with visibility $V=0.52 \pm 0.15$ after 1.43~km fiber delay. Lastly, we use two-photon interference to measure the temporal wavepackets of optomechanically generated single photons demonstrating narrow bandwidths as low as \qty{10}{MHz}. Our results pave the way for multinode quantum networks of mechanical oscillators and hybrid entanglement generation between mechanical oscillators and telecom quantum emitters.
\end{abstract}

\newpage
\maketitle

\section*{Introduction}

Quantum networks, used to distribute quantum information over long distances between many physical nodes, hold great promise for the realization of networked quantum computation, quantum communication, as well as distributed quantum sensing~\cite{kimble_quantum_2008}. A high-fidelity interface between a long-lived quantum memory and optical photons for long-distance distribution of entanglement forms the fundamental building block of any practical quantum network~\cite{sangouard_quantum_2011,azuma_quantum_2023}. Such light-matter interfaces have been realized in various physical platforms ranging from atomic vapors~\cite{matsukevich_entanglement_2006,chou_measurement-induced_2005}, trapped ions~\cite{krutyanskiy_telecom-wavelength_2023}, individual trapped atoms~\cite{ritter_elementary_2012} to color center defects in solid-state crystals~\cite{pompili_realization_2021,hermans_qubit_2022}. Alternatively, entangled photon pairs from spontaneous parametric downconversion sources can be stored in the collective excitations of rare-earth ion doped crystals to distribute entanglement~\cite{usmani_heralded_2012,lago-rivera_telecom-heralded_2021}.

\begin{figure}[ht!]
	\includegraphics[width = 1.0 \linewidth]{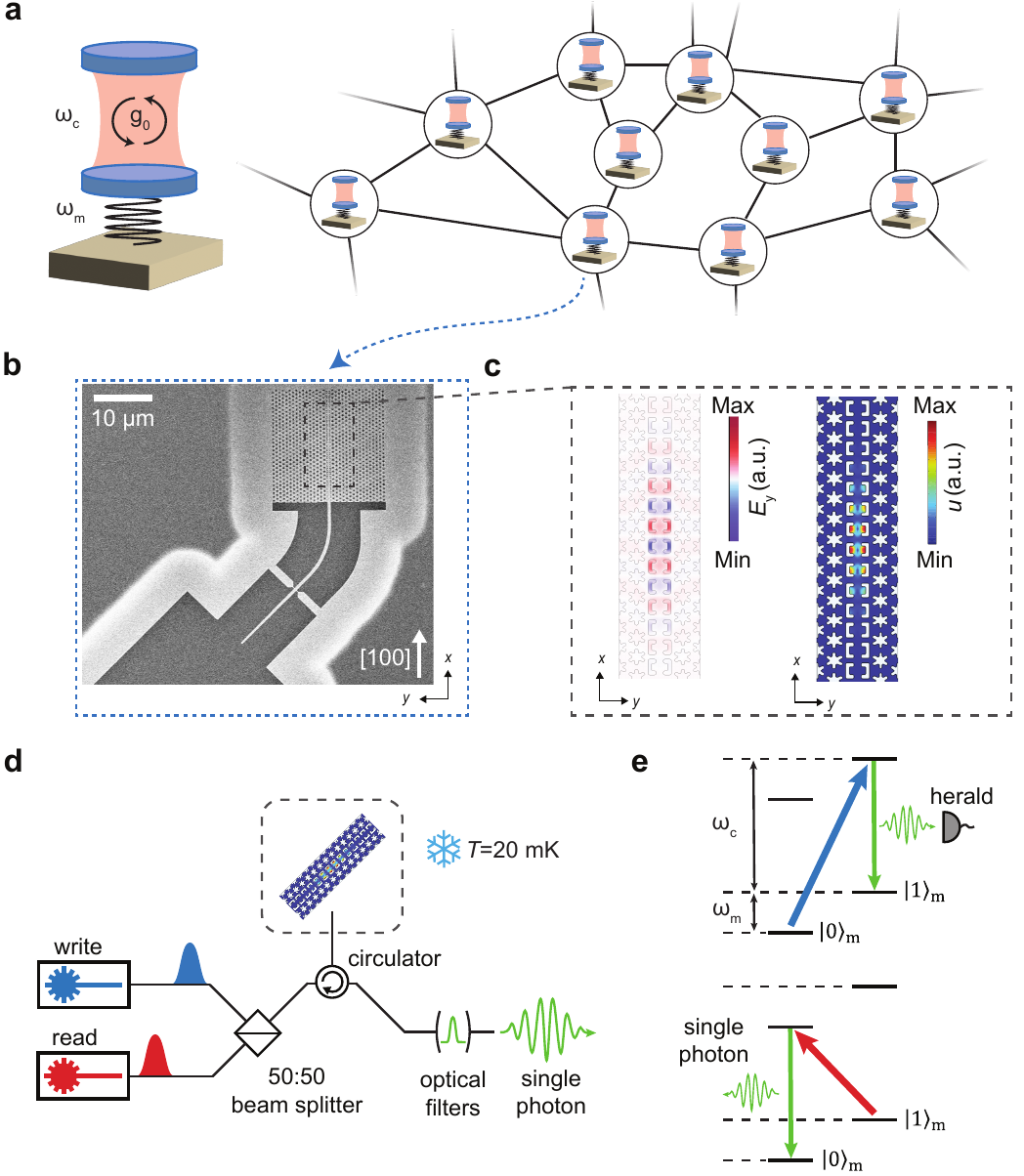} 
	\caption{\textbf{Quantum network based on optomechanical crystals.} \textbf{a} Schematic illustration of a quantum network consisting of cavity optomechanical systems as network nodes with optical cavity frequency $\omega_\mathrm{c}$, mechanical frequency $\omega_\mathrm{m}$, and single-photon optomechanical coupling strength $g_0$. \textbf{b} Scanning  electron microscope image of a 2D OMC device, which can form one of the telecom-wavelength quantum network nodes. The principal axis of the OMC cavity is aligned to the [100] direction of the silicon crystal lattice. \textbf{c} FEM simulations of the electric field (left) of the optical mode $E_\mathrm{y}$ at design wavelength $\lambda = \qty{1537.24}{nm}$ and the displacement field (right) of the mechanical mode at design frequency $\omega_\mathrm{m}/2\pi = \qty{10.18}{GHz}$. \textbf{d} Schematic illustration of the optical measurement setup. Write (read) laser pulses detuned to the blue (red) optomechanical sideband are sent to the OMC device inside a dilution refrigerator at base temperature $T=\qty{20}{mK}$ via lensed fiber coupling. Single photons created through optomechanical interaction are filtered from the reflected light using optical filters locked to the optical cavity resonance of the OMC. \textbf{e} Illustration of the optomechanical Stokes- and anti-Stokes scattering processes used for single-phonon generation and phonon-photon conversion. Top: photons from the blue-detuned write pulse at frequency $\omega_\mathrm{b} = \omega_\mathrm{c} + \omega_\mathrm{m}$ undergo a Stokes scattering process resulting in the probabilistic creation of a single photon at the optical cavity frequency $\omega_\mathrm{c}$ and a single phonon at the mechanical frequency $\omega_\mathrm{m}$. Detection of a single photon heralds the preparation of the mechanical mode in the Fock state $\ket{1}_\mathrm{m}$. Bottom: the red-detuned readout pulse at frequency $\omega_\mathrm{r} = \omega_\mathrm{c} - \omega_\mathrm{m}$ induces an anti-Stokes scattering process converting the single phonon in the mechanical mode to a single photon in the telecom band.}
	\label{Fig:1_OM_single_photon_source}
\end{figure}

In recent years, integrated optomechanical crystals (OMCs) have emerged as a promising and versatile platform for quantum technologies~\cite{aspelmeyer_cavity_2014,riedinger_non-classical_2016,barzanjeh_optomechanics_2022}. The flexible operating wavelength of OMCs -- including the telecom C-band where fiber transmission losses are minimized -- as well as the long lifetimes of their mechanical mode of up to $T_1\approx\qty{1}{s}$~\cite{maccabe_nano-acoustic_2020} and coherence times $T_2^*\approx\qty{100}{\micro s}$~\cite{wallucks_quantum_2020} make OMCs a natural candidate for storage and distribution of quantum information in long-distance quantum networks (see Fig.~\ref{Fig:1_OM_single_photon_source}a). Furthermore, phonons in the mechanical mode can be readily coupled to other quantum systems such as solid-state defects~\cite{kuruma_controlling_2025,joe_observation_2025}, quantum dots~\cite{spinnler_single-photon_2024}, and most prominently superconducting circuits~\cite{mirhosseini_superconducting_2020, forsch_microwave--optics_2020,van_thiel_optical_2023,meesala_quantum_2023,weaver_integrated_2024}, enabling hybrid quantum information architectures for quantum information processing and networked quantum computation~\cite{kurizki_quantum_2015}.

Using a heralded entanglement scheme based on the Duan-Lukin-Cirac-Zoller (DLCZ) protocol~\cite{duan_long-distance_2001}, entanglement between the mechanical modes of two OMCs has been demonstrated~\cite{riedinger2018remote} and such entanglement has been used as a resource to perform an optomechanical Bell test~\cite{marinkovic2018optomechanical} as well as optomechanical quantum teleportation~\cite{fiaschi_optomechanical_2021} -- demonstrating crucial steps towards quantum networks of mechanical oscillators. However, these initial demonstrations using one-dimensional nanobeam OMCs suffered from significant thermal noise due to optical absorption heating and weak thermal anchoring to the substrate~\cite{meenehan_pulsed_2015}, resulting in a relatively low purity of the optomechanically generated single photons~\cite{hong_hanbury_2017}, and thus inhibiting scaling up of the optomechanical DLCZ scheme to more complex quantum networks. To address this issue, quasi-two-dimensional (2D) OMCs have been developed, allowing for more efficient dissipation of generated thermal phonons into the cryogenic environment~\cite{Safavi-Naeini2014, ren_two-dimensional_2020, kersul_silicon_2023}. The improved thermal noise performance of such 2D OMCs has been verified in previous studies by measuring the thermal phonon occupancy of the mechanical mode~\cite{mayor_high_2025, sonar_high-efficiency_2025}. However, to use such devices in quantum information distribution as well as networked quantum computation, it is crucial to demonstrate the purity and coherence of optomechanically generated single photons. In particular, although previous experiments on entanglement of two mechanical oscillators imply the indistinguishability of optomechanically generated single photons~\cite{riedinger2018remote}, a direct demonstration of photon indistinguishability by two-photon interference has so far remained elusive due to prohibitively low optomechanical scattering rates limited by thermal noise.

In this work, we demonstrate heralded single-phonon generation and single phonon-photon conversion to telecom single-photons using integrated 2D OMCs with low thermal noise compatible with entanglement and quantum repeater schemes following the DLCZ protocol. We verify the low thermal noise of our device and demonstrate strong and robust quantum correlations between phonons and photons even at high mechanics-to-optics conversion efficiencies of up to $58\%$. We characterize the single-photon purity of the generated state through a Hanbury Brown-Twiss experiment. The obtained value $g^{(2)}(0)=0.35^{+0.10}_{-0.08}$ is the lowest measured for integrated OMC systems and violates the threshold of $g^{(2)}(0)=0.5$ for a genuine single-photon Fock state. The reduced thermal noise of our device enables operation at higher optomechanical scattering rates, allowing us to perform Hong-Ou-Mandel interference of subsequently emitted photons to quantify the coherence and indistinguishability of the generated single photons with a two-photon interference visibility of $V=0.52 \pm 0.15$. Lastly, we use two-photon interference to measure the temporal wavepacket envelope of the emitted photons, demonstrating narrow linewidths as low as $\qty{10}{MHz}$. The significant performance increase of 2D compared to 1D OMCs positions them as a promising platform to establish quantum networks of multiple mechanical oscillators. In addition, the narrow photon bandwidth as well as the (freely) designable operation wavelength make our system directly suitable for integration with narrow-linewidth telecom quantum emitters such as rare-earth ions~\cite{weiss_erbium_2021,gritsch2022narrow,ourari_indistinguishable_2023} or silicon T-centers~\cite{higginbottom_optical_2022,komza2024indistinguishable}, as well as telecom quantum memories based on ensembles of rare-earth ions~\cite{zhang2023, rancic_coherence_2018} or other optomechanical systems~\cite{kristensen_2024} in hybrid quantum networks.

\section*{Results}
\subsection*{Single photon generation based on quasi-2D optomechanical crystals}
Our device is a quasi two-dimensional suspended OMC structure (see Fig.~\ref{Fig:1_OM_single_photon_source}b) fabricated on a silicon-on-insulator material platform based on a design of~\cite{ren_two-dimensional_2020,kersul_silicon_2023}. The two-dimensional anchoring allows efficient dissipation of thermal phonons generated through optical absorption heating. The optical and mechanical modes of our structure (see Fig.~\ref{Fig:1_OM_single_photon_source}c) are confined along the $y$-direction by a snowflake crystal pattern exhibiting both photonic and phononic band gaps~\cite{Safavi-Naeini2014}. In between the snowflake areas, a periodic pattern of C-shaped holes modulated in size allows confinement along the $x$-direction. The $x$-direction of our device is oriented along the [100] crystal direction of the silicon lattice, which has been shown to lead to single-mode mechanical mode spectra robust against fabrication imperfections~\cite{kersul_silicon_2023}. Co-localization of the optical and mechanical mode in a small mode volume leads to optomechanical coupling with a measured (simulated) single-photon coupling strength $g_0/2\pi=\qty{1.0}{MHz}$ ($g_{0,\mathrm{sim}}/2\pi = \qty{997}{kHz}$) (see Methods). Our device operates far in the resolved-sideband regime of cavity optomechanics where $\kappa \ll \omega_\mathrm{m}$ with optical cavity linewidth $\kappa/2\pi = \qty{2.4}{GHz}$ and mechanical frequency $\omega_\mathrm{m}/2\pi = \qty{10.3699}{GHz}$.

We generate single photons through optomechanical heralded phonon generation and readout~\cite{riedinger_non-classical_2016}. A laser pulse detuned on the blue optomechanical sideband of the optical cavity resonance induces a two-mode squeezing interaction between the optical and mechanical mode (see Fig.~\ref{Fig:1_OM_single_photon_source}e). In the resulting Stokes scattering process, a single pump photon is probabilistically converted into a photon at the optical cavity resonance frequency and a phonon in the mechanical mode (see Fig.~\ref{Fig:1_OM_single_photon_source}e). The light coming back from the device is filtered by a series of optical filters to remove the strong blue-detuned pump and detect single optomechanically scattered photons on superconducting nanowire single-photon detectors (SNSPD). Detection of the optical photon from the Stokes process heralds the creation of a single-phonon Fock state $\ket{\psi_\mathrm{m}} \approx \ket{1}$. A second laser pulse detuned to the red optomechanical sideband transfers the phonon state onto the optical mode through a beam splitter interaction corresponding to an anti-Stokes scattering process (see Fig.~\ref{Fig:1_OM_single_photon_source}e). This mechanics-to-optics conversion process allows us to generate single photons after filtering out the strong red pump pulse. 

\subsection*{Thermal phonon occupancy and quantum cross-correlations}

\begin{figure}
	\centering
	\includegraphics[width = 1.0 \linewidth]{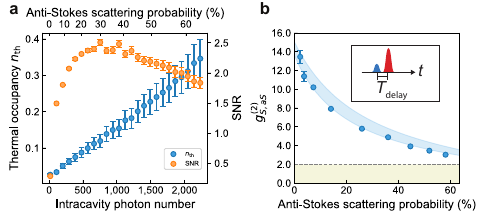}
	\caption{\textbf{Thermal performance and quantum cross-correlations.} \textbf{a} The blue dots show the thermal phonon occupancy $n_\mathrm{th}$ of the mechanical mode as a function of intracavity photon number (bottom) and anti-Stokes scattering probability (top). Error bars originate from errors in the calibration of the detection path efficiency (see Supplementary Information). The orange dots indicate the signal-to-noise (SNR) ratio in the conversion process. \textbf{b} Cross-correlation function $g^{(2)}_\mathrm{S, aS}$ between optomechanically scattered photons from the write and read pulse as a function of anti-Stokes scattering probability of the readout pulse. The insert shows the pulse sequence used for the cross-correlation measurement. A blue-detuned write pulse creates a single phonon, which is read out by a red-detuned read pulse after a delay time of $T_\mathrm{delay}=\qty{150}{ns}$. The pulse sequence is repeated with a repetition period of $T_\mathrm{rep}=\qty{10}{\micro s}$. The Stokes scattering probability of the write pulse is fixed at $p_\mathrm{S}=1.3\%$. The blue shaded area corresponds to the theoretically expected dependence $g^{(2)}_\mathrm{S,aS} = 1+e^{-T_\mathrm{delay}/\tau_\mathrm{m}}/(p_\mathrm{s} + n_\mathrm{th})$ (see Supplementary Information), where $n_\mathrm{th}$ is calibrated from the results in \textbf{a} and $\tau_\mathrm{m}=\qty{1.0}{\micro s}$ is the phonon lifetime of the mechanical mode (see Supplementary Information). The dashed horizontal line and shaded area underneath indicate the regime of classical correlations $g^{(2)}_\mathrm{S, aS} \leq 2$. The error bars are calculated from the photon counting statistics and correspond to the $68\%$ confidence interval of the binomial distribution.}
	\label{Fig:3_optomechanical_characterization}
\end{figure}

To verify the improved thermal anchoring of our device compared to conventional one-dimensional nanobeam OMCs, we measure the thermal phonon occupancy of the mechanical mode  $n_\mathrm{th}$ for varying intracavity photon number $n_\mathrm{c}$ in the optical cavity. We use $40$-ns-long optical pulses detuned to the red optomechanical sideband and measure the count rate of optomechanically scattered anti-Stokes photons $C_\mathrm{aS}$ after filtering out the strong pump pulse. By calibrating the efficiency of our detection path $\eta=0.05$ and the anti-Stokes scattering probability $p_\mathrm{aS}$ at various powers, we obtain the thermal phonon occupancy as $n_\mathrm{th} =C_\mathrm{aS}/(\eta p_\mathrm{aS})$, where $p_\mathrm{aS}$ is the anti-Stokes scattering probability at the respective pump power (see Methods). As shown in Fig.~\ref{Fig:3_optomechanical_characterization}a, the thermal phonon occupancy increases with increasing intracavity photon number, but remains close to the quantum ground state even at photon numbers exceeding $n_\mathrm{c} > 2,000$, corresponding to a readout probability in the anti-Stokes scattering process of $p_\mathrm{aS} > 60\%$. Compared to 1D structures, this represents a factor of three reduction of thermal occupancy of the mechanical mode at similar level of intracavity photon number~\cite{fiaschi_optomechanical_2021}. The highest anti-Stokes scattering probability achieved in these measurements is limited only by the power of our laser and residual losses in the optical setup. If we define the signal-to-noise ratio (SNR) as $\xi = p_\mathrm{as}/n_\mathrm{th}$, we obtain the highest SNR of $\xi=2.5$ at about $30\%$ of anti-Stokes scattering probabilities, a factor of $2.5$ increase compared to the highest value achieved in 1D structures~\cite{fiaschi_optomechanical_2021}. Based on the low thermal occupancy of our device, we demonstrate strong non-classical correlations between Stokes- and anti-Stokes-scattered photons detected on the SNSPDs during the write and read pulses~\cite{riedinger_non-classical_2016}. The measured correlations (see Fig.~\ref{Fig:3_optomechanical_characterization}b) exceed the classical limit by 51 standard deviations even at high anti-Stokes scattering probability up to $p_\mathrm{aS} = 58 \%$.

\begin{figure}
	\centering
	\includegraphics[width = 1.0 \linewidth]{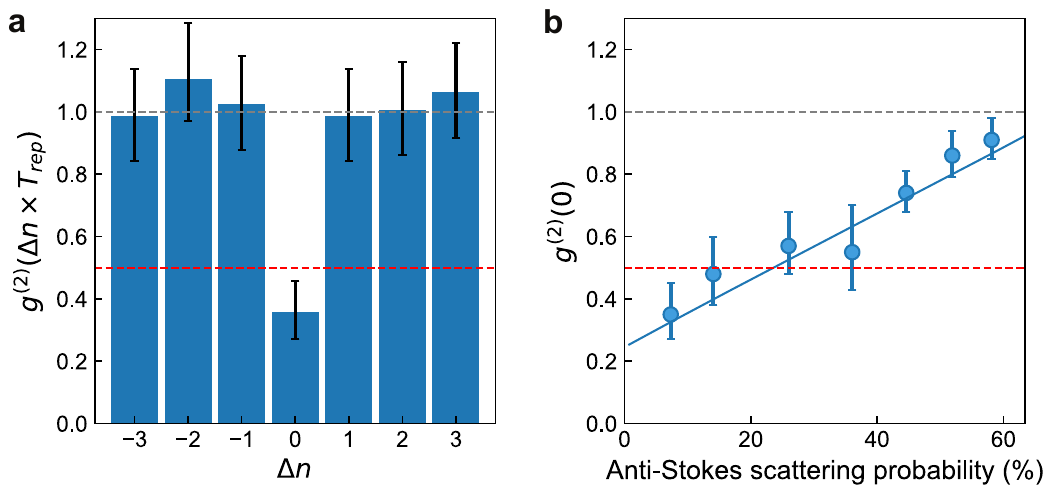}
	\caption{\textbf{Hanbury Brown-Twiss measurement.} \textbf{a} Measured second-order autocorrelation function $g^{(2)}$ of detection events from the read pulse conditioned on the detection of a Stokes-scattered photon during the blue-detuned write pulse. Each pulse sequence is labeled by a number $n$. Pulse sequences used for the  $g^{(2)}$ calculation are shifted by $\Delta n$. The Stokes-scattering (anti-Stokes-scattering) probability are $p_\mathrm{S}=1.3\%$ ($p_\mathrm{aS} = 7\%$) corresponding to pulse energies of $E_\mathrm{p,S}=\qty{0.1}{pJ}$ ($E_\mathrm{p,aS}=\qty{1.0}{pJ}$). \textbf{b} Second-order autocorrelation function $g^{(2)}(0)$ at fixed Stokes-scattering probability $p_\mathrm{S}=1.3\%$ as a function of anti-Stokes read out probability. Solid blue line shows the result of simulations of the quantum systems using the Python package QuTIP (see Supplementary Information)~\cite{johansson_qutip_2012, johansson_qutip_2013}. A value below unity demonstrates sub-Poissonian photon statistics (grey dashed line in \textbf{a} and \textbf{b}) whereas a value below $0.5$ unambiguously demonstrates a single-photon state (red dashed line in \textbf{a} and \textbf{b}). The error bars are calculated from the photon counting statistics using the exact binomial confidence interval (see Supplementary information). For all measurements, the optical pulse sequence is repeated with a repetition period of $T_\mathrm{rep}=\qty{10}{\micro s}$.}
	\label{Fig:4_hbt}
\end{figure}
\subsection*{Characterization of single photon purity}
The purity of the generated single photons can be characterized by performing a Hanbury Brown-Twiss (HBT) experiment on the read-out optical state conditioned on the detection of a Stokes-scattered photon during the write pulse. Figure~\ref{Fig:4_hbt}a shows the result of the HBT measurement for phonons read from the same ($\Delta n=0$) or different ($\Delta n \neq 0$) repetition periods (see Supplementary Information). We observe strong anti-bunching of the read out photons and determine a value of the conditional autocorrelation function of $g^{(2)}(0)=0.35^{+0.10}_{-0.08}$, which is significantly below the limit of $g^{(2)}(0)=0.5$ for a genuine single photon state~\cite{gerry2023introductory}. The measured value of $g^{(2)}(0)$ is mainly limited by residual absorption heating from the write and read pulse creating added thermal noise on the read out optical state. Dark counts account for $0.78\%$ of the total coincidences (see Supplementary Information). At higher anti-Stokes scattering probabilities, more thermal noise is added reducing the fidelity of the read out optical state (see Fig.~\ref{Fig:4_hbt}b). Nonetheless, we observe sub-Poissonian photon statistics with $g^{(2)}(0) < 1$ even for the highest anti-Stokes scattering probabilities used.

\begin{figure*}
	\centering
	\includegraphics[width = 1.0 \linewidth]{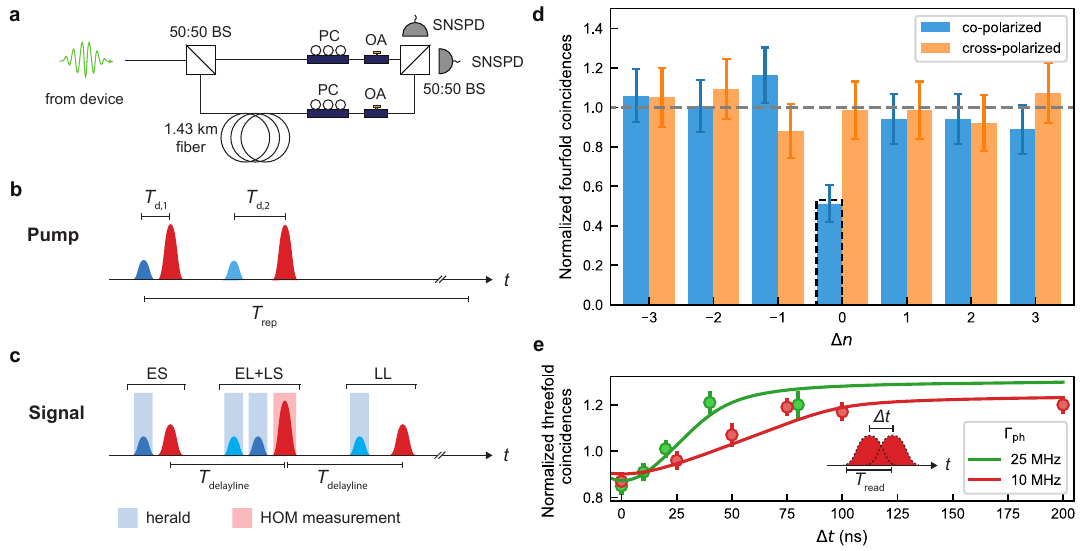}
	\caption{\textbf{Hong-Ou-Mandel interference.} \textbf{a} Unbalanced Mach-Zehnder interferometer used for Hong-Ou-Mandel (HOM) measurements. BS, beam splitter; PC, polarization controller; OA, optical attenuator; SNSPD, superconducting nanowire single-photon detector. \textbf{b} Pump pulse sequence used for HOM measurements. The delay between the blue and red pulses is $T_\mathrm{d,1}=\qty{105}{ns}$ and $T_\mathrm{d,2} =\qty{225}{ns}$. The delay between the two red pulses equals the time delay induced by the fiber delay line with $T_\mathrm{delayline}=\qty{7.146}{\micro s}$. \textbf{c} Schematic of the measured detection events on SNSPDs. Time bins are labeled according to when the photon was created (E, early; L, late) and which interferometer arm it passed through (S, short; L, long). Photons generated from the first (second) blue-detuned pump pulses are shown in dark (light) blue.	Simultaneous clicks during either combination of two dark and light blue-shaded time bins herald the generation of two phonons. The phonons are read out by the red pulses leading to two-photon interference during the time bin associated with the second red pulse (red shaded). \textbf{d} Number of four-fold coincidences measured on the two SNSPDs with co-polarized (blue) or cross-polarized (orange) arms of the interferometer during the same ($\Delta n = 0$) or different ($\Delta n \neq 0$) repetitions of the experiment normalized to the average value measured on the satellite peaks ($\Delta n \neq 0$). The error bars are calculated from the photon counting statistics and correspond to the $68\%$ confidence interval of the binomial distribution. The dashed bar at $\Delta n=0$ shows the value predicted from QuTiP simulations (see Supplementary Information). \textbf{e} Normalized number of threefold coincidences in co-polarized interferometer configuration during the same repetition of the experiment ($\Delta n = 0$) as a function of timing offset $\Delta t$ between the two red pulses for two different bandwidths $\Gamma_\mathrm{ph}$ of generated photons. The solid lines are fits to a phenomenological model based on the photon pulse shape (see Supplementary Information).}
	\label{Fig:5_HOM}
\end{figure*}
\subsection*{Hong-Ou-Mandel interference of optomechanically generated single photons}
Photon indistinguishability is crucial for many quantum information processing protocols including building quantum repeaters for long-distance quantum networks. We verify the coherence and indistinguishability of the single photons generated by our source by performing Hong-Ou-Mandel (HOM) interference. We pass two subsequently generated single photons through an unbalanced Mach-Zehnder interferometer with a $\qty{1.43}{km}$ fiber delay line in one arm corresponding to a time delay of $T_\mathrm{delayline} = \qty{7.146}{\micro s}$ (see Fig.~\ref{Fig:5_HOM}a). To generate two subsequent photons, we use the pulse sequence shown in Fig.~\ref{Fig:5_HOM}b.  Within one repetition period ($T_\mathrm{rep} = \qty{18}{\micro s}$), two pairs of blue and red detuned pulses (\qty{40}{ns}) are sent to the device to generate single phonons ($p_\mathrm{S} = 10~\%$) and read them out ($p_\mathrm{aS} = 45~\%$). The pulse energies of the write and readout pulses are $E_\mathrm{p,S}=\qty{1.1}{pJ}$  and $E_\mathrm{p,aS}=\qty{8.3}{pJ}$, respectively. The delay between the two red read out pulses in the two pulse groups is set to be equal to $T_\mathrm{delayline}$ leading to HOM interference at the second beam splitter (see Fig.~\ref{Fig:5_HOM}c). The long time delay $T_\mathrm{delayline} \gg \tau_\mathrm{m}$ between subsequently generated photons allows the mechanical mode to thermalize to the cryogenic environment before each photon is generated. For the HOM measurement, we measure four-fold coincidences between two clicks from the blue write pulses and two clicks from the red read out pulses. When the two interferometer arms are co-polarized, the photons arriving at the beam splitter during the same repetition are indistinguishable resulting in the characteristic dip in coincidence detection events (see Fig.~\ref{Fig:5_HOM}d). As a control experiment, we repeat the same measurement with cross-polarized arms of the interferometer observing no dip in coincidence events, as expected. From the two measurements, we calculate the HOM interference visibility $V_\mathrm{raw}= 0.48 \pm 0.14$. After correcting the power imbalance of the two arms (see Supplementary Information), we obtain a HOM visibility of $V= 0.52 \pm 0.15$. The visibility is reduced compared to the case of ideal single photons ($V=1$) due to the added thermal component of the optical state. We model the impact of the thermal component on the HOM visibility through numerical simulations and determine a simulated visibility of $V_\mathrm{sim}=0.53$ (see Supplementary Information) in good agreement with the measured value. Although, the HOM visibility does not violate the theoretical bound for non-classical states $V>0.5$, the observation of HOM interference nevertheless demonstrates the coherence and indistinguishability of optomechanically generated single-photons generated more than $\qty{7}{\micro s}$ apart in time.
\subsection*{Measurement of photon temporal wavepacket and bandwidth}
Finally, we use two-photon interference to measure the temporal wavepacket shape and thus the bandwidth of optomechanically generated single-photons, offering insight for interfacing with other components (e.g. quantum memories) in quantum networks. For short readout pulse durations used in our experiment $T_\mathrm{read} \ll 2\pi/\Gamma_\mathrm{om}$, the photon bandwidth is expected to closely follow the readout pulse wavepacket shape~\cite{galland_heralded_2014, riedinger2018single}. Here, $T_\mathrm{read}$ is the readout pulse length and $\Gamma_\mathrm{om} = \Gamma_\mathrm{m} + 4 n_\mathrm{c} g_0^2/\kappa$ is the optomechanically enhanced mechanical linewidth with the intrinsic mechanical linewidth $\Gamma_\mathrm{m}$. The photon bandwidth $\Gamma_\mathrm{ph}$ is given as the inverse of the temporal wavepacket shape of the generated photon. We verify the temporal wavepacket shape and hence the photon bandwidth by offsetting the two readout pulses by a time delay $\Delta t$, which reduces the overlap of two photons arriving at the beamsplitter leading to reduced HOM interference. To increase the statistics, we measure three-fold coincidences between one click from the blue write pulses and two clicks from the red readout pulses. Hence, only one single photon is generated while the other red-detuned pulse reads out an unheralded mechanical thermal state. This HOM interference between a single photon and a thermal state leads to a reduced depth of the observed HOM dip. Figure~\ref{Fig:5_HOM}e shows the normalized number of threefold coincidences as a function of timing offset $\Delta t$ for photon bandwidths of $\qty{25}{MHz}$ and $\qty{10}{MHz}$, corresponding to read out pulse lengths of $\qty{40}{ns}$ and $\qty{100}{ns}$, respectively. The HOM dip persists up to longer time delays for longer readout pulse length showcasing the tunability of the temporal shape of the photon wavepacket in agreement with a phenomenological model (solid lines). We note that as the pulse offset is increased, the normalized number of coincidences increases above the value of unity as a consequence of the bunched photon statistics of the thermal state (see Supplementary Information).

\section*{Discussion}

Our work demonstrates heralded single phonon creation and phonon-photon conversion at telecom wavelengths based on 2D OMC platform, paving the way for building quantum network based on nano-structured OMCs. In contrast to previous experiments, in which the measurement of the photon autocorrelation function was limited by thermal noise to greater than the threshold of $g^{(2)}(0)=0.5$ of a genuine single-photon Fock state, the reduced thermal noise of our 2D OMC enables single-photon emission with $g^{(2)}(0)=0.35^{+0.10}_{-0.08}$ in an HBT measurement, unambiguously demonstrating the single-quantum nature of the emitted state. Furthermore, the improved thermal performance enables operating the device at high optomechanical scattering probabilities and thus enables the realization of a HOM experiment, which requires the detection of four-photon coincidence events. This has so far proven elusive due to low rates caused by limited optomechanical scattering probability. Our experiment confirms the indistinguishable character of the generated photons through the observation of HOM interference with visibility $V= 0.52 \pm 0.15$, and by introducing a delay line for one photon of $T_\mathrm{delayline} = \qty{7.146}{\micro s}$ further demonstrates that the emitted photons are coherent over this timescale. Since the typical amplitude of mechanical frequency jittering which ultimately limits photon coherence $\Delta f_\mathrm{m}\approx\qty{10}{kHz}$ is small compared to the emitted photon bandwidth, we expect the coherence of photons emitted from our source to be preserved for long periods of time, ideal for long-distance quantum network applications.

Our 2D OMC device also showcases narrow photon linewidth as low as $\qty{10}{MHz}$ -- essential for interfacing with telecom quantum memories, limited only by the optical pulse length used for phonon-photon conversion. While longer optical readout pulse lengths would directly enable narrower optical linewidth, ultimately only limited by the intrinsic mechanical linewidth $\Gamma_\mathrm{m}/2\pi = \qty{119}{kHz}$ (see Supplementary Information), this would lead to an accumulation of more thermal phonons in the mechanical mode~\cite{meenehan_pulsed_2015} and thus currently still impede measurements in the quantum regime. Notably, compared to other systems that generate single photons at telecom C-band, the bandwidth of our system already surpasses the narrowest linewidths achieved with telecom quantum dots ($>$\SI{100}{MHz}~\cite{holewa_high-throughput_2024}), as well as with silicon-based on-chip sources based on spontaneous four-wave mixing ($\approx\qty{30}{MHz}$~\cite{chen_2024}). Moreover, since the temporal wavepacket shape of the emitted photons from our OMC device is determined by the pulse shape of the optical readout pulse, pulse shaping of the readout pulse can be used to generate shaped single photons. Shaped single-photons are essential to realize quantum interference with single photons from other quantum systems, such as rare-earth ions~\cite{weiss_erbium_2021,gritsch2022narrow,ourari_indistinguishable_2023} or quantum dots~\cite{holewa_high-throughput_2024}.

While thermal noise currently still limits the single-photon purity and HOM visibility as well as the attainable narrowest optical linewidths, further improvements to the design to reduce pump-induced heating through the use of evanescently coupled optomechanical cavities or non-suspended OMCs exist~\cite{kolvik_clamped_2023, sonar_high-efficiency_2025}, which will allow to further boost the purity and rate of generated single photons, enabling a more efficient quantum network of OMCs. The lifetime of the phonon mode in the device used in this work is chosen to be only $\tau_\mathrm{m}=\qty{1}{\micro s}$ to increase the repetition rate of the experiment. However, by engineering the phononic band structure of the OMC geometry, we have already fabricated devices with long phonon lifetimes up to $\tau_\mathrm{m}=\qty{9}{ms}$ (see Supplementary Information). This long phonon lifetime combined with high purity and long coherence time of optomechanically generated single photons at telecom wavelength, position 2D OMCs as a promising platform for the realization of long-distance quantum networks following the DLCZ protocol~\cite{duan_long-distance_2001}. With our current performance, we estimate that heralded entanglement between two 2D OMCs embedded in a phase-stabilized Mach-Zehnder interferometer can be generated at a heralding rate of $\qty{100}{Hz}$ and verified at a total event rate of $\qty{2.9e3}{h^{-1}}$ -- an improvement by more than two orders of magnitude over previous demonstrations~\cite{riedinger2018remote} (see Supplementary Information). The heralding rate is already surpassing demonstrations of heralded entanglement generation using nitrogen vacancy centers ($\qty{39}{Hz}$~\cite{humphreys_deterministic_2018}). In a next step, well-established technical improvements of optical setup efficiency will further boost the attainable entanglement generation rates to $\qty{300}{Hz}$ comparable with recent experiments based on atomic ensembles ($\qty{280}{Hz}$~\cite{yu_entanglement_2020}). Importantly, the generated single-photons in~\cite{humphreys_deterministic_2018} and~\cite{yu_entanglement_2020} are, unlike for our device, not at telecom wavelength and hence for any practical quantum network application wavelength conversion would reduce the attainable entanglement rate significantly. Further reduction of thermal noise by evanescent light coupling~\cite{sonar_high-efficiency_2025} will allow an additional boost of the entanglement rates to $\qty{1.2}{kHz}$ (see Supplementary Information), on par with leading quantum networking platforms based on cavity-enhanced parametric-down-conversion sources and atomic frequency comb quantum memories in solid-state crystals ($\qty{1.4}{kHz}$~\cite{lago-rivera_telecom-heralded_2021}), and in an on-chip platform compatible with scalable CMOS fabrication~\cite{benevides_ultrahigh_q_2017}. These rate improvements will enable scale-up of these rudimentary quantum networks by demonstrating entanglement swapping between multiple pairs of entangled mechanical oscillators.

Finally, low-noise 2D OMCs are ideally suited to explore hybrid quantum network architectures interfacing different physical platforms:\ the narrow linewidth and by-design controllable wavelength of optomechanically generated single-photons allow for hybrid entanglement creation through interference of single photons from an OMC and telecom quantum emitters, such as rare-earth ions~\cite{weiss_erbium_2021,gritsch2022narrow,ourari_indistinguishable_2023} or silicon T-centers~\cite{higginbottom_optical_2022,komza2024indistinguishable}. Alternatively, single-photons from OMC devices can be stored in telecom quantum memories with narrow acceptance bandwidth, such as those based on ensembles of rare-earth ions~\cite{zhang2023, rancic_coherence_2018} or other optomechanical systems~\cite{kristensen_2024}. A second unique property of these systems is that phonons can be coupled directly to a large variety of other quantum systems, for example to color centers~\cite{kuruma_controlling_2025,joe_observation_2025} or quantum dots~\cite{spinnler_single-photon_2024} through strain interaction. When combined with a piezomechanical element, high-efficiency microwave-to-optics conversion is also directly attainable, where the low thermal noise demonstrated in our device will enable coherent, low-noise on-demand entanglement between microwave superconducting quantum circuits, critical for future quantum operations involving distant quantum processors in networked quantum computing~\cite{mirhosseini_superconducting_2020, forsch_microwave--optics_2020,van_thiel_optical_2023,meesala_quantum_2023,weaver_integrated_2024,weaver_scalable_2025}.

\section*{Methods}
\subsection*{Nanofabrication}
\label{nanofab}
The quasi-2D optomechanical crystal structures are fabricated from a silicon-on-insulator wafer with a $\qty{250}{nm}$ device layer. Devices are patterned using electron beam lithography and HBr/Ar reactive ion etching. After etching, we perform cleaning using a piranha solution. The devices are released by wet etching of the buried oxide layer using hydrofluoric acid ($40\%$).

\subsection*{Calculation of $g_0$}
\label{g0}

We calibrate the vacuum optomechanical coupling rate $g_0$ by measuring the optomechanical scattering probabilities with calibrated input optical power and setup efficiency. We send either blue or red-detuned optical pulses with $\qty{40}{ns}$ pulse length to the device and measure the resulting count rates on the SNSPDs. The expected count rates for the Stokes and anti-Stokes process, $C_\mathrm{S}$ and $C_\mathrm{aS}$, are given by
\begin{align}
\label{asymmetry_count_rates_S}
    C_\mathrm{S} &= (\eta_1 + \eta_2) p_\mathrm{S} (1+n), \\
    C_\mathrm{aS} &= (\eta_1 + \eta_2) p_\mathrm{aS} n,\label{asymmetry_count_rates_aS}
\end{align}
where $\eta_1$ and $\eta_2$ are the efficiencies of the optical paths for SNSPD 1 or 2, respectively, and $n$ is the thermal phonon occupancy of the mechanical mode. The scattering probabilities for the Stokes and anti-Stokes process $p_\mathrm{S}$ and $p_\mathrm{aS}$ are given by~\cite{riedinger2018single}
\begin{align}
\label{asymmetry_scattering_probs_S}
    p_\mathrm{S} &= \exp \left[\frac{\kappa_\mathrm{e}}{\Delta^2 + \kappa^2/4} \frac{g_0^2 \kappa}{(\Delta - \omega_\mathrm{m})^2 + \kappa^2/4} N_\mathrm{p} \right] -1, \\ \label{asymmetry_scattering_probs_aS}
    p_\mathrm{aS} &= 1-\exp \left[-\frac{\kappa_\mathrm{e}}{\Delta^2 + \kappa^2/4} \frac{g_0^2 \kappa}{(\Delta + \omega_\mathrm{m})^2 + \kappa^2/4} N_\mathrm{p} \right],
\end{align}
where $\kappa$ and $\kappa_\mathrm{e}$ are the total optical linewidth and extrinsic optical coupling rate, and $N_\mathrm{p}$ is the number of photons in the excitation pulse.

For low optical power and laser pulses resonant with the blue or red optomechanical sideband ($\Delta = \pm \omega_\mathrm{m}$), we can approximate the exponential in the Stokes scattering probability in Eq.~\eqref{asymmetry_count_rates_S}. Furthermore, at low optical powers the thermal phonon occupancy is small so that $n \ll 1$ and thus $n+1 \approx 1$ in Eq.~\eqref{asymmetry_count_rates_S}. We obtain for the vacuum optomechanical coupling rate
\begin{align}
\label{g0}
    g_0 = \left( \frac{C_\mathrm{S}}
    {\eta_1 + \eta_2 }
    \frac{\omega_\mathrm{m}^2 + \kappa^2/4}{4 N_\mathrm{p}} \frac{\kappa}{\kappa_\mathrm{e}} \right)^{1/2}.
\end{align}
From the measured count rate of the Stokes process $C_\mathrm{S}$, we obtain $g_0/2\pi = \qty{1.0}{MHz}$.

\subsection*{Measurement of the thermal phonon occupancy from the readout pulse}
\label{n_th_measurement}

We can determine the phonon occupancy $n$ by detuning the laser on the red sideband of the optical cavity and measuring the click rates of optomechanically scattered photons in the anti-Stokes process. From Eq.~\eqref{asymmetry_count_rates_aS}, we calculate the thermal phonon occupancy as
\begin{align}
\label{n_th}
    n = \frac{C_\mathrm{aS}}{(\eta_1 + \eta_2) p_\mathrm{aS}}.
\end{align}
For each laser power used in the experiment, we use the previously calculated value of $g_0$ and Eq.~\eqref{asymmetry_scattering_probs_aS} to calculate the corresponding anti-Stokes scattering probability $p_\mathrm{aS}$. We estimate the systematic error of the measured efficiency of the line to be $15\%$, which results in an increased error bar in the measured thermal occupancy $n_\mathrm{th}$ as the measured thermal phonon number increases. This systematic error arises from fiber optical connectors in our optical detection path, which need to be disconnected and reconnected to measure the path efficiency. The losses of each fiber optical connector vary each time a connector is reconnected. We calibrate this error by performing the same calibration of the detection path efficiency multiple times. From the standard deviation of the measured efficiency, we estimate the systematic error.

\medskip

\section*{Data availability}
Source data for the plots are available on \href{https://zenodo.org/records/18030151}{Zenodo}.

\section*{Acknowledgements}
We would like to thank Radim Filip for helpful discussions. We further acknowledge assistance from the Kavli Nanolab Delft. This work is financially supported by the European Research Council (ERC CoG Q-ECHOS, 101001005) and is part of the research program of the Netherlands Organization for Scientific Research (NWO), supported by the NWO Frontiers of Nanoscience program, as well as through a Vrij Programma (680-92-18-04) grant. C.M.K., R.B, and T.P.M.A acknowledge support by São Paulo Research Foundation (FAPESP) through grants 18/15580-6, 18/25339-4, 19/01402-1, 20/06348-2, 22/14273-8, and Coordena\c{c}\~{a}o de Aperfei\c{c}oamento de Pessoal de N\'{i}vel Superior - Brasil (CAPES) (Finance Code 001), and Financiadora de Estudos e Projetos (Finep).

\section*{Author contributions}
L.C., A.R.K. and S.G. devised and planned the experiment. C.M.K., L.C. and R.B. worked on the device simulation and design. L.C. and A.R.K. fabricated the device. C.M.K. assisted in the device characterization. L.C. and A.R.K. executed the experiments and performed data analysis with assistance from Y.Y. A.R.K., L.C. and S.G. wrote the manuscript with input from all authors. S.G. and T.P.M.A. supervised the project.\\

\section*{Competing interests}
The other authors declare no competing interests.

\setcounter{figure}{0}
\renewcommand{\thefigure}{S\arabic{figure}}
\setcounter{equation}{0}
\renewcommand{\theequation}{S\arabic{equation}}

\clearpage
\newpage

\begin{center}
	\textsc{\Large Supplementary Information} 
\end{center}
\label{SI}

\subsection{Finite-element simulations}
\label{fem_simulations}
\begin{figure*}
	\centering
	\includegraphics[width = 18cm]{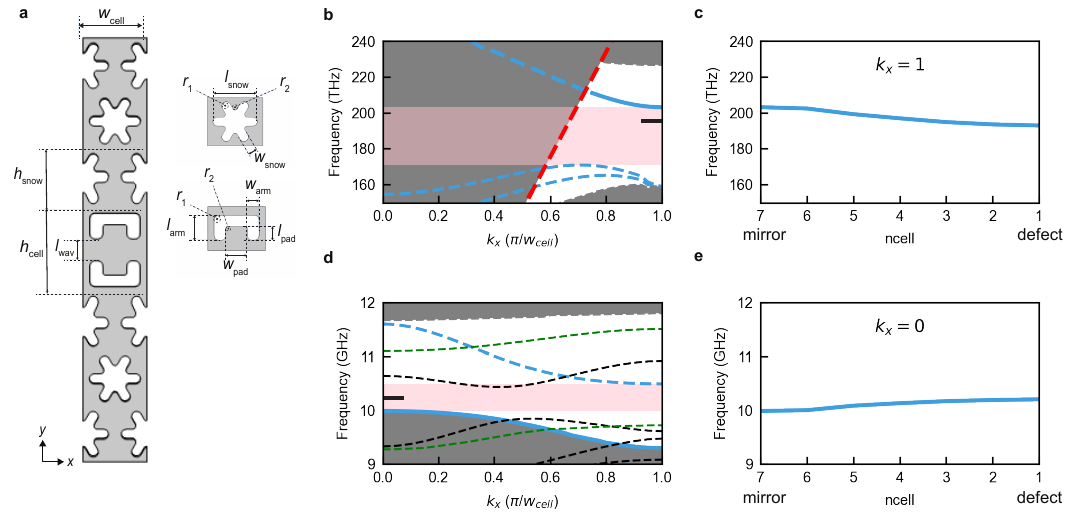}
	\caption{\textbf{Finite element simulation of the unit cell.} \textbf{a} Schematic of the quasi-2D optomechanical crystal mirror unit cell consisting of both snowflakes and C-shape holes, $w_\mathrm{cell} = \qty{502} {nm} $ is the unit cell width along $x$ direction. The parameters for mirror unit cells are:\ $\left(l_\mathrm{snow}, w_\mathrm{snow}\right) = \left(201, 80.4\right) $ nm, $\left(l_\mathrm{arm}, l_\mathrm{pad}, w_\mathrm{arm}, w_\mathrm{pad}\right) = \left(203, 112, 104, 188\right) $ nm. For the rows of snowflakes which are closest to C-shapes, the length of the second half of the snowflake holes along $y$ direction is scaled by $1.2$. \textbf{b} and \textbf{d} show the photonic and phononic band diagrams of the mirror unit cell, respectively. The shaded grey areas represent modes of continuum. In the photonic band diagram, only transverse-electric (TE) modes are shown since they are predominantly guided in the silicon slab with a thickness of $\qty{250}{nm}$. The dashed red line represents the light cone. The solid blue lines on the band gap edges are the C-shape modes of interest. The dashed blue lines show other guided modes in the structure. In the phononic band diagram, the dashed black lines are for modes with symmetry group $\sigma_z = -1 $, the dashed green lines are for modes with symmetry group $\left(\sigma_z = +1, \sigma_y = -1\right)$. The pink shaded areas indicate the optical and phononic band gaps formed from such mirror cells. The black ticks indicate the optical and mechanical frequency of the whole device, which are inside the band gaps. \textbf{c} and \textbf{e} indicate how the band edge modes of interest shift in frequency when we change the parameters from mirror cell (index $ncell = 7$) to defect center cell (index $ncell = 1$).}
	\label{Fig:S6_comsol_simulations}
\end{figure*}

The quasi-two-dimensional (2D) OMCs used in our work were first introduced in~\cite{Safavi-Naeini2014} and later optimized in~\cite{ren_two-dimensional_2020,kersul_silicon_2023}. The general idea is to incorporate a line defect of C-shaped holes replacing the snowflake lattice~\cite{ren_two-dimensional_2020}. With careful design, such structures can realize co-localization of both optical and mechanical modes. The effective mirror unit cell consists of both snowflakes and C-shape holes in the center. Figure~\ref{Fig:S6_comsol_simulations} shows the finite-element simulation of such a mirror unit cell when Floquet periodic boundary conditions are applied. The pink shaded areas in Fig.~\ref{Fig:S6_comsol_simulations}b and d indicate the photonic and phononic bandgaps, extending from $\qty{170.8}{THz}$ to $\qty{203}{THz}$, and from $\qty{9.99}{GHz}$ to $\qty{10.49}{GHz}$, respectively. The black ticks indicate the simulated optical resonance with $\omega_\mathrm{c}/2\pi = \qty{195.02}{THz}$ and mechanical resonance with $\omega_\mathrm{m}/2\pi = \qty{10.18}{GHz}$ for the whole device, both of which are inside the bandgaps. The solid blue lines at the bandgap edges indicate the modes which mainly live in the C-shape region as our modes of interest. Specifically for the phononic modes, only modes with symmetry group $\left(\sigma_z = +1, \sigma_y = +1\right)$ couple strongly to the optical mode of interest.

Similar to 1D OMCs, a defect region is created by adiabatically tapering the parameters of the C-shape holes. The whole defect region consists of $7\times 2 = 14$ cells. The transition function used between the mirror unit cell and defect center cell is~\cite{kersul_silicon_2023}:
\begin{align}
\label{mirror_defect_transition_function}
    y_n = y_\mathrm{min} + (y_\mathrm{max} - y_\mathrm{min})\cdot e^{-\frac{9\cdot(n - n_\mathrm{def})^2}{2\cdot n_\mathrm{def}^2}} 
\end{align}
where $n$ is the index of the defect cell ranging from $7$ to $1$, $n_\mathrm{def} = 7$ represents the number of cells in half of the defect region. The defect is created by specifically changing parameters $l_\mathrm{arm}$, $l_\mathrm{pad}$, and $w_\mathrm{arm}$, $w_\mathrm{pad}$. In the transition function above, $y_\mathrm{max}$ represents the values of said parameters in the mirror cell ($n = 7$) and $y_\mathrm{min}$ represent the values in the defect center cell ($n = 1$).

The remaining design parameters are provided below in Table~\ref{tab:design_parameters}:
\begin{table}[htbp]
  \centering
  \caption{Design Parameters}
    \begin{tabular}{cc}
        \hline
          Parameters         & Value ($\qty{}{nm}$)  \\ 
        \hline
          $h_\mathrm{cell}$        & 721  \\
          $h_\mathrm{snow}$        & 435 \\
          $l_\mathrm{wav}$         & 171.5      \\
          $l_\mathrm{arm, max}$      & 203       \\
          $l_\mathrm{arm, min}$      & 191.6    \\
          $w_\mathrm{arm, max}$   & 104          \\
          $w_\mathrm{arm, min}$   & 82            \\
          $l_\mathrm{pad, max}$      & 112       \\
          $l_\mathrm{pad, min}$      & 108       \\
          $w_\mathrm{pad, max}$   & 188      \\
          $w_\mathrm{pad, min}$   & 168      \\ 
          $r_1$   & 40              \\
          $r_2$   & 25             \\          
        \hline
    \end{tabular}
  \label{tab:design_parameters}
\end{table}
\newline
Alternative designs for two-dimensional OMC structures have recently been demonstrated~\cite{mayor_high_2025}. The choice of the ``boomerang" unit cell in~\cite{mayor_high_2025} was mainly motivated by bringing down the mechanical frequency ($\sim$$\SI{7.4}{GHz}$) to reduce the complexity of coupling to superconducting circuits. In addition, the boomerang unit cell has a larger filling factor ($74.5\%$) than the snowflake unit cells ($64.1\%$)~\cite{mayor_high_2025}, which could be beneficial for better thermalization. However, this is at the cost of smaller optical and mechanical band gaps from the waveguide unit cell, which are $\SI{27}{THz}$ wide for optical band gap and $\SI{0.15}{GHz}$ wide for mechanical band gap, compared to $\SI{32}{THz}$ optical and $\SI{0.5}{GHz}$ mechanical band gaps for the snowflake + Cshape unit cell used in our work. Moreover, the mechanical band of interest in~\cite{mayor_high_2025} is very flat compared to the snowflake + Cshape structure, which makes the former more susceptible to fabrication imperfections~\cite{ren_two-dimensional_2020}.

\subsection{Optical setup}
\label{optical_setup}

\begin{figure}
	\centering
	\includegraphics[width = 8.8cm]{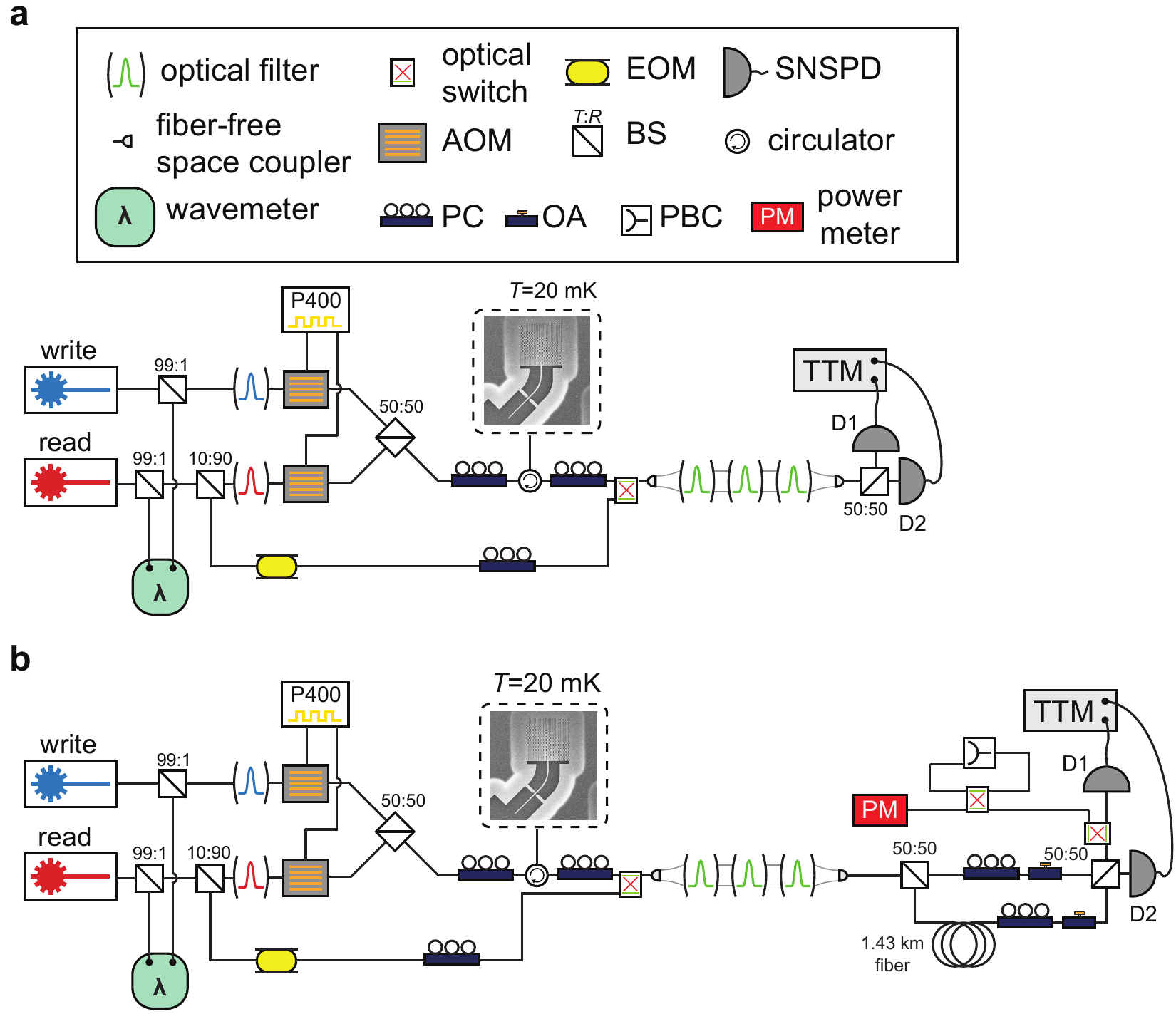}
	\caption{\textbf{Optical setup.} \textbf{a} Detailed scheme of the optical setup used for pulsed optomechanical measurements. BS, beam splitter with transmission (reflection) coefficient $T$ ($R$); AOM, acousto-optic modulator; EOM, electro-optic modulator; TTM, time-tagging module; PC, polarization controller; OA, optical attenuator; PM, optical power meter; PBC, polarizing beam combiner. \textbf{b} Modified scheme for HOM measurements.}
	\label{Fig:S3_optical_setup}
\end{figure}

Figure~\ref{Fig:S3_optical_setup}a shows the full experimental setup used to perform pulsed optomechanical measurements. Two tunable continuous-wave external diode lasers are locked to the blue and red optomechanical sideband of the optical cavity respectively. Both lasers are stabilized by active feedback using a wavemeter. We note that our laser stabilization is performed in an open-loop configuration with respect to the optical resonance of the OMC. This is possible because of the relatively broad optical linewidth $\kappa/2\pi = \qty{2.4}{GHz}$ and the consequently very stable optical resonance frequency. We did not observe any significant drifts of the optical cavity resonance frequency over the measurement time of three months.

The light from both lasers is filtered using one fiber-coupled filter cavity ($\qty{50}{MHz}$ linewidth) to suppress Gigahertz frequency laser noise. Pulses are created using $\qty{110}{MHz}$ acousto-optic modulators (AOMs) gated by a P400 pulse generator. The two paths for pulse generation are combined using a fiber beam splitter and routed to the OMC device inside the dilution refrigerator at base temperature $T=\qty{20}{mK}$ via a fiber-based circulator. The light coming back from the device is filtered using three consecutive home-built free-space Fabry-P\'erot cavities (each with linewidth $\qty{150}{MHz}$)~\cite{Vlassov_Filter_Cavity_Design_2022}. The filter cavities are locked to the frequency of the optical cavity of the OMC device, suppressing the pump light by $\qty{110}{dB}$. Locking is performed every $\qty{10}{s}$ during which the experiment is paused and continuous wave light is sent to the filter cavities via a fiber-based MEMS optical switch. Each locking procedure takes approximately $\qty{1.5}{s}$. The filtered signal from the device containing the optomechanically scattered photons is detected on two superconducting nanowire single-photon detectors (SNSPDs) with dark count rate $\sim$$\qty{30}{Hz}$. The electronic signal from the SNSPDs is amplified and recorded by a time-tagging module (TTM).

We calibrate the efficiencies throughout our optical setup, such as lensed fiber to device waveguide coupling ($\eta_\mathrm{fc}=0.5$), the coupling between the waveguide and the optical cavity ($\kappa_\mathrm{e}/\kappa = 0.45$), optical losses in the detection filter setup ($\eta_\mathrm{filters}=0.4$) and the efficiencies of the two SNSPDs ($\eta_\mathrm{SPD1}=0.61$, $\eta_\mathrm{SPD1}=0.55$). The fiber coupling efficiency is measured by measuring the input and output power of the fiber circulator in front of the device. This results in a total detection path efficiency of $\eta=0.05$.

Figure~\ref{Fig:S3_optical_setup}b shows the modified experimental setup of Hong-Ou-Mandel interference measurements. After the optical filters the light is sent through a Mach-Zehnder interferometer with a delay line of $\qty{1.43}{km}$ length in one arm. Single-photon detectors are placed at the two outputs of the interferometer. To maximize the interference visibility, the polarization of the two interferometer arms have to be aligned. For polarization alignment, we place two optical switches at one of the interferometer outputs, which can be used to send the light to an optical power meter (PM) and optionally through a polarizing beam combiner (PBC). This PBC features two polarization maintaining fibers at the input which can be used as a reference for polarization alignment. Furthermore, optical attenuators (OA) are used to balance the optical power in both arms. In the following, we describe the detailed procedure for polarization alignment.

\begin{itemize}
\item \textbf{Co-polarized interferometer arms:} We first minimize the transmission of the long interferometer arm via the OA and send the output light of the interferometer through the PBC. We use the polarization controller (PC) in the short arm to minimize the signal on the PM. We fully attenuate the transmission through the short arm and fully unattenuate the transmission through the long arm. We use the PC in the long arm to again minimize the signal on the PM. We send the output light from the interferometer directly to the PM without going through the PBC by switching the optical switch and measure the output power $P_\mathrm{long}$ of the long arm. To balance the optical power in the two arms, we fully attenuate the transmission of the long arm again and unattenuate the transmission of the short arm until the measured power $P_\mathrm{short}=P_\mathrm{long}$. Afterwards, we fully unattenuate the transmission of the long arm. Light passing through the two arms has the same polarization at the second 50:50 BS (co-polarized).
\item \textbf{Cross-polarized interferometer arms:} The procedure to align the two inteferometer arms to have orthogonal polarization at the second 50:50 BS (cross-polarized) is identical to the alignment procedure for co-polarized measurement except that in between the polarization alignment steps of the two arms we manually change the input fiber of the PBC used as polarization reference. As the two input fibers of the PBC have orthogonal polarizations, after this procedure the two interferometer arms are cross-polarized.
\end{itemize}

\subsection{Extended device characterization}
\label{extended_device_characterization}

\subsubsection*{Mechanical linewidth measurement}
Optomechanically induced transparency (OMIT), resulting from the quantum destructive interference between anti-Stokes scattered field and a weak intracavity pump field~\cite{weis2010optomechanically}, can be used to measure the mechanical spectrum as well as mechanical decay rate ($\Gamma_\mathrm{m}$). The induced transparency window (or OMIT linewidth) scales linearly with intracavity photon number $n_c$. When extrapolating to zero photon number, we obtain $\Gamma_\mathrm{m}/2\pi = \qty{119.7}{kHz} $, which matches well with further characterization of the mechanical life time (see below).

\subsubsection*{Mechanical lifetime measurement}

\label{lifetime_measurements}

\begin{figure}
	\centering
	\includegraphics[width = 8.8cm]{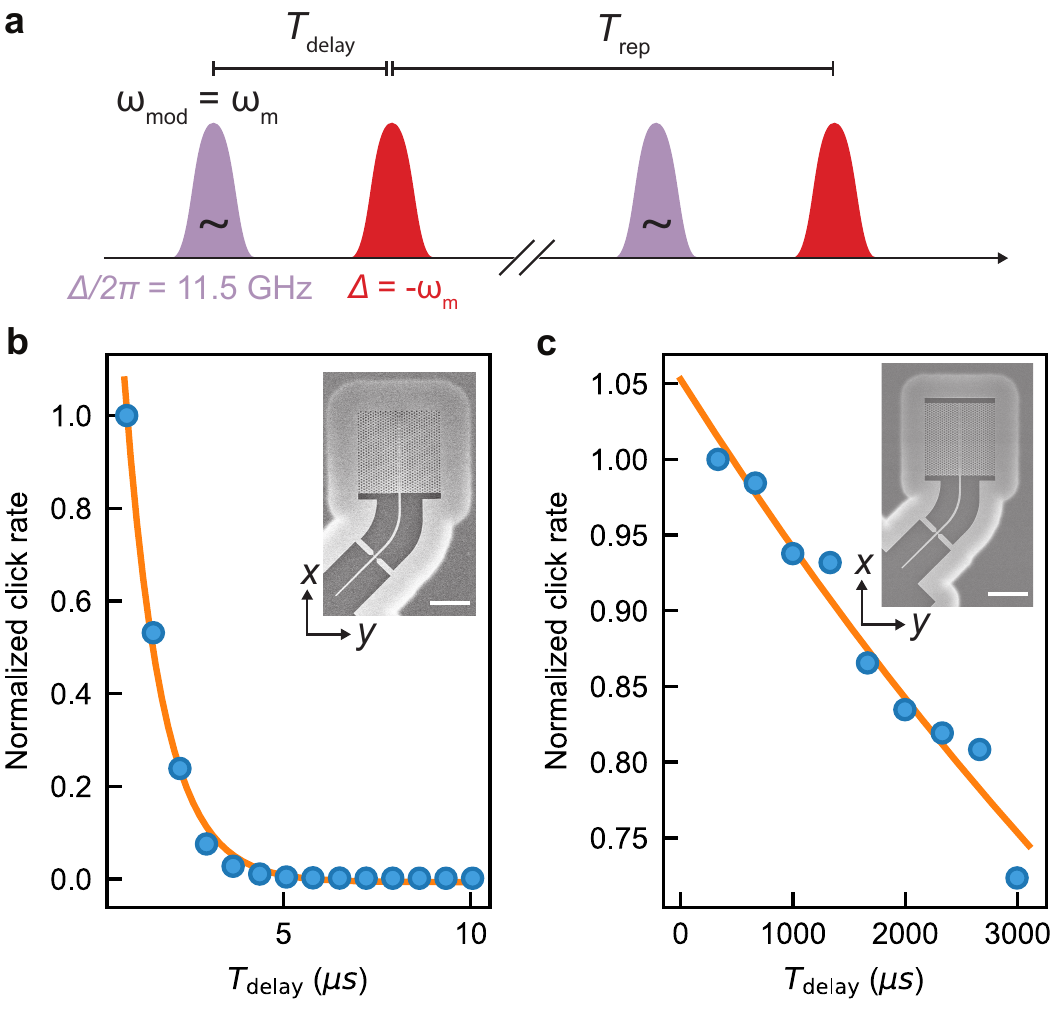}
	\caption{\textbf{Measurement of phonon lifetime.} \textbf{a} Pulse sequence used to measure the phonon lifetime. A first pulse detuned from the optical cavity resonance by $\Delta/2\pi = \qty{11.5}{GHz}$ is modulated by an electro-optical phase modulator at the mechanical frequency $\omega_\mathrm{m}$. After a variable delay time of $T_\mathrm{delay}$, a readout pulse detuned to the red optomechanical sideband reads out the phonon population in the mechanical mode. The pulse sequence is repeated with repetition period $T_\mathrm{rep}$. \textbf{b} Click rate measured on SNSPDs during the readout pulse normalized to the first data point as a function of delay time $T_\mathrm{delay}$. Solid orange line is a fit to an exponential decay yielding a lifetime of phonons in the mechanical mode of $\tau = \qty{1.0}{\micro s}$. \textbf{c} Normalised click rate as a function of delay time $T_\mathrm{delay}$ from another device with a much longer lifetime. The exponential fit gives a lifetime $\tau = \qty{9.0\pm0.7}{\milli s}$. Insets in \textbf{b} and \textbf{c} show SEM images of the respective devices with scale bars corresponding to $\qty{10}{\micro m}$.}
	\label{Fig:S2_phonon_lifetime}
\end{figure}

We measure the lifetime of the phonons in the mechanical mode of interest by creating a large coherent phonon population and monitoring its decay over time. Figure~\ref{Fig:S2_phonon_lifetime}a shows the pulse sequence used for measuring the lifetime of the mechanical mode. First, a laser pulse (frequency $\omega_\mathrm{l}$) of duration $\qty{40}{ns}$ detuned from the optical cavity resonance at $\omega_\mathrm{c}$ by $\Delta/2\pi = (\omega_\mathrm{l} - \omega_\mathrm{c})/2\pi = \qty{11.5}{GHz}$ and modulated by an electro-optic phase modulator at the mechanical frequency ($\omega_\mathrm{mod} = \omega_\mathrm{m}$) is sent to the device. The beating between the sideband and carrier frequency of the optical field coherently drives the mechanical mode creating a large phonon population. After a delay time $T_\mathrm{delay}$, we read out the phonon population by a second $\qty{40}{ns}$ long laser pulse detuned on the red optomechanical sideband of the optical cavity. We detect optomechanically scattered photons from the readout pulse on the SNSPDs and vary the delay time between the two pulses to measure the decay of the coherent phonon population over time. Figure~\ref{Fig:S2_phonon_lifetime}b shows the normalized click rate measured on the SNSPDs as a function of $T_\mathrm{delay}$. From an exponential fit to the experimental data, we extract the phonon lifetime $\tau = \qty{1.0}{\micro s}$. Figure~\ref{Fig:S2_phonon_lifetime}c shows the lifetime measurement of another device with a much longer phonon lifetime of $\tau = \qty{9.0}{\milli s}$. The significantly increased lifetime of the device in Fig.~\ref{Fig:S2_phonon_lifetime}c is achieved by adding an air gap at the end of the device as shown in the inset to reduce radiation loss through the connection to the substrate. For the experiments in the main text, we picked the device in Figure~\ref{Fig:S2_phonon_lifetime}b in order to obtain a higher repetition rate.

Higher optical power in general will decrease the mechanical quality factor of the quasi-2D OMC structure. This effect has been studied in earlier works~\cite{mayor_high_2025}, where the mechanical linewidth ($\gamma/(2\pi)$) increased from around $\SI{100}{kHz}$ to $\SI{160}{kHz}$ when the intracavity photon number increased from $10^3$ to $10^4$. Since we only use relatively small intracavity photon numbers in our experiments ($n_\mathrm{c}\lesssim 2\times 10^3$), we expect the modification of the mechanical quality factor in our experiments to be small. Furthermore, for each repetition cycle, $T_\mathrm{rep} \gg \tau$, i.e.\ we wait long enough so the light-induced heat is fully dissipated and the device is fully thermalized to the cold environment ($\SI{20}{mK}$) before the start of the next repetition cycle. Therefore, a slightly lower mechanical quality factor for higher optical power inside the cavity does not significantly affect our experiments.

\subsubsection*{Cavity resoanance shift measurement}
\label{n_th_measurement}

Importantly, high-power optical laser pulses can lead to shifts in the optical resonance frequency of the OMC device either due to the photothermal effect or free carrier dispersion~\cite{barclay_nonlinear_2005, chan_laser_2011}. A shift in the optical cavity resonance by $\delta \omega_\mathrm{c}$ can modify the measured value of Stokes and anti-Stokes scattering probabilities and thus impact the calculated thermal phonon occupancy. We correct the values of anti-Stokes scattering probabilities for this shift by replacing $\Delta \rightarrow \Delta - \delta \omega_\mathrm{c}$ in main text Eq.~(4).

\begin{figure}
	\centering
	\includegraphics[width = 8.8cm]{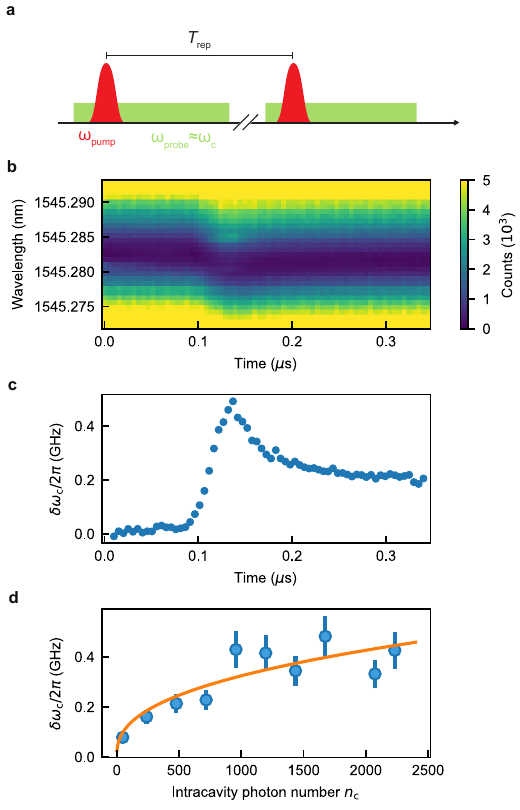}
	\caption{\textbf{Calibration of the optical cavity shift.} \textbf{a} Pulse sequence used for optical cavity shift calibration measurement. \textbf{b} Integrated counts on SNSPDs as a function of time while varying the wavelength of the probe laser at input laser power of $\qty{1400}{nW}$. \textbf{c} Time-resolved shift of the optical cavity resonance frequency $\delta \omega_\mathrm{c}$ extracted from the data in \textbf{b}. \textbf{d} Maximum shift of the optical cavity resonance frequency $\delta \omega_\mathrm{c}$ measured at various intracavity photon number $n_\mathrm{c}$. Solid orange line is a fit to a power law $f(x) = a x^A$ with coefficients $a = 0.025$, $A = 0.39$.}
	\label{Fig:S4_cavity_shift_calibration}
\end{figure}

To calibrate the shift $\delta \omega_\mathrm{c}$ in our device, we use a pump-probe measurement of the optical cavity resonance frequency. The pulse sequence used for this experiment is shown in Fig.~\ref{Fig:S4_cavity_shift_calibration}a. We send optical pump pulses detuned to the red optomechanical sideband of the optical cavity with pulse length $\qty{40}{ns}$ and, simultaneously, a second strongly attenuated probe pulse with pulse length $\qty{5}{\micro s}$ to the device. This pulse sequence is repeated with a repetition period of $T_\mathrm{rep}=\qty{10}{\micro s}$. We lock the free-space optical filters to the wavelength of the probe pulse to remove the strong pump light and measure the reflected probe signal from the optical cavity on the SNSPDs.

By varying the wavelength of the probe pulse we can measure the optical cavity resonance frequency as a function of time (see Fig.~\ref{Fig:S4_cavity_shift_calibration}b). When the pump pulse arrives after approximately $\qty{0.1}{\micro s}$, a blue shift of the cavity resonance is observed. While the exact mechanism of this blue shift warrants further investigation in the future, it is likely due to free carrier dispersion effects routinely observed in silicon photonic crystal devices~\cite{barclay_nonlinear_2005}. We extract the value of the cavity resonance shift $\delta \omega_\mathrm{c}$ by binning the data in Fig.~\ref{Fig:S4_cavity_shift_calibration}b in $\qty{5}{ns}$ bins and fitting Lorentzian lineshapes to extract the time-resolved cavity resonance frequency as shown in Fig.~\ref{Fig:S4_cavity_shift_calibration}c. We vary the power of the pump pulse over the typical range used in the measurements presented in the main text (see Fig.~\ref{Fig:S4_cavity_shift_calibration}d). We use the data in Fig.~\ref{Fig:S4_cavity_shift_calibration}d to correct the measured values of the Stokes and anti-Stokes scattering probability as well as the obtained value of the thermal phonon occupancy in Eq.~(6) in main text.\\

Aside from cavity resonance shift, changes of the optical quality factor in response to different powers has been studied in~\cite{barclay2005nonlinear}, where different non-linear effects (such as two-photon absorption and free carrier absorption) due to the high electromagnetic energy stored in the OMC structure have been incorporated. When the optical power loaded to the cavity is increased to hundreds of $\SI{}{\micro W}$, it will eventually lead to a broadening of the optical linewidth. However, even for the highest power loaded into the cavity used in our experiment ($\sim$$\SI{1.6}{\femto W}$), the change in optical quality factor is negligible according to the simulation in~\cite{barclay2005nonlinear}, which is consistent with our observations in the experiment. Moreover, considering we are deep in the sideband-resolved regime (mechanical frequency $\omega_\mathrm{m}/2\pi \sim \SI{10}{GHz}$ and optical linewidth $\kappa/2\pi\sim\SI{2.4}{GHz}$), the resulting change from optical linewidth in scattering probabilities is even more negligible.

\subsubsection*{Measurement of the thermal phonon occupancy due to delayed heating from the write pulse}
\label{prepulse_asymmetry}

\begin{figure}
	\centering
	\includegraphics[width = 8.8cm]{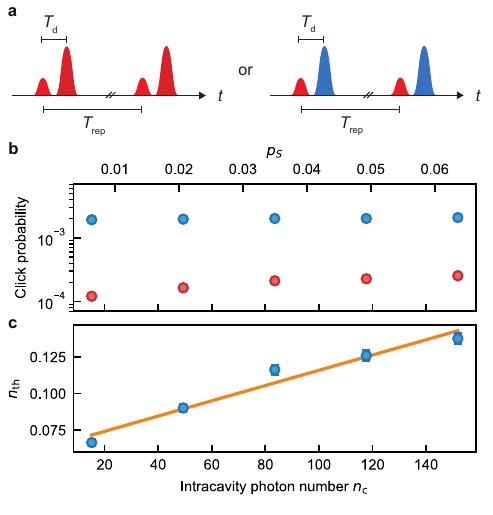}
	\caption{\textbf{Calibration of the added thermal phonons from write pulse.} \textbf{a} Pulse sequence for sideband asymmetry measurement with a prepulse to calibrate the thermal phonon occupancy added by the prepulse. The delay between the two pulses is $T_\mathrm{d}=\qty{150}{ns}$ as is used as the delay between the write and read pulse in the cross-correlation and HBT measurements presented in the main text. The prepulse is detuned on the red optomechanical sideband ($\Delta=-\omega_\mathrm{m}$). We repeat the experiment with the second pulse for sideband asymmetry either blue- ($\Delta=\omega_\mathrm{m}$) or red-detuned ($\Delta=-\omega_\mathrm{m}$). The power of the second pulse is fixed to an intracavity photon number of $n_\mathrm{c} = 300$. \textbf{b} Measured click probability during the second pulse when the laser is detuned to the blue or red optomechanical sideband and \textbf{c} calculated thermal phonon occupancy $n_\mathrm{th}$ as a function of intracavity photon number (Stokes scattering probability) of the prepulse on the bottom (top) axis. The solid orange line is a linear fit to the data.}
	\label{Fig:S7_sideband_asymmetry}
\end{figure}

For measuring quantum correlations between phonons and photons as well as measurements of the photon autocorrelation function and Hong-Ou-Mandel interference as presented in the main text, we use a pulse sequence consisting of a blue-detuned write pulse followed by a red-detuned readout pulse. We calibrate the additional thermal phonon population induced by the write pulse by using a pulse sequence consisting of two pulses as shown in Fig.~\ref{Fig:S7_sideband_asymmetry}a:\ the first red-detuned pulse creates thermal phonons. Note that we use a red pulse for this, whereas the write pulse in the experiments in the main text is blue-detuned. The thermal phonon occupancy induced by a blue- or red-detuned pulse is expected to be the same considering the same amount of intracavity photon number for blue- and red-detuned pulses at the same power. Hence, we choose to use a red-detuned pulse to avoid additional phonon occupancy by optomechanical driving through the two-mode squeezing interaction of a blue-detuned pulse. The second pulse is either blue- or red-detuned and is used to measure the phonon occupancy created by the first pulse through sideband asymmetry. For small optomechanical scattering probabilities such that $p_\mathrm{aS} \approx p_\mathrm{S}$ in Eqs.~(3) and (4) in main text, it follows from Eqs.~(1) and (2) in main text that the thermal phonon occupancy $n$ during the second pulse can be calculated from the asymmetry in count rates during the second pulse when using either a blue-detuned ($C_\mathrm{S}$) or red-detuned pulse ($C_\mathrm{aS}$) as  
\begin{align}
\label{n_th_asymmetry}
    n_\mathrm{th} = \frac{1}{C_\mathrm{S}/C_\mathrm{aS}-1}.
\end{align}
Figure~\ref{Fig:S7_sideband_asymmetry}b shows the measured click probability on the SNSPDs during the second pulse with the laser detuned on the blue or red optomechanical sideband as a function of optical power or equivalently Stokes scattering probability of the first pulse. From Eq.~\eqref{n_th_asymmetry}, we calculate the corresponding thermal phonon occupancy during the second pulse as shown in Fig.~\ref{Fig:S7_sideband_asymmetry}c. The value of $n_\mathrm{th}$ obtained in this way contains both contribution from thermal phonons created due to the first ($n_\mathrm{p1}$) and second pulse ($n_\mathrm{p2}$). The optical power of the second pulse is fixed to $\qty{100}{nW}$. Therefore, we can isolate the contribution of the first pulse by subtracting the contribution of the second pulse which is known from the calibration of the thermal phonon occupancy shown in the main text to be $n_\mathrm{p2} = 0.047$. Both the cross-correlation as well as the HBT measurements shown in the main text use a write pulse power of intracavity photon number $n_\mathrm{c} = 30$ (Stokes scattering probability $p_\mathrm{S} = 1.3\%$). For this write pulse power, we obtain $n_\mathrm{write} = n_\mathrm{p1} = n_\mathrm{th}(n_\mathrm{c} = 30) - n_\mathrm{p2} = 0.039$.

\subsection{Extended discussion of improved thermalization of 2D optomechanical crystal cavities}

A phenomenological model of optical absorption heating and the generation of thermal noise in silicon OMCs has been established in earlier work~\cite{meenehan_pulsed_2015,meenehan_silicon_2014}:\ due to the existence of dangling bonds at the silicon/silicon oxide interface, discrete interface defect states below and above the silicon electronic midgap can be formed due to unreacted Si dangling bonds or Si dangling bonds interacting with oxygen atoms~\cite{kobayashi1995interface,yamashita1996spectroscopic,sakurai1981theory}. It is believed that upon absorption of telecom photons through the interface states, a high-frequency (THz) thermal phonon bath is generated, which will go through three-phonon scattering processes involving the mechanical mode of interest and the local thermal bath, where phonons of GHz frequency will be generated through the interaction between two high-frequency phonon modes~\cite{srivastava_physics_2022,meenehan2014silicon}. The general design strategy behind the 2D OMC structure is to make use of the higher density of phononic states resulting from a large contact area with the substrate that can carry away heat before the GHz mechanical mode of interest is populated~\cite{ren_two-dimensional_2020}. In~\cite{ren_two-dimensional_2020}, the difference in thermal conductance between 1D nanobeam and 2D OMC structures has been simulated, and a factor of 42 enhancement was expected in the thermal conductance coefficient for the quasi-2D OMC, which was believed to be the main reason for the improvement in thermal performance.

Based on the calibration of thermal occupancy as well as the optomechanical sideband thermometry measurement presented in the main text, it is evident that the 2D OMC devices used in this work exhibit reduced thermal noise at the same intracavity photon numbers compared to 1-dimensional nanobeam OMCs. While we observe a factor of three reduction in thermal noise in the devices in our work, we compare the thermal performance of different 2D OMC structures here in more detail. For direct comparison, according to the data in Figure 4(c) in~\cite{mayor_high_2025}, with scattering probability of approx.\ $5\%$ and photon-phonon pair generation rate of \SI{5}{kHz}, the estimated phonon occupation is between $0.02$ to $0.03$. For our Figure 2(a) in the main text, with scattering probability of $5\%$, and a photon-phonon pair generation rate of \SI{5}{kHz} (since our repetition rate is \SI{100}{kHz}), the measured thermal phonon occupancy is $0.029$, which is very comparable to~\cite{mayor_high_2025}. We would like to note that there can still be some variation in thermal performance between individual devices due to fabrication imperfections.

For the side-coupled snowflake + Cshape structure investigated in~\cite{sonar_high-efficiency_2025} with a repetition rate of \SI{250}{Hz}, at $10\%$ transduction efficiency, the phonon occupancy is around $0.045$, while at $90\%$ transduction efficiency the phonon occupancy is $0.25$. For our end-coupled snowflake + Cshape structure, with a repetition rate of \SI{100}{kHz}, at $10\%$ transduction efficiency, the phonon occupation is approx.\ $0.041$; at $60\%$ transduction efficiency, the phonon occupation is $0.3$. We can see that at high power and high transduction efficiency, the side-coupled structure performs better than our device used in the experiments (despite the different repetition rates used here). This is mainly due to the side-coupled geometry that is used in~\cite{sonar_high-efficiency_2025}, which mechanically detaches the optical cavity from the coupling waveguide which could act as a source of thermal noise~\cite{sonar_high-efficiency_2025}. We note that alternative approaches to achieving such evanescent coupling to 2D OMC devices exist, for example by using pick-and-place techniques to transfer photonic coupling waveguide structures on top of the 2D OMC~\cite{guo_integrated_2022}.

\subsection{Optomechanical single-photon generation protocol}

 As a first step, we generate single phonons in the mechanical mode through a pulsed optomechanical interaction, in combination with single-photon detection~\cite{riedinger_non-classical_2016}. A laser pulse detuned on the blue optomechanical sideband of the optical cavity resonance induces a two-mode squeezing interaction between the optical and mechanical mode with interaction Hamiltonian $H_\mathrm{tms} = \sqrt{n_\mathrm{c}} g_0 (a^\dag b^\dag + a b)$, where $a$ ($b$) are the annihilation operators of the optical (mechanical) mode and $n_\mathrm{c}$ is the intracavity photon number. This corresponds to a Stokes scattering process, in which the optomechanical interaction converts a photon from the blue-detuned pump into a photon at the optical cavity resonance and a phonon in the mechanical mode. After the interaction, the joint system is in the state
\begin{align}
\label{state_after_tms}
    \ket{\psi}=\sqrt{1-p_\mathrm{S}}\sum_{n=0}^{\infty} \sqrt{p_\mathrm{S}}^n \ket{n n},
\end{align}
where $p_\mathrm{S}$ is the optomechanical scattering probability for the Stokes process. The light coming back from the device is filtered by a series of optical filters to remove the strong blue-detuned pump and detect single optomechanically scattered photons on superconducting nanowire single-photon detectors (SNSPD). The detection of a single-photon projects the mechanical mode into the state 
\begin{align}
\label{state_after_tms_heralded}
    \ket{\psi_\mathrm{m}} = \sqrt{\frac{1-p_\mathrm{S}}{p_\mathrm{S}}}\sum_{n=1}^{\infty} \sqrt{p_\mathrm{S}}^n \ket{n},
\end{align}
which for small scattering probability $p_\mathrm{S} \ll 1$ is close to a single phonon Fock state $\ket{\psi_\mathrm{m}} \approx \ket{1}$. Thus, the emitted photon from the Stokes process heralds the creation of the single phonon state. After the phonon creation, a second laser pulse detuned to the red optomechanical sideband transfers the phonon state onto the optical mode through a beam-splitter interaction with interaction Hamiltonian $H_\mathrm{bs} = \sqrt{n_\mathrm{c}} g_0 (a^\dag b + a b^\dag)$. This corresponds to an anti-Stokes scattering process in which a photon from the red-detuned pump and a phonon in the mechanical mode scatter into a photon at the optical cavity resonance. This mechanics-to-optics conversion process allows us to generate single photons after filtering out the strong red pump pulse.

\subsection{Hanbury Brown-Twiss measurements}
\label{extended_information_hbt}

%\subsubsection{Measurement of the auto-correlation function}
\subsubsection*{Definition and measurement of the second-order intensity autocorrelation function $ g^{(2)}(\tau)$}

In the Hanbury Brown-Twiss (HBT) measurement in the main text, we measure the intensity autocorrelation function of the optical anti-Stokes field generated by the optomechanical beam splitter interaction during the red-detuned readout pulse. The evaluation of the autocorrelation function is conditioned on the detection of a Stokes-scattered photon during the blue-detuned write pulse, which heralds the creation of a single phonon in the mechanical mode and thus represents a third-order correlation function. After heralding, the conditional second-order autocorrelation function used to characterize the purity of the generated single photons from the optomechanical anti-Stokes scattering process is defined as~\cite{hong_hanbury_2017}
\begin{align}
\label{g_2_theory}
    g^{(2)}(\tau) = \frac{\langle  a^{\dag}(0) a^{\dag}(\tau) a(\tau) a(0)  \rangle }
    {\langle a^{\dag}(\tau) a(\tau) \rangle
    \langle a^{\dag}(0) a(0) \rangle},
\end{align}
 where $\tau$ is the time delay between a coincidence detection on detector \num{1} and detector \num{2} in the HBT setup.

In the regime of low photon detection probability~\cite{stevens_third-order_2014} and for $\tau=0$, the theoretical definition of the photon autocorrelation function in Eq.~\eqref{g_2_theory} can be connected to the probability of detecting coincidence clicks on the two outputs of a beam splitter in a HBT configuration as discussed in the main text. In this case, Eq.~\eqref{g_2_theory} reduces to the cross-correlation between the two detectors~\cite{hong_hanbury_2017}
\begin{align}
\label{g_2_measure}
    g^{(2)}(0) = \frac{P(\mathrm{D_1} \cap \mathrm{D_2})}{P(\mathrm{D_1}) P(\mathrm{D_2})}, 
\end{align}
where $P(\mathrm{X})$ describes the probability of detection event $X$. Event $\mathrm{D_1} \cap \mathrm{D_2}$ represents the coincidence detection on both detectors heralded on a Stokes photon click, and event  $\mathrm{D_1} (\mathrm{D_2})$ represents a single click on detector  $\mathrm{D_1} (\mathrm{D_2})$ heralded on a Stokes photon click.

To illustrate the anti-bunching behavior of optomechanically generated single photons, we also calculate the value of the conditional autocorrelation function $g^{(2)}(\tau)$ for detection events in different repetitions of the experiment where $\tau = T_\mathrm{rep} \Delta n$ with the repetition time of the experiment $T_\mathrm{rep}=\qty{10}{\micro s}$. The integer offset $\Delta n$ corresponds to the number of repetitions of the experiment between two clicks on detectors D1 and D2 for which heralding was successful. Hence, $\Delta n$ represents the stochastically varying time delay between two such detection events. Such detection events are fully uncorrelated and are thus expected to yield $g^{(2)}(\Delta n \neq 0)=1$ as shown in the main text.

The error bars of the conditional autocorrelation function are calculated from the single-photon counting statistics by evaluating the exact binomial confidence interval using the Clopper-Pearson method to account for the asymmetry of the binomial distribution due to the low number of successful detection events.

\subsubsection*{Criteria for non-classicality and single-photon generation}

The autocorrelation function in Eq.~\eqref{g_2_theory} characterizes the statistical properties of the anti-Stokes scattered light field described by the annihilation operator $a$. The value of the autocorrelation function at $\tau=0$ serves as a metric to proof the non-classical nature of an optical state: for classical light fields, the autocorrelation function is bound to $g^{(2)}(0) \geq1$ and hence a value of $g^{(2)}(0)<1$ indicates a non-classical state of light~\cite{davidovich_sub-poissonian_1996}. An even more stringent bound can be derived to proof the single-photon nature of an optical state: for a photon number state $\ket{n}$, Eq.~\eqref{g_2_theory} yields
 \begin{align}
\label{g_2_theory_fock}
    g^{(2)}(0) = \frac{\langle  n(n-1)  \rangle }{\langle n \rangle^2} = \frac{n-1}{n}.
\end{align}
Equation~\eqref{g_2_theory_fock} shows that $g^{(2)}(0)=0$ for an ideal single-photon Fock state $\ket{n=1}$, whereas $g^{(2)}(0)\geq0.5$ for any higher Fock state $n \geq 2$. Therefore, $g^{(2)}(0)<0.5$ has become a standard criterion to demonstrate the single-photon nature of an optical state~\cite{fishman_photon-emission-correlation_2023}. Since the anti-Stokes photons are created by mechanics-to-optics conversion through an optomechanical beam splitter interaction, the autocorrelation function in Eq.~\eqref{g_2_theory} also directly measures the second-order intensity correlation of the mechanical mode. Therefore, demonstrations of $g^{(2)}(0)<1$ and $g^{(2)}(0)<0.5$ imply non-classical and single-phonon states in the mechanical mode, respectively.

\subsubsection*{Impact of dark counts}

The impact of dark counts on the measured $g^{(2)}(0)$ value is evaluated by choosing a time window of same length but $\qty{80}{ns}$ after the optomechanical signal. The rate of the dark counts is in general less than $2\%$ of both the Stokes scattered photon count rate and the anti-Stokes scattered photon count rate, which results in $0.78\%$ of coincidences.

\subsubsection*{Impact of SNSPD dead time}

\begin{figure}
	\centering
	\includegraphics[width = 8.8cm]{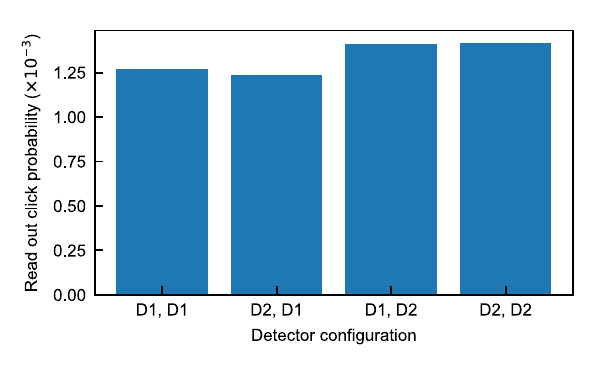}
	\caption{\textbf{Impact of SNSPD dead time on HBT measurements.} Click probability on SNSPDs during the read out pulse after heralding for different detector configurations. Detector configurations are labeled as $\mathrm{D}_i$, $\mathrm{D}_j$, where $\mathrm{D}_i$ ($\mathrm{D}_j$) indicates the detector used for heralding (readout).}
	\label{Fig:S5_deadtime_check}
\end{figure}

If the pulse delay between the write and read pulse is shorter than the dead time of the superconducting nanowire single-photon detectors (SNSPD), this can lead to incorrect values of the observed cross- and auto-correlation functions. In particular, in Hanbury Brown-Twiss measurements this would lead to a measured value of the auto-correlation function $g^{(2)}(0)$ that is lower than the true value. To rule out the impact of the SNSPD dead time, we consider the probability to detect a photon during the read out pulse on one detector after heralding for different detector configurations (see Fig.~\ref{Fig:S5_deadtime_check}). The different detector configurations are labeled as $\mathrm{D}_i$, $\mathrm{D}_j$, where $\mathrm{D}_i$ ($\mathrm{D}_j$) indicates the detector used for heralding (readout). The probability to detect a photon on $\mathrm{D}_1$ does not depend significantly on whether $\mathrm{D}_1$ or $\mathrm{D}_2$ was used for heralding. This implies that after the heralding photon is detected on $\mathrm{D}_1$, the detector has already recovered to its full detection efficiency before the read out photon arrives. The same observation can be made for the remaining two detector configurations in which the readout photon is detected on $\mathrm{D}_2$. The slightly lower detection probability between the configurations where the read out photon is detected on $\mathrm{D}_1$ compared to the configurations where is detected on $\mathrm{D}_2$ can be explained by the slightly higher detection efficiency $\eta_1 = 0.55$ compared to $\eta_2 = 0.61$.

\subsection{Hong-Ou-Mandel interference}

\subsubsection*{Calibration of interference visibility using weak coherent light}

\begin{figure*}
	\centering
	\includegraphics[width = 12cm]{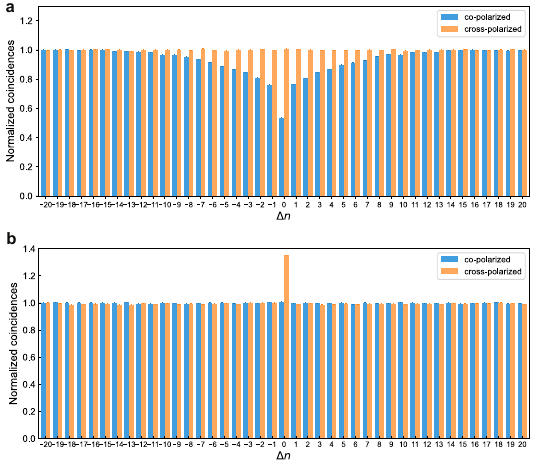}
	\caption{\textbf{HOM interference of weak coherent light and thermal light.} HOM interference measurement of weak coherent light (\textbf{a}) and thermal light (\textbf{b}). In the weak coherent light measurement, visibility $V=0.464$, the deviation from $0.5$ visibility is mainly attributed to the power imbalance of the two arms. In the thermal light analysis, the ratio between the peak and the side bars is $1.35$ in the cross-polarized case, which matches well with the Qutip simulation considering the amount of thermal component introduced in our pulse scheme (see QuTiP Simulations below).}
	\label{Fig:S9_HOM_wcs_and_thermal}
\end{figure*}

To calibrate for the imperfect power balance of the two arms in the Hong-Ou-Mandel (HOM) setup, we measure the HOM interference of weak coherent light in the co- and cross-polarized cases. Weak coherent light pulses at optical cavity resonance are generated and sent through the same HOM setup for interference. In~\ref{Fig:S9_HOM_wcs_and_thermal}a, we can see the dip in the co-polarized case. It represents a visibility of $V=0.464$. The deviation from $0.5$ visibility is mainly attributed to the power imbalance of the two arms. This ratio is used to correct the measured HOM visibility of optomechanically generated single photons (see main text). The smaller dips on both sides result from the relatively long coherence time of the coherent laser light used in this measurement.

In the HOM measurement of single photons (see main text), if we do not perform any heralding on the Stokes photons from the blue-detuned pulse and directly analyze the HOM interference of anti-Stokes photons from the red-detuned pulse, this corresponds to the HOM interference of thermal light. In ~\ref{Fig:S9_HOM_wcs_and_thermal}b, we can see there is a peak in the cross-polarized case. The ratio between the peak and the side bars is $1.35$, which matches well with the Qutip simulation considering the amount of thermal component introduced in our pulse scheme (see QuTiP Simulations below).

\subsubsection*{Modeling of temporal wavepacket shape}

In the main text we measure the HOM interference between an optomechanically generated single photon and a thermal photon state as a function of timing offset $\Delta t$ of the red-detuned read out pulses. Here, we introduce a phenomenological model to describe the dependence of the HOM dip on $\Delta t$. The depth of the HOM dip depends on the overlap of the two wavepackets of the photons impinging on the beam splitter, which is given by a convolution of the wavepacket envelops $P_1(t)$ and $P_2(t)$. We model the normalized number of coincidences $g_\mathrm{HOM}(\Delta t)$ by
\begin{align}
\label{g_HOM_three}
    g_\mathrm{HOM}(\Delta t) = A \left(1-B \int_{-\infty}^{\infty} P_1(t) P_2(t-\Delta t) \right),
\end{align}
where $A$ and $B$ are fitting parameters to model the relative depth of the HOM dip. The pulse shape of the photon wave packet follows the pulse shape of the red readout pulse, which is created by applying a square gating voltage pulse of length $T_\mathrm{read}$ to an acousto-optic modulator with $f_\mathrm{AOM} = \qty{110}{MHz}$ operating frequency. The relatively slow cutoff time constant of the AOM ($\tau_\mathrm{AOM} = 1/f_\mathrm{AOM} \approx \qty{9.1}{ns}$) leads to significant broadening of the pulse length compared to the nominal values of $\qty{40}{ns}$ or $\qty{100}{ns}$ used in the experiment. Thus, we model the pulse shape by convoluting the square pulse with a Lorentzian with width $\tau_\mathrm{AOM}$ in the time domain.

As mentioned in the main text, for increasing pulse offset $\Delta t$ we measure a normalized number of coincidences larger than unity. The number of coincidences is normalized to the value measured on the satellite peaks with uncorrelated coincidences where $\Delta n \neq 0$. When, the timing offset increases, the overlap of the two readout pulses is reduced. The parts of the two pulses that are not overlapping in time cannot interfere. Therefore, the coincidence events detected in the non-overlapping part of the pulses measure the photon statistics of the incoming photon state in an HBT-type experiment. As the measured coincidence number is measured within the whole pulse window of both readout pulses, all of these coincidence events overlap. When the two pulses are completely separated in time, this measurement measures the overlapping coincidences from two simultaneous HBT experiments on a thermal state and an optomechanically generated single photon giving rise to a normalized coincidence number larger than $1$ due to the bunched photon statistics of the thermal state.

\subsection{QuTiP simulations}
\label{qutip_simulations}

\begin{figure}
	\centering
	\includegraphics[width = 8.8cm]{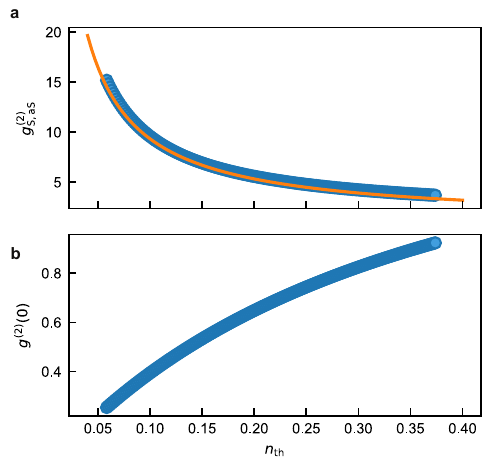}
	\caption{\textbf{Qutip simulations of cross- and auto-correlation functions.} \textbf{a} Simulated cross-correlation function $g^{(2)}_\mathrm{S,aS}$ between write and read pulse as a function of total thermal population $n_\mathrm{th}$. In the simulation, the Stokes scattering probability is fixed at $p_\mathrm{S}=1.3\%$ and the anti-Stokes scattering probability during the readout pulse $p_\mathrm{aS}$ is varied. The orange line corresponds to the phenomenological expression in Eq.~\eqref{gcc_heating}. \textbf{b} Simulated autocorrelation function $g^{(2)}(0)$ of the optical readout field as a function of $n_\mathrm{th}$.}
	\label{Fig:S1_qutip_simulations}
\end{figure}

We use the Python package QuTiP to simulate the expected values of the cross-correlation, auto-correlation functions~\cite{johansson_qutip_2012, johansson_qutip_2013}, and the results of HOM interference measurements.

\subsubsection*{Simulation procedure}
\label{sim_procedure}

We keep track of the full evolution of the density matrix of the mechanical and optical mode of the system throughout a pulse sequence consisting of first a blue-detuned and secondly a red-detuned pulse as used for the measurements in the main text with respective optomechanical scattering probabilities $p_\mathrm{S}$ and $p_\mathrm{aS}$ for the induced Stokes and anti-Stokes scattering processes. More details of the simulation procedure can be found in~\cite{fiaschi_optomechanical_2021}. We include the following imperfections and sources of loss in our simulation:

\begin{enumerate}
\item We include the thermal occupancy of the mechanical mode both originating from the write and readout pulse. We include the thermal occupancy due to the write pulse by initiating the mechanical mode in a thermal state with thermal occupancy $n_\mathrm{write}$ in the beginning of the simulation.
\item We include the additional thermal occupancy added through the readout pulse by coupling the system to a thermal bath with phonon occupancy $n_\mathrm{read}$. The coupling is modeled by a two-mode squeezing interaction with squeezing angle $\sqrt{n_\mathrm{read}}$ between the mechanical mode and an auxiliary vacuum mode in the environment and then tracing out the auxiliary mode. The coupling is assumed to occur right before the readout pulse.
\item In the time period between the write and read pulses, phonons in the mechanical mode may decay due to the limited lifetime of the mechanical mode $\tau = \qty{1.0}{\micro s}$. We account for this by applying an amplitude damping channel with loss probability $p_\mathrm{loss} = e^{-\tau/T_\mathrm{delay}}$, where $T_\mathrm{delay}$ is the delay between the write and read pulses.
\item After the mechanical state is transferred to the optical state by optomechanical readout, we apply an amplitude damping channel to account for optical losses on the detection path with loss probability $p_\mathrm{loss} = 1-\eta$, where $\eta$ is the efficiency of the detection path.
%\item We include dark counts (dark count rate $\qty{30}{Hz}$) on the single-photon detectors that lead to additional uncorrelated detection events.
\end{enumerate}

We denote the optical output state after following this simulation procedure as $\rho_\mathrm{om, out}(p_\mathrm{S}, p_\mathrm{aS}, n_\mathrm{write}, n_\mathrm{read})$.

\subsubsection*{Simulation of cross- and auto-correlation functions}
\label{sim_gcc_g2}

We evaluate the resulting cross-correlation between detection events during the write and readout pulse $g^{(2)}_\mathrm{S,aS}$ and autocorrelation function of the readout optical field $g^{(2)}(0)$. Figure \ref{Fig:S1_qutip_simulations}a shows the simulated cross-correlation function as a function of total thermal phonon occupancy during the readout pulse. The total thermal phonon occupancy is given by 
\begin{align}
\label{total_n_th}
    n_\mathrm{th} = n_\mathrm{write} + n_\mathrm{read},
\end{align}
where $n_\mathrm{write}$ and $n_\mathrm{read}$ are the thermal phonon populations induced due to heating during the write pulse and read pulse. To compare the simulations to the measurements we fix the Stokes scattering probability during the write pulse to $p_\mathrm{S}=1.3\%$ as used in the cross-correlation and HBT measurements. As calibrated above, the write pulse creates a thermal phonon population of $n_\mathrm{write}=0.039$. We vary the value of anti-Stokes scattering probability during the red-detuned readout pulse and use the calibrated thermal phonon occupancy shown in the main text for $n_\mathrm{read}$.

The maximum observable value of the cross-correlation function is limited by higher-order phonon creation events from the two-mode squeezing interaction during the write pulse limiting $g^{(2)}_\mathrm{S,aS}$ as 
\begin{align}
\label{gcc_no_heating}
    g^{(2)}_\mathrm{S,aS} = 1+\frac{1}{p_\mathrm{S}},
\end{align}
where $p_\mathrm{S}$ is the Stokes scattering probability. In accordance with previous works \cite{riedinger_non-classical_2016}, we find that the presence of an additional thermal phonon population $n_\mathrm{th}$ reduces the observable cross-correlation as
\begin{align}
\label{gcc_heating}
    g^{(2)}_\mathrm{S,aS} = 1+\frac{e^{-T_\mathrm{delay}/\tau}}{p_\mathrm{s} + n_\mathrm{th}}.
\end{align}
The factor $e^{-T_\mathrm{delay}/\tau}$ accounts for phonon decay due to the finite phonon lifetime. Equation~\eqref{gcc_heating} is shown as an orange line in Fig.~\ref{Fig:S1_qutip_simulations} and shows excellent agreement with the simulated values of expected $g^{(2)}_\mathrm{S,aS}$.

Figure~\ref{Fig:S1_qutip_simulations}b shows the simulated autocorrelation function $g^{(2)}(0)$ as a function of thermal phonon occupancy $n_\mathrm{th}$. We estimate that the measured value of the photon autocorrelation function can be improved to $g^{(2)}(0)=0.09$ using OMC devices with improved device geometry that reduce the thermal phonon occupancy by a factor of six at equal scattering probabilities~\cite{sonar_high-efficiency_2025}.

\subsubsection*{Simulation of HOM interference}
\label{sim_hom}

To simulate the result of the HOM interference measurements, we define four optical modes corresponding to horizontal (H) and vertical polarizations (V) in the two arms of the Mach-Zehnder interferometer (long:\ L, short:\ S) each with Hilbert space size $N$. The joint density matrix of all modes is defined as $\rho_\mathrm{in} = \rho_\mathrm{L,H} \otimes \rho_\mathrm{L,V} \otimes \rho_\mathrm{S,H} \otimes \rho_\mathrm{S,V}$. Since our detection scheme is insensitive to the polarization of the detected photons, we mix the two polarization modes in each of the arms by applying a beam splitter interaction and then interfere modes of the same polarization by applying another beam splitter interaction. After the interference, we simulate polarization-insensitive detection of photons in the two output modes 1 and 2 by implementing the following projection operators:
\begin{align}
\label{sim_hom_projection_operators}
    P_{\neq \ket{0}}^{(N^2)} &= I^{(N)} \otimes I^{(N)} - \left(\ket{0}\bra{0} \otimes \ket{0}\bra{0}\right)^{(N)} \\
    P_{\neq \ket{0}, 1/2}^{(N^4)} &= P_{\neq \ket{0}}^{(N^2)} \otimes I^{(N)} \otimes I^{(N)}\\
    P_{\neq \ket{0}, \mathrm{both}}^{(N^4)} &= P_{\neq \ket{0}}^{(N^2)} \otimes P_{\neq \ket{0}}^{(N^2)}, 
\end{align}
where the superscript indicates the dimension of the Hilbert space the operator is acting on. The operator $P_{\neq \ket{0}, 1/2}^{(N^4)}$ projects the system into the state after detection of a single click at the detector at output 1 or 2 of the interferometer, whereas $P_{\neq \ket{0}, \mathrm{both}}^{(N^4)}$ projects onto the state after a coincidence detection at both detectors. The corresponding probabilities for single click ($p_{\mathrm{D1},\mathrm{D2}}$) or coincidence detection ($p_\mathrm{c}$) are calculated as:
\begin{align}
\label{sim_hom_probabilities}
    p_{\mathrm{D1},\mathrm{D2}} &= \Tr{\left(P_{\neq \ket{0}, 1/2}^{(N^4)} \rho_\mathrm{HOM, out}\right)} \\
    p_{\mathrm{c}} &= \Tr{\left(P_{\neq \ket{0}, \mathrm{both}}^{(N^4)} \rho_\mathrm{HOM, out} \right)},
\end{align}
where $\rho_\mathrm{HOM, out}$ is the joint density matrix of all four output modes. We define the second-order correlation function of detection events on the single-photon detectors as
\begin{align}
\label{g_hom}
    g^{(2)}_\mathrm{HOM} = \frac{p_\mathrm{c}}{p_\mathrm{D1}p_\mathrm{D2}},
\end{align}
which corresponds to the normalized number of coincidence clicks as presented for the measurement result in the main text.

\begin{figure}
	\centering
	\includegraphics[width = 8.8cm]{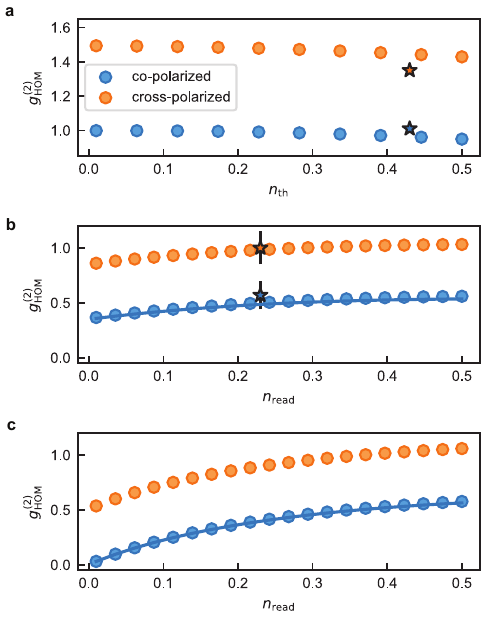}
	\caption{\textbf{Qutip simulations of HOM interference.} \textbf{a} Second order correlation function for the HOM measurement $g^{(2)}_\mathrm{HOM}$ for thermal states as a function of thermal phonon occupancy $n_\mathrm{th}$.  \textbf{b} Second order correlation function for the HOM measurement $g^{(2)}_\mathrm{HOM}$ for optomechanically generated single photons with density matrix $\rho_\mathrm{om, out}(p_\mathrm{S}, p_\mathrm{aS}, n_\mathrm{write}, n_\mathrm{read})$ as a function of thermal phonon occupancy added during the readout pulse $n_\mathrm{read}$ with other simulation parameters fixed at the parameters used for the experiment presented in the main text $p_\mathrm{S} = 0.1$, $n_\mathrm{write} = 0.2$. Stars in \textbf{a} and \textbf{b} indicate experimental data points from the HOM measurement using two thermal states (see Fig.~\ref{Fig:S9_HOM_wcs_and_thermal}b) and optomechanically generated single photons (see main text). For the measured data point in \textbf{a}, the errorbar is smaller than the marker size. \textbf{c} Second order correlation function for the HOM measurement $g^{(2)}_\mathrm{HOM}$ for optomechanically generated single photons as a function of thermal phonon occupancy added during the readout pulse $n_\mathrm{read}$ for the ideal case of low Stokes scattering probability $p_\mathrm{S} = 0.001$ and no initial thermal phonon occupancy from the write pulse $n_\mathrm{write} = 0$. The solid line in \textbf{b} and \textbf{c} is the analytical approximation in Eq.~\eqref{g2_hom} based on the auto-correlation function $g^{(2)}(0)$ of the state $\rho_\mathrm{om, out}$.}
	\label{Fig:S8_qutip_simulations_hom}
\end{figure}

Figure~\ref{Fig:S8_qutip_simulations_hom}a shows the result of QuTiP simulations using the procedure described above with both mechanical modes initialized in thermal states with varying thermal phonon occupancy $n_\mathrm{th}$ either in co-polarized or cross-polarized configuration. We observe $g^{(2)}_\mathrm{HOM} > 1$ for the cross-polarized measurement configuration irrespective of $n_\mathrm{th}$. The slight decrease of $g^{(2)}_\mathrm{HOM}$ with increasing $n_\mathrm{th}$ in the simulations is an artefact of the truncated Hilbert space dimension ($N=7$) limiting the accuracy of the simulation for large thermal states. In this case, interference of the modes in the two interferometer arms occurs for each polarization independently. As our detection scheme is not sensitive to the photon polarization, this case corresponds to simultaneous HBT measurements on each of the input modes. Detection events of the two HBT measurements overlap giving rise to $g^{(2)}_\mathrm{HOM} > 1$ due to the thermal nature of the input state. For co-polarized measurement configuration, the modes in the two arms interfere reducing the probability of coincidence detection events. The result of the QuTiP simulations are in reasonable agreement with the experimental data point (indicated by a star in Fig.~\ref{Fig:S8_qutip_simulations_hom}a) from the HOM measurement of two thermal states (see Fig.~\ref{Fig:S9_HOM_wcs_and_thermal})b). The slightly lower value of $g^{(2)}_\mathrm{HOM}$ measured in the experiment compared to the numerical simulations may be due to imperfect balancing of the optical power in the two interferometer arms as discussed above.

To model the realistic case of HOM interference of optomechanically generated single-photons, we initialize the modes in the two interferometer arms in a state with density matrix $\rho_\mathrm{om, out}(p_\mathrm{S}, p_\mathrm{aS}, n_\mathrm{write}, n_\mathrm{read})$ obtained through the simulation procedure described above in Sec.~\ref{sim_procedure}. Figure~\ref{Fig:S8_qutip_simulations_hom}b shows the resulting second-order correlation function as a function of added thermal phonon occupancy during the readout pulse. The scattering probability during the write pulse ($p_\mathrm{S} = 0.1$) and initial phonon occupancy ($n_\mathrm{write} = 0.2$) due to heating from the write pulse are fixed at the values used in the HOM experiment presented in the main text. Again, the result of the QuTiP simulations agrees well with the experimental data point (indicated by a star in Fig.~\ref{Fig:S8_qutip_simulations_hom}b) from the HOM measurement of optomechanically generated single photons as presented in the main text.

In Fig.~\ref{Fig:S8_qutip_simulations_hom}b, the relatively high scattering probability of the write pulse leads to higher-order excitations of the mechanical mode during the two-mode squeezing interaction. Combined with the thermal phonon occupancy due to heating from the write pulse, this leads to deviations of the mechanical state before readout from the ideal single-phonon Fock state. Figure~\ref{Fig:S8_qutip_simulations_hom}c shows the result of the same simulation for very low optomechanical scattering probability of the write pulse ($p_\mathrm{S} = 0.001$) and no initial thermal phonon occupancy $n_\mathrm{write} = 0$. For low added thermal phonon occupancy during the readout pulse, we obtain perfect HOM interference $g^{(2)}_\mathrm{HOM}=0$.

\subsubsection*{Analytical model of HOM interference}

We compare the results of our QuTiP simulations to a simple analytical model relating the value of $g^{(2)}_\mathrm{HOM}$ to the value of the auto-correlation function $g^{(2)}(0)$ as measured through an HBT experiment. We consider a general optical input state $\rho_\mathrm{in} = p_0 \ket{0}\bra{0} + p_1 \ket{1}\bra{1} + p_2 \ket{2}\bra{2} + \orderof (p_3)$. In an HBT measurement, assuming low click rates with $p_0 \approx 1 \gg p_1 \gg p_2$; the probability for a coincidence detection $p^\mathrm{HBT}_\mathrm{c}$ and for single click detection on detectors 1 and 2 $p^\mathrm{HBT}_{\mathrm{D1},\mathrm{D2}}$ approximated to second-order are then
\begin{align}
\label{ps_g2}
    p^\mathrm{HBT}_\mathrm{c} &= \frac{1}{2}p_2\\
    p^\mathrm{HBT}_{\mathrm{D1},\mathrm{D2}} &= \frac{1}{2}p_1 + \frac{3}{4}p_2 \approx \frac{1}{2} p_1.
\end{align}
The value of the second order auto-correlation function is then obtained as
\begin{align}
\label{g2_p}
    g^{(2)}(0) &= \frac{p^\mathrm{HBT}_\mathrm{c}}{p^\mathrm{HBT}_{\mathrm{D1}} p^\mathrm{HBT}_{\mathrm{D2}}},\\
    &\approx \frac{2 p_2}{p_1^2}.
\end{align}
In an HOM measurement, we analogously obtain
\begin{align}
\label{g_hom_p}
    p^\mathrm{HOM}_\mathrm{c} &= p_2,\\
    p^\mathrm{HOM}_{\mathrm{D1},\mathrm{D2}} &= p_1 + \frac{1}{2}p_1^2 + \frac{3}{8}p_2 \approx p_1,\\
    g^{(2)}_\mathrm{HOM}&= \frac{p^\mathrm{HOM}_\mathrm{c}}{p^\mathrm{HOM}_{\mathrm{D1}} p^\mathrm{HOM}_{\mathrm{D2}}},\\
    &= \frac{p_2}{p_1^2}.
\end{align}
Finally, we can relate the value of the auto-correlation function $g^{(2)}(0)$ to the value of the HOM correlation function $g^{(2)}_{\mathrm{HOM}}$
\begin{align}
\label{g2_hom}
    g^{(2)}_{\mathrm{HOM}} = \frac{1}{2} g^{(2)}(0).
\end{align}
For the simulations in Figs.~\ref{Fig:S8_qutip_simulations_hom}b and c, we also simulate $g^{(2)}(0)$ of the state $\rho_\mathrm{om, out}$ and plot the result of Eq.~\eqref{g2_hom} as a solid line showing good agreement with the QuTiP simulation.

We observe in Fig.~\ref{Fig:S8_qutip_simulations_hom}c that for the cross-polarized case and low added thermal phonon occupancy, the value of the HOM correlation function $g^{(2)}_\mathrm{HOM} < 1$. This observation is a result of low optical readout efficiency $\eta \ll 1$ and so large vacuum component in the optical states interfering at the beam splitter. We recover this result analytically by considering the optical input state at each port of the beam splitter approximated to first order $\rho_\mathrm{in} = p_0 \ket{0}\bra{0} + p_1 \ket{1}\bra{1} + \orderof (p_2)$. For small $p_2$, we approximate $\rho_\mathrm{in} \approx (1-p_1) \ket{0}\bra{0} + p_1 \ket{1}\bra{1}$. As above, we consider the coincidence and single click probabilities without approximating $p_0\approx 1$ for now
\begin{align}
\label{g_hom_p_losses}
    p^\mathrm{HOM}_\mathrm{c} &= \frac{1}{2} p_1^2,\\
    p^\mathrm{HOM}_{\mathrm{D1},\mathrm{D2}} &= \frac{3}{4} p_1^2 + p_1 (1-p_1),\\
    g^{(2)}_\mathrm{HOM}(p_1)&= \frac{p^\mathrm{HOM}_\mathrm{c}}{p^\mathrm{HOM}_{\mathrm{D1}} p^\mathrm{HOM}_{\mathrm{D2}}},\\
    &= \frac{\frac{1}{2} p_1^2}{\left[\frac{3}{4} p_1^2 + p_1 (1-p_1) \right]^2}.
\end{align}
In the limit of high optical losses we obtain
\begin{align}
\label{g_hom_p_losses_limit}
    \lim_{p_1 \to 0} g^{(2)}_\mathrm{HOM}(p_1) = \frac{1}{2}
\end{align}
in good agreement with the result obtained from the QuTiP simulation in Fig.~\ref{Fig:S8_qutip_simulations_hom}c.

\subsection{Single-photon and entanglement generation efficiency}

\subsubsection*{Single-photon rate in the HBT measurement}

The total efficiency $\eta_\mathrm{s}=\eta_\mathrm{herald}\eta_\mathrm{conv}$ of our single-photon generation protocol during the HBT experiment is comprised of the efficiency of the phonon heralding ($\eta_\mathrm{herald}$) and the phonon-photon conversion process ($\eta_\mathrm{conv}$). The efficiency of the heralding process including the intrinsic optomechanical scattering probability of the Stokes scattering process $p_\mathrm{S}=0.013$, optical cavity impedance ratio $\eta_\mathrm{cav}=\kappa_\mathrm{e}/\kappa = 0.45$, lensed fiber coupling efficiency $\eta_\mathrm{fc} = 0.5$, efficiency of the optical filter setup $\eta_\mathrm{filters}=0.4$, and single-photon detector efficiency (averaged over the two detectors) $\eta_\mathrm{SPD}=(\eta_\mathrm{SPD1}+\eta_\mathrm{SPD2})/2=0.58$ is then $\eta_\mathrm{herald}=p_\mathrm{S}\eta_\mathrm{cav}\eta_\mathrm{fc}\eta_\mathrm{filters}\eta_\mathrm{SPD}=7\times 10^{-4}$. The efficiency of the conversion step with anti-Stokes scattering probability $p_\mathrm{aS}=0.07$ is similarly obtained as $\eta_\mathrm{conv}=p_\mathrm{aS}\eta_\mathrm{cav}\eta_\mathrm{fc}\eta_\mathrm{filters}\eta_\mathrm{SPD}=4\times 10^{-3}$. The corresponding rate of generated single photons with the repetition period of the pulse sequence $T_\mathrm{rep} = \qty{10}{\micro s}$ (repetition rate $R_\mathrm{rep}=\qty{100}{kHz}$) is $R_\mathrm{ph} = \qty{0.23}{Hz}$.

\subsubsection*{Projected entanglement rates}

\begin{figure}
	\centering
	\includegraphics[width = 8.8cm]{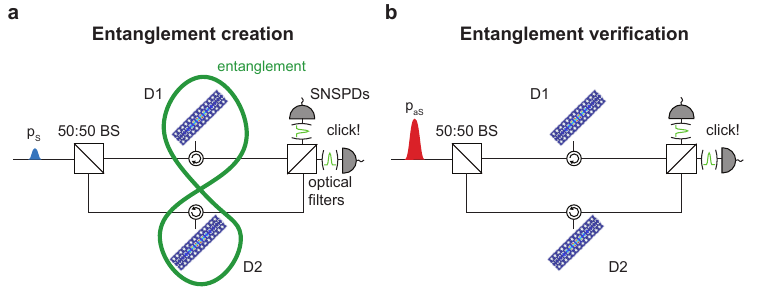}
	\caption{\textbf{Protocol for remote entanglement generation and verification.} \textbf{a} Heralded creation of entanglement between two mechanical oscillators embedded in a phase-stabilized Mach-Zehnder interferometer. A blue-detuned pump pulse generates a phonon in one of the two mechanical modes. The second beam splitter erases the which-path information of the generated Stokes photons. Detection of a single photon at either of the two outputs of the interferometer projects the two mechanical modes into a maximally entangled state. \textbf{b} The created entangled state can be verified by readout of the mechanical states through a red-detuned readout pulse.}
	\label{Fig:S11_entanglement}
\end{figure}

Remote entanglement between the mechanical modes of two OMCs can be created by embedding two such devices with matching mechanical and optical resonance frequencies into a phase-stabilized Mach-Zehnder interferometer~\cite{riedinger2018remote}. This is an implementation of the Duan-Lukin-Cirac-Zoller (DLCZ) protocol for entanglement distribution in quantum networks~\cite{duan_long-distance_2001}. By sending a blue-detuned pump pulse with Stokes scattering probability $p_\mathrm{S}$ into the interferometer and detecting a click on either of two single photon detectors at the two output ports of the interferometer, the joint state of the two mechanical modes is projected into a maximally entangled state $\ket{\Psi}=(\ket{10}+\ket{01})/\sqrt{2}$. To verify the generated entanglement, a red-detuned readout pulse with anti-Stokes scattering probability $p_\mathrm{aS}$ is sent into the interferometer. An entanglement witness can be defined for such an optomechanical entanglement experiment $R_\mathrm{m}$~\cite{borkje_proposal_2011}, which can be expressed in terms of the cross-correlation functions between the Stokes and anti-Stokes click events as
\begin{align}
\label{entanglement_witness}
    R_\mathrm{m}(\theta,j)=4\frac{g_{\mathrm{r}_1,\mathrm{w}_j}^{(2)}(\theta)g_{\mathrm{r}_2,\mathrm{w}_j}^{(2)}(\theta)-1}{\left[g_{\mathrm{r}_1,\mathrm{w}_j}^{(2)}(\theta)-g_{\mathrm{r}_2,\mathrm{w}_j}^{(2)}(\theta)\right]^2},
\end{align}
where $\theta$ is a phase given by the relative phase between the two interferometer arms and the second-order cross-correlation functions between the write (Stokes field) and read (anti-Stokes field) pulses on the two detectors are defined as $g_{\mathrm{r}_i,\mathrm{w}_j}^{(2)}=\langle a_{\mathrm{aS},i}^\dagger a_{\mathrm{S},j}^\dagger a_{\mathrm{S},i} a_{\mathrm{aS},j} \rangle/(\langle a_{\mathrm{aS},i}^\dagger a_{\mathrm{aS},i}\rangle \langle a_{\mathrm{S},j}^\dagger a_{\mathrm{S},j}\rangle)$. A value $R_\mathrm{m}<1$ demonstrates entanglement of the two mechanical modes.

Thermal noise degrades the cross-correlation between the write and read pulse and leads to an increased value of $R_\mathrm{m}$. A theoretical analysis shows that, in general, a total thermal phonon occupancy during the read pulse of $n_\mathrm{th} < 0.26$ is required to obtain $R_\mathrm{m}<1$~\cite{borkje_proposal_2011}. To achieve a statistically significant violation of this bound, previous demonstrations of remote entanglement between two one-dimensional nanobeam optomechanical crystals operated at reduced thermal phonon occupancy of $n_\mathrm{th} = 0.12$ limiting the Stokes (anti-Stokes) scattering probabilities to $p_\mathrm{S} = 0.007$ ($p_\mathrm{aS} = 0.034$)~\cite{riedinger2018remote}. Combined with the repetition period of the experiment $R_\mathrm{rep} = \qty{20}{kHz}$ and the total efficiency of the detection path $\eta=0.04$, entanglement of the two mechanical oscillators was heralded at a rate of $R_\mathrm{ent}=p_\mathrm{S}\eta R_\mathrm{rep}=\qty{6}{Hz}$. Coincidence events between the write and read pulse required for verification of entanglement occurred at a rate of $R_\mathrm{coinc}=p_\mathrm{S} p_\mathrm{aS} \eta^2 R_\mathrm{rep}=\qty{27}{h^{-1}}$.

Using our 2D OMC device, we can achieve similar levels of thermal noise with increased optomechanical scattering probabilities $p_\mathrm{S} = 0.02$ ($p_\mathrm{aS} = 0.16$). At these scattering probabilities the added thermal noise from the blue-detuned write pulse is $n_{\mathrm{th},1}=0.043$ as calibrated from the measurements shown in Fig.~\ref{Fig:S7_sideband_asymmetry}. The red-detuned readout pulse adds an additional thermal noise of $n_{\mathrm{th},2}=0.075$ as deduced from the thermometry measurement presented in the main text. The total added thermal noise is thus $n_\mathrm{tot} = 0.118$. Combined with the repetition rate of our experiment $R_\mathrm{rep} = \qty{100}{kHz}$ and a total detection efficiency of $\eta = 0.05$, we can project a heralded entanglement generation rate in a remote entanglement experiment of $R_\mathrm{ent}=\qty{100}{Hz}$ and a rate of coincidence events of $R_\mathrm{coinc}=\qty{2.9e3}{h^{-1}}$ \textemdash an improvement by almost two orders of magnitude over previous demonstrations~\cite{riedinger2018remote}.

Through technical improvements to our optical setup by using adiabatically tapered fiber coupling with $\eta_\mathrm{fc}=0.85$~\cite{groblacher_highly_2013}, using improved single-photon detectors with state-of-the-art detection efficiency $\eta_\mathrm{SPD}=0.9$, as well as straightforward adjustment of the cavity impedance ratio to operate in the ideal configuration of critical coupling $\eta_\mathrm{cav}=0.5$ (corresponding to a total detection path efficiency of $\eta = 0.15$), the entanglement generation and verification rate can be boosted to $R_\mathrm{ent}=\qty{300}{Hz}$ and $R_\mathrm{coinc}=\qty{2.5e4}{h^{-1}}$, respectively.

More elaborate improvements to the device geometry, such as using an evanescent coupling geometry~\cite{sonar_high-efficiency_2025}, can be used to further reduce the thermal noise by a factor of six and thus enable higher entanglement rates by using larger optomechanical scattering rates. We can estimate the improvement to the entanglement rate attainable using such side-coupled 2D OMC devices by limiting the thermal noise to $n_\mathrm{th,max}=0.12$, the same level as in previous demonstrations of remote entanglement~\cite{riedinger2018remote}, and assuming fixed optomechanical phonon-photon conversion efficiency on the readout pulse for entanglement verification as above $p_\mathrm{aS} = 0.16$. The thermal noise from the readout pulse would be reduced to $n_{\mathrm{th},2}=0.013$, allowing a maximum added noise from the blue write pulse of $n_\mathrm{noise,1,max}=n_\mathrm{th,max}-n_{\mathrm{th},2}=0.107$. We can estimate the added noise from the write pulse from the measurement results in Fig.~\ref{Fig:S7_sideband_asymmetry} by dividing the thermal occupation from the prepulse by a factor of six to account for the reduction of thermal noise with an improved device geometry. The reduced thermal noise would allow for a substantial increase of Stokes scattering probability $p_\mathrm{S}$. Importantly, the thermal noise would be so low that the Stokes scattering probability is comparable or even larger than the added thermal noise from the prepulse. In this case, it is important to account for additional noise in the heralding process due to higher-order excitations from the two-mode squeezing interaction of the write pulse that are on the order of $p_\mathrm{S}$. The total noise originating write pulse is then $n_\mathrm{noise,1}=n_{\mathrm{th},1}+p_\mathrm{S}$. We find that $n_\mathrm{noise,1}<n_\mathrm{noise,1,max}$ up to a Stokes scattering probability of $p_\mathrm{S} \approx 0.08$ with a projected added thermal noise $n_{\mathrm{th},1} \approx 0.02$. This four-fold increase in Stokes scattering probability would translate to a four-fold increase in the entanglement rate $R_\mathrm{ent}=\qty{1.2}{kHz}$.

We note that while the outcoupling efficiency of photons from our device can, in principle, be further increased by using an overcoupled optical cavity, the corresponding increase in total optical cavity linewidth $\kappa$ would lead to a reduction in optomechanical scattering probability at the same intracavity photon number. For the pulse scheme used in this experiment, a critically coupled cavity provides the best trade-off between scattering probability and photon outcoupling efficiency~\cite{riedinger2018single}.

\subsubsection*{Comparison to other quantum network platforms}

We compare the projected entanglement rates attainable using 2D OMCs to the entanglement rates demonstrated in recent state-of-the-art quantum network experiments. Firstly, we compare to an experiment where a single-photon interference scheme is used to generate remote entanglement between two nitrogen vacancy centers in diamond at a heralded entanglement rate of $\qty{39}{Hz}$~\cite{humphreys_deterministic_2018}.

Secondly, we compare to an emissive quantum memory formed by an atomic ensemble~\cite{yu_entanglement_2020}. In this work, the highest entanglement rate is achieved in an experiment in which two atomic quantum memories are connected by a \qty{10}{m} long fiber and entanglement is heralded by a single-photon interference scheme. The experiment runs with a repetition time of $T_\mathrm{rep}=\qty{20}{ms}$ out of which $T_\mathrm{load}=\qty{18}{ms}$ are allocated to load and cool atoms and $T_\mathrm{ent}=\qty{2}{ms}$ are allocated to perform entanglement trials with a repetition period of $\tau_\mathrm{ent}=\qty{5}{\micro s}$. The authors achieve an entanglement probability per trial of $P_\mathrm{ent}=0.014$ yielding a total entanglement rate of $R_\mathrm{ent}=P_\mathrm{ent}T_\mathrm{ent}/(\tau_\mathrm{ent}T_\mathrm{rep})=\qty{280}{Hz}$.

In both~\cite{humphreys_deterministic_2018} and~\cite{yu_entanglement_2020}, the photons used for heralding the entanglement are not at telecom wavelength. For practical applications in remote quantum networks, wavelength conversion would be required and would lead to a substantial reduction in entanglement rate.

Lastly, we compare to an experiment in which single-photons generated by spontaneous parametric down-conversion are stored in atomic frequency comb quantum memories in solid-state crystals~\cite{lago-rivera_telecom-heralded_2021}. Entanglement between two quantum memories is established in a heralded fashion in a single-photon interference scheme in which the two memories are connected by \qty{25}{m} of optical fiber. The authors achieve a record-high entanglement rate of $\qty{1.4}{kHz}$.


\begin{thebibliography}{73}%
\makeatletter
\providecommand \@ifxundefined [1]{%
 \@ifx{#1\undefined}
}%
\providecommand \@ifnum [1]{%
 \ifnum #1\expandafter \@firstoftwo
 \else \expandafter \@secondoftwo
 \fi
}%
\providecommand \@ifx [1]{%
 \ifx #1\expandafter \@firstoftwo
 \else \expandafter \@secondoftwo
 \fi
}%
\providecommand \natexlab [1]{#1}%
\providecommand \enquote  [1]{``#1''}%
\providecommand \bibnamefont  [1]{#1}%
\providecommand \bibfnamefont [1]{#1}%
\providecommand \citenamefont [1]{#1}%
\providecommand \href@noop [0]{\@secondoftwo}%
\providecommand \href [0]{\begingroup \@sanitize@url \@href}%
\providecommand \@href[1]{\@@startlink{#1}\@@href}%
\providecommand \@@href[1]{\endgroup#1\@@endlink}%
\providecommand \@sanitize@url [0]{\catcode `\\12\catcode `\$12\catcode
  `\&12\catcode `\#12\catcode `\^12\catcode `\_12\catcode `\%12\relax}%
\providecommand \@@startlink[1]{}%
\providecommand \@@endlink[0]{}%
\providecommand \url  [0]{\begingroup\@sanitize@url \@url }%
\providecommand \@url [1]{\endgroup\@href {#1}{\urlprefix }}%
\providecommand \urlprefix  [0]{URL }%
\providecommand \Eprint [0]{\href }%
\providecommand \doibase [0]{https://doi.org/}%
\providecommand \selectlanguage [0]{\@gobble}%
\providecommand \bibinfo  [0]{\@secondoftwo}%
\providecommand \bibfield  [0]{\@secondoftwo}%
\providecommand \translation [1]{[#1]}%
\providecommand \BibitemOpen [0]{}%
\providecommand \bibitemStop [0]{}%
\providecommand \bibitemNoStop [0]{.\EOS\space}%
\providecommand \EOS [0]{\spacefactor3000\relax}%
\providecommand \BibitemShut  [1]{\csname bibitem#1\endcsname}%
\let\auto@bib@innerbib\@empty
%</preamble>
\bibitem [{\citenamefont {Kimble}(2008)}]{kimble_quantum_2008}%
  \BibitemOpen
  \bibfield  {author} {\bibinfo {author} {\bibfnamefont {H.~J.}\ \bibnamefont
  {Kimble}},\ }\href {https://doi.org/10.1038/nature07127} {\bibfield
  {journal} {\bibinfo  {journal} {Nature}\ }\textbf {\bibinfo {volume} {453}},\
  \bibinfo {pages} {1023} (\bibinfo {year} {2008})}\BibitemShut {NoStop}%
\bibitem [{\citenamefont {Sangouard}\ \emph {et~al.}(2011)\citenamefont
  {Sangouard}, \citenamefont {Simon}, \citenamefont {De~Riedmatten},\ and\
  \citenamefont {Gisin}}]{sangouard_quantum_2011}%
  \BibitemOpen
  \bibfield  {author} {\bibinfo {author} {\bibfnamefont {N.}~\bibnamefont
  {Sangouard}}, \bibinfo {author} {\bibfnamefont {C.}~\bibnamefont {Simon}},
  \bibinfo {author} {\bibfnamefont {H.}~\bibnamefont {De~Riedmatten}},\ and\
  \bibinfo {author} {\bibfnamefont {N.}~\bibnamefont {Gisin}},\ }\href
  {https://doi.org/10.1103/RevModPhys.83.33} {\bibfield  {journal} {\bibinfo
  {journal} {Rev. Mod. Phys.}\ }\textbf {\bibinfo {volume} {83}},\ \bibinfo
  {pages} {33} (\bibinfo {year} {2011})}\BibitemShut {NoStop}%
\bibitem [{\citenamefont {Azuma}\ \emph {et~al.}(2023)\citenamefont {Azuma},
  \citenamefont {Economou}, \citenamefont {Elkouss}, \citenamefont {Hilaire},
  \citenamefont {Jiang}, \citenamefont {Lo},\ and\ \citenamefont
  {Tzitrin}}]{azuma_quantum_2023}%
  \BibitemOpen
  \bibfield  {author} {\bibinfo {author} {\bibfnamefont {K.}~\bibnamefont
  {Azuma}}, \bibinfo {author} {\bibfnamefont {S.~E.}\ \bibnamefont {Economou}},
  \bibinfo {author} {\bibfnamefont {D.}~\bibnamefont {Elkouss}}, \bibinfo
  {author} {\bibfnamefont {P.}~\bibnamefont {Hilaire}}, \bibinfo {author}
  {\bibfnamefont {L.}~\bibnamefont {Jiang}}, \bibinfo {author} {\bibfnamefont
  {H.-K.}\ \bibnamefont {Lo}},\ and\ \bibinfo {author} {\bibfnamefont
  {I.}~\bibnamefont {Tzitrin}},\ }\href
  {https://doi.org/10.1103/RevModPhys.95.045006} {\bibfield  {journal}
  {\bibinfo  {journal} {Rev. Mod. Phys.}\ }\textbf {\bibinfo {volume} {95}},\
  \bibinfo {pages} {045006} (\bibinfo {year} {2023})}\BibitemShut {NoStop}%
\bibitem [{\citenamefont {Matsukevich}\ \emph {et~al.}(2006)\citenamefont
  {Matsukevich}, \citenamefont {Chanelière}, \citenamefont {Jenkins},
  \citenamefont {Lan}, \citenamefont {Kennedy},\ and\ \citenamefont
  {Kuzmich}}]{matsukevich_entanglement_2006}%
  \BibitemOpen
  \bibfield  {author} {\bibinfo {author} {\bibfnamefont {D.~N.}\ \bibnamefont
  {Matsukevich}}, \bibinfo {author} {\bibfnamefont {T.}~\bibnamefont
  {Chanelière}}, \bibinfo {author} {\bibfnamefont {S.~D.}\ \bibnamefont
  {Jenkins}}, \bibinfo {author} {\bibfnamefont {S.-Y.}\ \bibnamefont {Lan}},
  \bibinfo {author} {\bibfnamefont {T.~A.~B.}\ \bibnamefont {Kennedy}},\ and\
  \bibinfo {author} {\bibfnamefont {A.}~\bibnamefont {Kuzmich}},\ }\href
  {https://doi.org/10.1103/PhysRevLett.96.030405} {\bibfield  {journal}
  {\bibinfo  {journal} {Phys. Rev. Lett.}\ }\textbf {\bibinfo {volume} {96}},\
  \bibinfo {pages} {030405} (\bibinfo {year} {2006})}\BibitemShut {NoStop}%
\bibitem [{\citenamefont {Chou}\ \emph {et~al.}(2005)\citenamefont {Chou},
  \citenamefont {de~Riedmatten}, \citenamefont {Felinto}, \citenamefont
  {Polyakov}, \citenamefont {van Enk},\ and\ \citenamefont
  {Kimble}}]{chou_measurement-induced_2005}%
  \BibitemOpen
  \bibfield  {author} {\bibinfo {author} {\bibfnamefont {C.~W.}\ \bibnamefont
  {Chou}}, \bibinfo {author} {\bibfnamefont {H.}~\bibnamefont {de~Riedmatten}},
  \bibinfo {author} {\bibfnamefont {D.}~\bibnamefont {Felinto}}, \bibinfo
  {author} {\bibfnamefont {S.~V.}\ \bibnamefont {Polyakov}}, \bibinfo {author}
  {\bibfnamefont {S.~J.}\ \bibnamefont {van Enk}},\ and\ \bibinfo {author}
  {\bibfnamefont {H.~J.}\ \bibnamefont {Kimble}},\ }\href
  {https://doi.org/10.1038/nature04353} {\bibfield  {journal} {\bibinfo
  {journal} {Nature}\ }\textbf {\bibinfo {volume} {438}},\ \bibinfo {pages}
  {828} (\bibinfo {year} {2005})}\BibitemShut {NoStop}%
\bibitem [{\citenamefont {Krutyanskiy}\ \emph {et~al.}(2023)\citenamefont
  {Krutyanskiy}, \citenamefont {Canteri}, \citenamefont {Meraner},
  \citenamefont {Bate}, \citenamefont {Krcmarsky}, \citenamefont {Schupp},
  \citenamefont {Sangouard},\ and\ \citenamefont
  {Lanyon}}]{krutyanskiy_telecom-wavelength_2023}%
  \BibitemOpen
  \bibfield  {author} {\bibinfo {author} {\bibfnamefont {V.}~\bibnamefont
  {Krutyanskiy}}, \bibinfo {author} {\bibfnamefont {M.}~\bibnamefont
  {Canteri}}, \bibinfo {author} {\bibfnamefont {M.}~\bibnamefont {Meraner}},
  \bibinfo {author} {\bibfnamefont {J.}~\bibnamefont {Bate}}, \bibinfo {author}
  {\bibfnamefont {V.}~\bibnamefont {Krcmarsky}}, \bibinfo {author}
  {\bibfnamefont {J.}~\bibnamefont {Schupp}}, \bibinfo {author} {\bibfnamefont
  {N.}~\bibnamefont {Sangouard}},\ and\ \bibinfo {author} {\bibfnamefont
  {B.}~\bibnamefont {Lanyon}},\ }\href
  {https://doi.org/10.1103/PhysRevLett.130.213601} {\bibfield  {journal}
  {\bibinfo  {journal} {Phys. Rev. Lett.}\ }\textbf {\bibinfo {volume} {130}},\
  \bibinfo {pages} {213601} (\bibinfo {year} {2023})}\BibitemShut {NoStop}%
\bibitem [{\citenamefont {Ritter}\ \emph {et~al.}(2012)\citenamefont {Ritter},
  \citenamefont {Nölleke}, \citenamefont {Hahn}, \citenamefont {Reiserer},
  \citenamefont {Neuzner}, \citenamefont {Uphoff}, \citenamefont {Mücke},
  \citenamefont {Figueroa}, \citenamefont {Bochmann},\ and\ \citenamefont
  {Rempe}}]{ritter_elementary_2012}%
  \BibitemOpen
  \bibfield  {author} {\bibinfo {author} {\bibfnamefont {S.}~\bibnamefont
  {Ritter}}, \bibinfo {author} {\bibfnamefont {C.}~\bibnamefont {Nölleke}},
  \bibinfo {author} {\bibfnamefont {C.}~\bibnamefont {Hahn}}, \bibinfo {author}
  {\bibfnamefont {A.}~\bibnamefont {Reiserer}}, \bibinfo {author}
  {\bibfnamefont {A.}~\bibnamefont {Neuzner}}, \bibinfo {author} {\bibfnamefont
  {M.}~\bibnamefont {Uphoff}}, \bibinfo {author} {\bibfnamefont
  {M.}~\bibnamefont {Mücke}}, \bibinfo {author} {\bibfnamefont
  {E.}~\bibnamefont {Figueroa}}, \bibinfo {author} {\bibfnamefont
  {J.}~\bibnamefont {Bochmann}},\ and\ \bibinfo {author} {\bibfnamefont
  {G.}~\bibnamefont {Rempe}},\ }\href {https://doi.org/10.1038/nature11023}
  {\bibfield  {journal} {\bibinfo  {journal} {Nature}\ }\textbf {\bibinfo
  {volume} {484}},\ \bibinfo {pages} {195} (\bibinfo {year}
  {2012})}\BibitemShut {NoStop}%
\bibitem [{\citenamefont {Pompili}\ \emph {et~al.}(2021)\citenamefont
  {Pompili}, \citenamefont {Hermans}, \citenamefont {Baier}, \citenamefont
  {Beukers}, \citenamefont {Humphreys}, \citenamefont {Schouten}, \citenamefont
  {Vermeulen}, \citenamefont {Tiggelman}, \citenamefont {dos Santos~Martins},
  \citenamefont {Dirkse}, \citenamefont {Wehner},\ and\ \citenamefont
  {Hanson}}]{pompili_realization_2021}%
  \BibitemOpen
  \bibfield  {author} {\bibinfo {author} {\bibfnamefont {M.}~\bibnamefont
  {Pompili}}, \bibinfo {author} {\bibfnamefont {S.~L.~N.}\ \bibnamefont
  {Hermans}}, \bibinfo {author} {\bibfnamefont {S.}~\bibnamefont {Baier}},
  \bibinfo {author} {\bibfnamefont {H.~K.~C.}\ \bibnamefont {Beukers}},
  \bibinfo {author} {\bibfnamefont {P.~C.}\ \bibnamefont {Humphreys}}, \bibinfo
  {author} {\bibfnamefont {R.~N.}\ \bibnamefont {Schouten}}, \bibinfo {author}
  {\bibfnamefont {R.~F.~L.}\ \bibnamefont {Vermeulen}}, \bibinfo {author}
  {\bibfnamefont {M.~J.}\ \bibnamefont {Tiggelman}}, \bibinfo {author}
  {\bibfnamefont {L.}~\bibnamefont {dos Santos~Martins}}, \bibinfo {author}
  {\bibfnamefont {B.}~\bibnamefont {Dirkse}}, \bibinfo {author} {\bibfnamefont
  {S.}~\bibnamefont {Wehner}},\ and\ \bibinfo {author} {\bibfnamefont
  {R.}~\bibnamefont {Hanson}},\ }\href
  {https://doi.org/10.1126/science.abg1919} {\bibfield  {journal} {\bibinfo
  {journal} {Science}\ }\textbf {\bibinfo {volume} {372}},\ \bibinfo {pages}
  {259} (\bibinfo {year} {2021})}\BibitemShut {NoStop}%
\bibitem [{\citenamefont {Hermans}\ \emph {et~al.}(2022)\citenamefont
  {Hermans}, \citenamefont {Pompili}, \citenamefont {Beukers}, \citenamefont
  {Baier}, \citenamefont {Borregaard},\ and\ \citenamefont
  {Hanson}}]{hermans_qubit_2022}%
  \BibitemOpen
  \bibfield  {author} {\bibinfo {author} {\bibfnamefont {S.~L.~N.}\
  \bibnamefont {Hermans}}, \bibinfo {author} {\bibfnamefont {M.}~\bibnamefont
  {Pompili}}, \bibinfo {author} {\bibfnamefont {H.~K.~C.}\ \bibnamefont
  {Beukers}}, \bibinfo {author} {\bibfnamefont {S.}~\bibnamefont {Baier}},
  \bibinfo {author} {\bibfnamefont {J.}~\bibnamefont {Borregaard}},\ and\
  \bibinfo {author} {\bibfnamefont {R.}~\bibnamefont {Hanson}},\ }\href
  {https://doi.org/10.1038/s41586-022-04697-y} {\bibfield  {journal} {\bibinfo
  {journal} {Nature}\ }\textbf {\bibinfo {volume} {605}},\ \bibinfo {pages}
  {663} (\bibinfo {year} {2022})}\BibitemShut {NoStop}%
\bibitem [{\citenamefont {Usmani}\ \emph {et~al.}(2012)\citenamefont {Usmani},
  \citenamefont {Clausen}, \citenamefont {Bussières}, \citenamefont
  {Sangouard}, \citenamefont {Afzelius},\ and\ \citenamefont
  {Gisin}}]{usmani_heralded_2012}%
  \BibitemOpen
  \bibfield  {author} {\bibinfo {author} {\bibfnamefont {I.}~\bibnamefont
  {Usmani}}, \bibinfo {author} {\bibfnamefont {C.}~\bibnamefont {Clausen}},
  \bibinfo {author} {\bibfnamefont {F.}~\bibnamefont {Bussières}}, \bibinfo
  {author} {\bibfnamefont {N.}~\bibnamefont {Sangouard}}, \bibinfo {author}
  {\bibfnamefont {M.}~\bibnamefont {Afzelius}},\ and\ \bibinfo {author}
  {\bibfnamefont {N.}~\bibnamefont {Gisin}},\ }\href
  {https://doi.org/10.1038/nphoton.2012.34} {\bibfield  {journal} {\bibinfo
  {journal} {Nat. Photonics}\ }\textbf {\bibinfo {volume} {6}},\ \bibinfo
  {pages} {234} (\bibinfo {year} {2012})}\BibitemShut {NoStop}%
\bibitem [{\citenamefont {Lago-Rivera}\ \emph {et~al.}(2021)\citenamefont
  {Lago-Rivera}, \citenamefont {Grandi}, \citenamefont {Rakonjac},
  \citenamefont {Seri},\ and\ \citenamefont
  {de~Riedmatten}}]{lago-rivera_telecom-heralded_2021}%
  \BibitemOpen
  \bibfield  {author} {\bibinfo {author} {\bibfnamefont {D.}~\bibnamefont
  {Lago-Rivera}}, \bibinfo {author} {\bibfnamefont {S.}~\bibnamefont {Grandi}},
  \bibinfo {author} {\bibfnamefont {J.~V.}\ \bibnamefont {Rakonjac}}, \bibinfo
  {author} {\bibfnamefont {A.}~\bibnamefont {Seri}},\ and\ \bibinfo {author}
  {\bibfnamefont {H.}~\bibnamefont {de~Riedmatten}},\ }\href
  {https://doi.org/10.1038/s41586-021-03481-8} {\bibfield  {journal} {\bibinfo
  {journal} {Nature}\ }\textbf {\bibinfo {volume} {594}},\ \bibinfo {pages}
  {37} (\bibinfo {year} {2021})}\BibitemShut {NoStop}%
\bibitem [{\citenamefont {Aspelmeyer}\ \emph {et~al.}(2014)\citenamefont
  {Aspelmeyer}, \citenamefont {Kippenberg},\ and\ \citenamefont
  {Marquardt}}]{aspelmeyer_cavity_2014}%
  \BibitemOpen
  \bibfield  {author} {\bibinfo {author} {\bibfnamefont {M.}~\bibnamefont
  {Aspelmeyer}}, \bibinfo {author} {\bibfnamefont {T.~J.}\ \bibnamefont
  {Kippenberg}},\ and\ \bibinfo {author} {\bibfnamefont {F.}~\bibnamefont
  {Marquardt}},\ }\href {https://doi.org/10.1103/RevModPhys.86.1391} {\bibfield
   {journal} {\bibinfo  {journal} {Rev. Mod. Phys.}\ }\textbf {\bibinfo
  {volume} {86}},\ \bibinfo {pages} {1391} (\bibinfo {year}
  {2014})}\BibitemShut {NoStop}%
\bibitem [{\citenamefont {Riedinger}\ \emph {et~al.}(2016)\citenamefont
  {Riedinger}, \citenamefont {Hong}, \citenamefont {Norte}, \citenamefont
  {Slater}, \citenamefont {Shang}, \citenamefont {Krause}, \citenamefont
  {Anant}, \citenamefont {Aspelmeyer},\ and\ \citenamefont
  {Gröblacher}}]{riedinger_non-classical_2016}%
  \BibitemOpen
  \bibfield  {author} {\bibinfo {author} {\bibfnamefont {R.}~\bibnamefont
  {Riedinger}}, \bibinfo {author} {\bibfnamefont {S.}~\bibnamefont {Hong}},
  \bibinfo {author} {\bibfnamefont {R.~A.}\ \bibnamefont {Norte}}, \bibinfo
  {author} {\bibfnamefont {J.~A.}\ \bibnamefont {Slater}}, \bibinfo {author}
  {\bibfnamefont {J.}~\bibnamefont {Shang}}, \bibinfo {author} {\bibfnamefont
  {A.~G.}\ \bibnamefont {Krause}}, \bibinfo {author} {\bibfnamefont
  {V.}~\bibnamefont {Anant}}, \bibinfo {author} {\bibfnamefont
  {M.}~\bibnamefont {Aspelmeyer}},\ and\ \bibinfo {author} {\bibfnamefont
  {S.}~\bibnamefont {Gröblacher}},\ }\href
  {https://doi.org/10.1038/nature16536} {\bibfield  {journal} {\bibinfo
  {journal} {Nature}\ }\textbf {\bibinfo {volume} {530}},\ \bibinfo {pages}
  {313} (\bibinfo {year} {2016})}\BibitemShut {NoStop}%
\bibitem [{\citenamefont {Barzanjeh}\ \emph {et~al.}(2022)\citenamefont
  {Barzanjeh}, \citenamefont {Xuereb}, \citenamefont {Gröblacher},
  \citenamefont {Paternostro}, \citenamefont {Regal},\ and\ \citenamefont
  {Weig}}]{barzanjeh_optomechanics_2022}%
  \BibitemOpen
  \bibfield  {author} {\bibinfo {author} {\bibfnamefont {S.}~\bibnamefont
  {Barzanjeh}}, \bibinfo {author} {\bibfnamefont {A.}~\bibnamefont {Xuereb}},
  \bibinfo {author} {\bibfnamefont {S.}~\bibnamefont {Gröblacher}}, \bibinfo
  {author} {\bibfnamefont {M.}~\bibnamefont {Paternostro}}, \bibinfo {author}
  {\bibfnamefont {C.~A.}\ \bibnamefont {Regal}},\ and\ \bibinfo {author}
  {\bibfnamefont {E.~M.}\ \bibnamefont {Weig}},\ }\href
  {https://doi.org/10.1038/s41567-021-01402-0} {\bibfield  {journal} {\bibinfo
  {journal} {Nat. Phys.}\ }\textbf {\bibinfo {volume} {18}},\ \bibinfo {pages}
  {15} (\bibinfo {year} {2022})}\BibitemShut {NoStop}%
\bibitem [{\citenamefont {MacCabe}\ \emph {et~al.}(2020)\citenamefont
  {MacCabe}, \citenamefont {Ren}, \citenamefont {Luo}, \citenamefont {Cohen},
  \citenamefont {Zhou}, \citenamefont {Sipahigil}, \citenamefont
  {Mirhosseini},\ and\ \citenamefont {Painter}}]{maccabe_nano-acoustic_2020}%
  \BibitemOpen
  \bibfield  {author} {\bibinfo {author} {\bibfnamefont {G.~S.}\ \bibnamefont
  {MacCabe}}, \bibinfo {author} {\bibfnamefont {H.}~\bibnamefont {Ren}},
  \bibinfo {author} {\bibfnamefont {J.}~\bibnamefont {Luo}}, \bibinfo {author}
  {\bibfnamefont {J.~D.}\ \bibnamefont {Cohen}}, \bibinfo {author}
  {\bibfnamefont {H.}~\bibnamefont {Zhou}}, \bibinfo {author} {\bibfnamefont
  {A.}~\bibnamefont {Sipahigil}}, \bibinfo {author} {\bibfnamefont
  {M.}~\bibnamefont {Mirhosseini}},\ and\ \bibinfo {author} {\bibfnamefont
  {O.}~\bibnamefont {Painter}},\ }\href
  {https://doi.org/10.1126/science.abc7312} {\bibfield  {journal} {\bibinfo
  {journal} {Science}\ }\textbf {\bibinfo {volume} {370}},\ \bibinfo {pages}
  {840} (\bibinfo {year} {2020})}\BibitemShut {NoStop}%
\bibitem [{\citenamefont {Wallucks}\ \emph {et~al.}(2020)\citenamefont
  {Wallucks}, \citenamefont {Marinković}, \citenamefont {Hensen},
  \citenamefont {Stockill},\ and\ \citenamefont
  {Gröblacher}}]{wallucks_quantum_2020}%
  \BibitemOpen
  \bibfield  {author} {\bibinfo {author} {\bibfnamefont {A.}~\bibnamefont
  {Wallucks}}, \bibinfo {author} {\bibfnamefont {I.}~\bibnamefont
  {Marinković}}, \bibinfo {author} {\bibfnamefont {B.}~\bibnamefont {Hensen}},
  \bibinfo {author} {\bibfnamefont {R.}~\bibnamefont {Stockill}},\ and\
  \bibinfo {author} {\bibfnamefont {S.}~\bibnamefont {Gröblacher}},\ }\href
  {https://doi.org/10.1038/s41567-020-0891-z} {\bibfield  {journal} {\bibinfo
  {journal} {Nat. Phys.}\ }\textbf {\bibinfo {volume} {16}},\ \bibinfo {pages}
  {772} (\bibinfo {year} {2020})}\BibitemShut {NoStop}%
\bibitem [{\citenamefont {Kuruma}\ \emph {et~al.}(2025)\citenamefont {Kuruma},
  \citenamefont {Pingault}, \citenamefont {Chia}, \citenamefont {Haas},
  \citenamefont {Joe}, \citenamefont {Assumpcao}, \citenamefont {Ding},
  \citenamefont {Jin}, \citenamefont {Xin}, \citenamefont {Yeh}, \citenamefont
  {Sinclair},\ and\ \citenamefont {Loncar}}]{kuruma_controlling_2025}%
  \BibitemOpen
  \bibfield  {author} {\bibinfo {author} {\bibfnamefont {K.}~\bibnamefont
  {Kuruma}}, \bibinfo {author} {\bibfnamefont {B.}~\bibnamefont {Pingault}},
  \bibinfo {author} {\bibfnamefont {C.}~\bibnamefont {Chia}}, \bibinfo {author}
  {\bibfnamefont {M.}~\bibnamefont {Haas}}, \bibinfo {author} {\bibfnamefont
  {G.~D.}\ \bibnamefont {Joe}}, \bibinfo {author} {\bibfnamefont {D.~R.}\
  \bibnamefont {Assumpcao}}, \bibinfo {author} {\bibfnamefont {S.~W.}\
  \bibnamefont {Ding}}, \bibinfo {author} {\bibfnamefont {C.}~\bibnamefont
  {Jin}}, \bibinfo {author} {\bibfnamefont {C.~J.}\ \bibnamefont {Xin}},
  \bibinfo {author} {\bibfnamefont {M.}~\bibnamefont {Yeh}}, \bibinfo {author}
  {\bibfnamefont {N.}~\bibnamefont {Sinclair}},\ and\ \bibinfo {author}
  {\bibfnamefont {M.}~\bibnamefont {Loncar}},\ }\href
  {https://doi.org/10.1038/s41567-024-02697-5} {\bibfield  {journal} {\bibinfo
  {journal} {Nat. Phys.}\ }\textbf {\bibinfo {volume} {21}},\ \bibinfo {pages}
  {77} (\bibinfo {year} {2025})}\BibitemShut {NoStop}%
\bibitem [{\citenamefont {Joe}\ \emph {et~al.}(2025)\citenamefont {Joe},
  \citenamefont {Haas}, \citenamefont {Kuruma}, \citenamefont {Jin},
  \citenamefont {Kang}, \citenamefont {Ding}, \citenamefont {Chia},
  \citenamefont {Warner}, \citenamefont {Pingault}, \citenamefont {Machielse},
  \citenamefont {Meesala},\ and\ \citenamefont
  {Loncar}}]{joe_observation_2025}%
  \BibitemOpen
  \bibfield  {author} {\bibinfo {author} {\bibfnamefont {G.}~\bibnamefont
  {Joe}}, \bibinfo {author} {\bibfnamefont {M.}~\bibnamefont {Haas}}, \bibinfo
  {author} {\bibfnamefont {K.}~\bibnamefont {Kuruma}}, \bibinfo {author}
  {\bibfnamefont {C.}~\bibnamefont {Jin}}, \bibinfo {author} {\bibfnamefont
  {D.~D.}\ \bibnamefont {Kang}}, \bibinfo {author} {\bibfnamefont
  {S.}~\bibnamefont {Ding}}, \bibinfo {author} {\bibfnamefont {C.}~\bibnamefont
  {Chia}}, \bibinfo {author} {\bibfnamefont {H.}~\bibnamefont {Warner}},
  \bibinfo {author} {\bibfnamefont {B.}~\bibnamefont {Pingault}}, \bibinfo
  {author} {\bibfnamefont {B.}~\bibnamefont {Machielse}}, \bibinfo {author}
  {\bibfnamefont {S.}~\bibnamefont {Meesala}},\ and\ \bibinfo {author}
  {\bibfnamefont {M.}~\bibnamefont {Loncar}},\ }\href@noop {} {\bibinfo {title}
  {Observation of the acoustic {Purcell} effect with a color-center and a
  nanomechanical resonator}} (\bibinfo {year} {2025}),\ \Eprint
  {https://arxiv.org/abs/2503.09946} {arXiv:2503.09946} \BibitemShut {NoStop}%
\bibitem [{\citenamefont {Spinnler}\ \emph {et~al.}(2024)\citenamefont
  {Spinnler}, \citenamefont {Nguyen}, \citenamefont {Wang}, \citenamefont
  {Zhai}, \citenamefont {Javadi}, \citenamefont {Erbe}, \citenamefont {Scholz},
  \citenamefont {Wieck}, \citenamefont {Ludwig}, \citenamefont {Lodahl},
  \citenamefont {Midolo},\ and\ \citenamefont
  {Warburton}}]{spinnler_single-photon_2024}%
  \BibitemOpen
  \bibfield  {author} {\bibinfo {author} {\bibfnamefont {C.}~\bibnamefont
  {Spinnler}}, \bibinfo {author} {\bibfnamefont {G.~N.}\ \bibnamefont
  {Nguyen}}, \bibinfo {author} {\bibfnamefont {Y.}~\bibnamefont {Wang}},
  \bibinfo {author} {\bibfnamefont {L.}~\bibnamefont {Zhai}}, \bibinfo {author}
  {\bibfnamefont {A.}~\bibnamefont {Javadi}}, \bibinfo {author} {\bibfnamefont
  {M.}~\bibnamefont {Erbe}}, \bibinfo {author} {\bibfnamefont {S.}~\bibnamefont
  {Scholz}}, \bibinfo {author} {\bibfnamefont {A.~D.}\ \bibnamefont {Wieck}},
  \bibinfo {author} {\bibfnamefont {A.}~\bibnamefont {Ludwig}}, \bibinfo
  {author} {\bibfnamefont {P.}~\bibnamefont {Lodahl}}, \bibinfo {author}
  {\bibfnamefont {L.}~\bibnamefont {Midolo}},\ and\ \bibinfo {author}
  {\bibfnamefont {R.~J.}\ \bibnamefont {Warburton}},\ }\href
  {https://doi.org/10.1038/s41467-024-53882-2} {\bibfield  {journal} {\bibinfo
  {journal} {Nat. Commun.}\ }\textbf {\bibinfo {volume} {15}},\ \bibinfo
  {pages} {9509} (\bibinfo {year} {2024})}\BibitemShut {NoStop}%
\bibitem [{\citenamefont {Mirhosseini}\ \emph {et~al.}(2020)\citenamefont
  {Mirhosseini}, \citenamefont {Sipahigil}, \citenamefont {Kalaee},\ and\
  \citenamefont {Painter}}]{mirhosseini_superconducting_2020}%
  \BibitemOpen
  \bibfield  {author} {\bibinfo {author} {\bibfnamefont {M.}~\bibnamefont
  {Mirhosseini}}, \bibinfo {author} {\bibfnamefont {A.}~\bibnamefont
  {Sipahigil}}, \bibinfo {author} {\bibfnamefont {M.}~\bibnamefont {Kalaee}},\
  and\ \bibinfo {author} {\bibfnamefont {O.}~\bibnamefont {Painter}},\ }\href
  {https://doi.org/10.1038/s41586-020-3038-6} {\bibfield  {journal} {\bibinfo
  {journal} {Nature}\ }\textbf {\bibinfo {volume} {588}},\ \bibinfo {pages}
  {599} (\bibinfo {year} {2020})}\BibitemShut {NoStop}%
\bibitem [{\citenamefont {Forsch}\ \emph {et~al.}(2020)\citenamefont {Forsch},
  \citenamefont {Stockill}, \citenamefont {Wallucks}, \citenamefont
  {Marinković}, \citenamefont {Gärtner}, \citenamefont {Norte}, \citenamefont
  {van Otten}, \citenamefont {Fiore}, \citenamefont {Srinivasan},\ and\
  \citenamefont {Gröblacher}}]{forsch_microwave--optics_2020}%
  \BibitemOpen
  \bibfield  {author} {\bibinfo {author} {\bibfnamefont {M.}~\bibnamefont
  {Forsch}}, \bibinfo {author} {\bibfnamefont {R.}~\bibnamefont {Stockill}},
  \bibinfo {author} {\bibfnamefont {A.}~\bibnamefont {Wallucks}}, \bibinfo
  {author} {\bibfnamefont {I.}~\bibnamefont {Marinković}}, \bibinfo {author}
  {\bibfnamefont {C.}~\bibnamefont {Gärtner}}, \bibinfo {author}
  {\bibfnamefont {R.~A.}\ \bibnamefont {Norte}}, \bibinfo {author}
  {\bibfnamefont {F.}~\bibnamefont {van Otten}}, \bibinfo {author}
  {\bibfnamefont {A.}~\bibnamefont {Fiore}}, \bibinfo {author} {\bibfnamefont
  {K.}~\bibnamefont {Srinivasan}},\ and\ \bibinfo {author} {\bibfnamefont
  {S.}~\bibnamefont {Gröblacher}},\ }\href
  {https://doi.org/10.1038/s41567-019-0673-7} {\bibfield  {journal} {\bibinfo
  {journal} {Nat. Phys.}\ }\textbf {\bibinfo {volume} {16}},\ \bibinfo {pages}
  {69} (\bibinfo {year} {2020})}\BibitemShut {NoStop}%
\bibitem [{\citenamefont {van Thiel}\ \emph {et~al.}(2025)\citenamefont {van
  Thiel}, \citenamefont {Weaver}, \citenamefont {Berto}, \citenamefont
  {Duivestein}, \citenamefont {Lemang}, \citenamefont {Schuurman},
  \citenamefont {{\v{Z}}emli{\v{c}}ka}, \citenamefont {Hijazi}, \citenamefont
  {Bernasconi}, \citenamefont {Ferrer}, \citenamefont {Cataldo}, \citenamefont
  {Lachman}, \citenamefont {Field}, \citenamefont {Mohan}, \citenamefont
  {de~Vries}, \citenamefont {Bultink}, \citenamefont {van Oven}, \citenamefont
  {Mutus}, \citenamefont {Stockill},\ and\ \citenamefont
  {Gröblacher}}]{van_thiel_optical_2023}%
  \BibitemOpen
  \bibfield  {author} {\bibinfo {author} {\bibfnamefont {T.~C.}\ \bibnamefont
  {van Thiel}}, \bibinfo {author} {\bibfnamefont {M.~J.}\ \bibnamefont
  {Weaver}}, \bibinfo {author} {\bibfnamefont {F.}~\bibnamefont {Berto}},
  \bibinfo {author} {\bibfnamefont {P.}~\bibnamefont {Duivestein}}, \bibinfo
  {author} {\bibfnamefont {M.}~\bibnamefont {Lemang}}, \bibinfo {author}
  {\bibfnamefont {K.~L.}\ \bibnamefont {Schuurman}}, \bibinfo {author}
  {\bibfnamefont {M.}~\bibnamefont {{\v{Z}}emli{\v{c}}ka}}, \bibinfo {author}
  {\bibfnamefont {F.}~\bibnamefont {Hijazi}}, \bibinfo {author} {\bibfnamefont
  {A.~C.}\ \bibnamefont {Bernasconi}}, \bibinfo {author} {\bibfnamefont
  {C.}~\bibnamefont {Ferrer}}, \bibinfo {author} {\bibfnamefont
  {E.}~\bibnamefont {Cataldo}}, \bibinfo {author} {\bibfnamefont {E.~O.}\
  \bibnamefont {Lachman}}, \bibinfo {author} {\bibfnamefont {M.}~\bibnamefont
  {Field}}, \bibinfo {author} {\bibfnamefont {Y.}~\bibnamefont {Mohan}},
  \bibinfo {author} {\bibfnamefont {F.~K.}\ \bibnamefont {de~Vries}}, \bibinfo
  {author} {\bibfnamefont {C.~C.}\ \bibnamefont {Bultink}}, \bibinfo {author}
  {\bibfnamefont {J.~C.}\ \bibnamefont {van Oven}}, \bibinfo {author}
  {\bibfnamefont {J.~Y.}\ \bibnamefont {Mutus}}, \bibinfo {author}
  {\bibfnamefont {R.}~\bibnamefont {Stockill}},\ and\ \bibinfo {author}
  {\bibfnamefont {S.}~\bibnamefont {Gröblacher}},\ }\href
  {https://doi.org/10.1038/s41567-024-02742-3} {\bibfield  {journal} {\bibinfo
  {journal} {Nat. Phys.}\ }\textbf {\bibinfo {volume} {21}},\ \bibinfo {pages}
  {401} (\bibinfo {year} {2025})}\BibitemShut {NoStop}%
\bibitem [{\citenamefont {Meesala}\ \emph {et~al.}(2023)\citenamefont
  {Meesala}, \citenamefont {Lake}, \citenamefont {Wood}, \citenamefont
  {Chiappina}, \citenamefont {Zhong}, \citenamefont {Beyer}, \citenamefont
  {Shaw}, \citenamefont {Jiang},\ and\ \citenamefont
  {Painter}}]{meesala_quantum_2023}%
  \BibitemOpen
  \bibfield  {author} {\bibinfo {author} {\bibfnamefont {S.}~\bibnamefont
  {Meesala}}, \bibinfo {author} {\bibfnamefont {D.}~\bibnamefont {Lake}},
  \bibinfo {author} {\bibfnamefont {S.}~\bibnamefont {Wood}}, \bibinfo {author}
  {\bibfnamefont {P.}~\bibnamefont {Chiappina}}, \bibinfo {author}
  {\bibfnamefont {C.}~\bibnamefont {Zhong}}, \bibinfo {author} {\bibfnamefont
  {A.~D.}\ \bibnamefont {Beyer}}, \bibinfo {author} {\bibfnamefont {M.~D.}\
  \bibnamefont {Shaw}}, \bibinfo {author} {\bibfnamefont {L.}~\bibnamefont
  {Jiang}},\ and\ \bibinfo {author} {\bibfnamefont {O.}~\bibnamefont
  {Painter}},\ }\href@noop {} {\bibinfo {title} {Quantum entanglement between
  optical and microwave photonic qubits}} (\bibinfo {year} {2023}),\ \Eprint
  {https://arxiv.org/abs/2312.13559} {arXiv:2312.13559} \BibitemShut {NoStop}%
\bibitem [{\citenamefont {Weaver}\ \emph {et~al.}(2024)\citenamefont {Weaver},
  \citenamefont {Duivestein}, \citenamefont {Bernasconi}, \citenamefont
  {Scharmer}, \citenamefont {Lemang}, \citenamefont {Thiel}, \citenamefont
  {Hijazi}, \citenamefont {Hensen}, \citenamefont {Gröblacher},\ and\
  \citenamefont {Stockill}}]{weaver_integrated_2024}%
  \BibitemOpen
  \bibfield  {author} {\bibinfo {author} {\bibfnamefont {M.~J.}\ \bibnamefont
  {Weaver}}, \bibinfo {author} {\bibfnamefont {P.}~\bibnamefont {Duivestein}},
  \bibinfo {author} {\bibfnamefont {A.~C.}\ \bibnamefont {Bernasconi}},
  \bibinfo {author} {\bibfnamefont {S.}~\bibnamefont {Scharmer}}, \bibinfo
  {author} {\bibfnamefont {M.}~\bibnamefont {Lemang}}, \bibinfo {author}
  {\bibfnamefont {T.~C.~v.}\ \bibnamefont {Thiel}}, \bibinfo {author}
  {\bibfnamefont {F.}~\bibnamefont {Hijazi}}, \bibinfo {author} {\bibfnamefont
  {B.}~\bibnamefont {Hensen}}, \bibinfo {author} {\bibfnamefont
  {S.}~\bibnamefont {Gröblacher}},\ and\ \bibinfo {author} {\bibfnamefont
  {R.}~\bibnamefont {Stockill}},\ }\href
  {https://doi.org/10.1038/s41565-023-01515-y} {\bibfield  {journal} {\bibinfo
  {journal} {Nat. Nanotechnol.}\ }\textbf {\bibinfo {volume} {19}},\ \bibinfo
  {pages} {166} (\bibinfo {year} {2024})}\BibitemShut {NoStop}%
\bibitem [{\citenamefont {Kurizki}\ \emph {et~al.}(2015)\citenamefont
  {Kurizki}, \citenamefont {Bertet}, \citenamefont {Kubo}, \citenamefont
  {Mølmer}, \citenamefont {Petrosyan}, \citenamefont {Rabl},\ and\
  \citenamefont {Schmiedmayer}}]{kurizki_quantum_2015}%
  \BibitemOpen
  \bibfield  {author} {\bibinfo {author} {\bibfnamefont {G.}~\bibnamefont
  {Kurizki}}, \bibinfo {author} {\bibfnamefont {P.}~\bibnamefont {Bertet}},
  \bibinfo {author} {\bibfnamefont {Y.}~\bibnamefont {Kubo}}, \bibinfo {author}
  {\bibfnamefont {K.}~\bibnamefont {Mølmer}}, \bibinfo {author} {\bibfnamefont
  {D.}~\bibnamefont {Petrosyan}}, \bibinfo {author} {\bibfnamefont
  {P.}~\bibnamefont {Rabl}},\ and\ \bibinfo {author} {\bibfnamefont
  {J.}~\bibnamefont {Schmiedmayer}},\ }\href
  {https://doi.org/10.1073/pnas.1419326112} {\bibfield  {journal} {\bibinfo
  {journal} {Proc. Natl Acad. Sci.}\ }\textbf {\bibinfo {volume} {112}},\
  \bibinfo {pages} {3866} (\bibinfo {year} {2015})}\BibitemShut {NoStop}%
\bibitem [{\citenamefont {Duan}\ \emph {et~al.}(2001)\citenamefont {Duan},
  \citenamefont {Lukin}, \citenamefont {Cirac},\ and\ \citenamefont
  {Zoller}}]{duan_long-distance_2001}%
  \BibitemOpen
  \bibfield  {author} {\bibinfo {author} {\bibfnamefont {L.-M.}\ \bibnamefont
  {Duan}}, \bibinfo {author} {\bibfnamefont {M.~D.}\ \bibnamefont {Lukin}},
  \bibinfo {author} {\bibfnamefont {J.~I.}\ \bibnamefont {Cirac}},\ and\
  \bibinfo {author} {\bibfnamefont {P.}~\bibnamefont {Zoller}},\ }\href
  {https://doi.org/10.1038/35106500} {\bibfield  {journal} {\bibinfo  {journal}
  {Nature}\ }\textbf {\bibinfo {volume} {414}},\ \bibinfo {pages} {413}
  (\bibinfo {year} {2001})}\BibitemShut {NoStop}%
\bibitem [{\citenamefont {Riedinger}\ \emph {et~al.}(2018)\citenamefont
  {Riedinger}, \citenamefont {Wallucks}, \citenamefont {Marinković},
  \citenamefont {Löschnauer}, \citenamefont {Aspelmeyer}, \citenamefont
  {Hong},\ and\ \citenamefont {Gröblacher}}]{riedinger2018remote}%
  \BibitemOpen
  \bibfield  {author} {\bibinfo {author} {\bibfnamefont {R.}~\bibnamefont
  {Riedinger}}, \bibinfo {author} {\bibfnamefont {A.}~\bibnamefont {Wallucks}},
  \bibinfo {author} {\bibfnamefont {I.}~\bibnamefont {Marinković}}, \bibinfo
  {author} {\bibfnamefont {C.}~\bibnamefont {Löschnauer}}, \bibinfo {author}
  {\bibfnamefont {M.}~\bibnamefont {Aspelmeyer}}, \bibinfo {author}
  {\bibfnamefont {S.}~\bibnamefont {Hong}},\ and\ \bibinfo {author}
  {\bibfnamefont {S.}~\bibnamefont {Gröblacher}},\ }\href
  {https://doi.org/10.1038/s41586-018-0036-z} {\bibfield  {journal} {\bibinfo
  {journal} {Nature}\ }\textbf {\bibinfo {volume} {556}},\ \bibinfo {pages}
  {473} (\bibinfo {year} {2018})}\BibitemShut {NoStop}%
\bibitem [{\citenamefont {Marinkovi{\'c}}\ \emph {et~al.}(2018)\citenamefont
  {Marinkovi{\'c}}, \citenamefont {Wallucks}, \citenamefont {Riedinger},
  \citenamefont {Hong}, \citenamefont {Aspelmeyer},\ and\ \citenamefont
  {Gr{\"o}blacher}}]{marinkovic2018optomechanical}%
  \BibitemOpen
  \bibfield  {author} {\bibinfo {author} {\bibfnamefont {I.}~\bibnamefont
  {Marinkovi{\'c}}}, \bibinfo {author} {\bibfnamefont {A.}~\bibnamefont
  {Wallucks}}, \bibinfo {author} {\bibfnamefont {R.}~\bibnamefont {Riedinger}},
  \bibinfo {author} {\bibfnamefont {S.}~\bibnamefont {Hong}}, \bibinfo {author}
  {\bibfnamefont {M.}~\bibnamefont {Aspelmeyer}},\ and\ \bibinfo {author}
  {\bibfnamefont {S.}~\bibnamefont {Gr{\"o}blacher}},\ }\href@noop {}
  {\bibfield  {journal} {\bibinfo  {journal} {Physical review letters}\
  }\textbf {\bibinfo {volume} {121}},\ \bibinfo {pages} {220404} (\bibinfo
  {year} {2018})}\BibitemShut {NoStop}%
\bibitem [{\citenamefont {Fiaschi}\ \emph {et~al.}(2021)\citenamefont
  {Fiaschi}, \citenamefont {Hensen}, \citenamefont {Wallucks}, \citenamefont
  {Benevides}, \citenamefont {Li}, \citenamefont {Alegre},\ and\ \citenamefont
  {Gröblacher}}]{fiaschi_optomechanical_2021}%
  \BibitemOpen
  \bibfield  {author} {\bibinfo {author} {\bibfnamefont {N.}~\bibnamefont
  {Fiaschi}}, \bibinfo {author} {\bibfnamefont {B.}~\bibnamefont {Hensen}},
  \bibinfo {author} {\bibfnamefont {A.}~\bibnamefont {Wallucks}}, \bibinfo
  {author} {\bibfnamefont {R.}~\bibnamefont {Benevides}}, \bibinfo {author}
  {\bibfnamefont {J.}~\bibnamefont {Li}}, \bibinfo {author} {\bibfnamefont
  {T.~P.~M.}\ \bibnamefont {Alegre}},\ and\ \bibinfo {author} {\bibfnamefont
  {S.}~\bibnamefont {Gröblacher}},\ }\href
  {https://doi.org/10.1038/s41566-021-00866-z} {\bibfield  {journal} {\bibinfo
  {journal} {Nat. Photon.}\ }\textbf {\bibinfo {volume} {15}},\ \bibinfo
  {pages} {817} (\bibinfo {year} {2021})}\BibitemShut {NoStop}%
\bibitem [{\citenamefont {Meenehan}\ \emph {et~al.}(2015)\citenamefont
  {Meenehan}, \citenamefont {Cohen}, \citenamefont {MacCabe}, \citenamefont
  {Marsili}, \citenamefont {Shaw},\ and\ \citenamefont
  {Painter}}]{meenehan_pulsed_2015}%
  \BibitemOpen
  \bibfield  {author} {\bibinfo {author} {\bibfnamefont {S.~M.}\ \bibnamefont
  {Meenehan}}, \bibinfo {author} {\bibfnamefont {J.~D.}\ \bibnamefont {Cohen}},
  \bibinfo {author} {\bibfnamefont {G.~S.}\ \bibnamefont {MacCabe}}, \bibinfo
  {author} {\bibfnamefont {F.}~\bibnamefont {Marsili}}, \bibinfo {author}
  {\bibfnamefont {M.~D.}\ \bibnamefont {Shaw}},\ and\ \bibinfo {author}
  {\bibfnamefont {O.}~\bibnamefont {Painter}},\ }\href
  {https://doi.org/10.1103/PhysRevX.5.041002} {\bibfield  {journal} {\bibinfo
  {journal} {Phys. Rev. X}\ }\textbf {\bibinfo {volume} {5}},\ \bibinfo {pages}
  {041002} (\bibinfo {year} {2015})}\BibitemShut {NoStop}%
\bibitem [{\citenamefont {Hong}\ \emph {et~al.}(2017)\citenamefont {Hong},
  \citenamefont {Riedinger}, \citenamefont {Marinković}, \citenamefont
  {Wallucks}, \citenamefont {Hofer}, \citenamefont {Norte}, \citenamefont
  {Aspelmeyer},\ and\ \citenamefont {Gröblacher}}]{hong_hanbury_2017}%
  \BibitemOpen
  \bibfield  {author} {\bibinfo {author} {\bibfnamefont {S.}~\bibnamefont
  {Hong}}, \bibinfo {author} {\bibfnamefont {R.}~\bibnamefont {Riedinger}},
  \bibinfo {author} {\bibfnamefont {I.}~\bibnamefont {Marinković}}, \bibinfo
  {author} {\bibfnamefont {A.}~\bibnamefont {Wallucks}}, \bibinfo {author}
  {\bibfnamefont {S.~G.}\ \bibnamefont {Hofer}}, \bibinfo {author}
  {\bibfnamefont {R.~A.}\ \bibnamefont {Norte}}, \bibinfo {author}
  {\bibfnamefont {M.}~\bibnamefont {Aspelmeyer}},\ and\ \bibinfo {author}
  {\bibfnamefont {S.}~\bibnamefont {Gröblacher}},\ }\href
  {https://doi.org/10.1126/science.aan7939} {\bibfield  {journal} {\bibinfo
  {journal} {Science}\ }\textbf {\bibinfo {volume} {358}},\ \bibinfo {pages}
  {203} (\bibinfo {year} {2017})}\BibitemShut {NoStop}%
\bibitem [{\citenamefont {Safavi-Naeini}\ \emph {et~al.}(2014)\citenamefont
  {Safavi-Naeini}, \citenamefont {Hill}, \citenamefont {Meenehan},
  \citenamefont {Chan}, \citenamefont {Gr\"oblacher},\ and\ \citenamefont
  {Painter}}]{Safavi-Naeini2014}%
  \BibitemOpen
  \bibfield  {author} {\bibinfo {author} {\bibfnamefont {A.~H.}\ \bibnamefont
  {Safavi-Naeini}}, \bibinfo {author} {\bibfnamefont {J.~T.}\ \bibnamefont
  {Hill}}, \bibinfo {author} {\bibfnamefont {S.}~\bibnamefont {Meenehan}},
  \bibinfo {author} {\bibfnamefont {J.}~\bibnamefont {Chan}}, \bibinfo {author}
  {\bibfnamefont {S.}~\bibnamefont {Gr\"oblacher}},\ and\ \bibinfo {author}
  {\bibfnamefont {O.}~\bibnamefont {Painter}},\ }\href
  {https://doi.org/10.1103/PhysRevLett.112.153603} {\bibfield  {journal}
  {\bibinfo  {journal} {Phys.\ Rev.\ Lett.}\ }\textbf {\bibinfo {volume}
  {112}},\ \bibinfo {pages} {153603} (\bibinfo {year} {2014})}\BibitemShut
  {NoStop}%
\bibitem [{\citenamefont {Ren}\ \emph {et~al.}(2020)\citenamefont {Ren},
  \citenamefont {Matheny}, \citenamefont {MacCabe}, \citenamefont {Luo},
  \citenamefont {Pfeifer}, \citenamefont {Mirhosseini},\ and\ \citenamefont
  {Painter}}]{ren_two-dimensional_2020}%
  \BibitemOpen
  \bibfield  {author} {\bibinfo {author} {\bibfnamefont {H.}~\bibnamefont
  {Ren}}, \bibinfo {author} {\bibfnamefont {M.~H.}\ \bibnamefont {Matheny}},
  \bibinfo {author} {\bibfnamefont {G.~S.}\ \bibnamefont {MacCabe}}, \bibinfo
  {author} {\bibfnamefont {J.}~\bibnamefont {Luo}}, \bibinfo {author}
  {\bibfnamefont {H.}~\bibnamefont {Pfeifer}}, \bibinfo {author} {\bibfnamefont
  {M.}~\bibnamefont {Mirhosseini}},\ and\ \bibinfo {author} {\bibfnamefont
  {O.}~\bibnamefont {Painter}},\ }\href
  {https://doi.org/10.1038/s41467-020-17182-9} {\bibfield  {journal} {\bibinfo
  {journal} {Nat. Commun.}\ }\textbf {\bibinfo {volume} {11}},\ \bibinfo
  {pages} {3373} (\bibinfo {year} {2020})}\BibitemShut {NoStop}%
\bibitem [{\citenamefont {Kersul}\ \emph {et~al.}(2023)\citenamefont {Kersul},
  \citenamefont {Benevides}, \citenamefont {Moraes}, \citenamefont {De~Aguiar},
  \citenamefont {Wallucks}, \citenamefont {Gröblacher}, \citenamefont
  {Wiederhecker},\ and\ \citenamefont {Mayer~Alegre}}]{kersul_silicon_2023}%
  \BibitemOpen
  \bibfield  {author} {\bibinfo {author} {\bibfnamefont {C.~M.}\ \bibnamefont
  {Kersul}}, \bibinfo {author} {\bibfnamefont {R.}~\bibnamefont {Benevides}},
  \bibinfo {author} {\bibfnamefont {F.}~\bibnamefont {Moraes}}, \bibinfo
  {author} {\bibfnamefont {G.~H.~M.}\ \bibnamefont {De~Aguiar}}, \bibinfo
  {author} {\bibfnamefont {A.}~\bibnamefont {Wallucks}}, \bibinfo {author}
  {\bibfnamefont {S.}~\bibnamefont {Gröblacher}}, \bibinfo {author}
  {\bibfnamefont {G.~S.}\ \bibnamefont {Wiederhecker}},\ and\ \bibinfo {author}
  {\bibfnamefont {T.~P.}\ \bibnamefont {Mayer~Alegre}},\ }\href
  {https://doi.org/10.1063/5.0135407} {\bibfield  {journal} {\bibinfo
  {journal} {APL Photonics}\ }\textbf {\bibinfo {volume} {8}},\ \bibinfo
  {pages} {056112} (\bibinfo {year} {2023})}\BibitemShut {NoStop}%
\bibitem [{\citenamefont {Mayor}\ \emph {et~al.}(2025)\citenamefont {Mayor},
  \citenamefont {Malik}, \citenamefont {Primo}, \citenamefont {Gyger},
  \citenamefont {Jiang}, \citenamefont {Alegre},\ and\ \citenamefont
  {Safavi-Naeini}}]{mayor_high_2025}%
  \BibitemOpen
  \bibfield  {author} {\bibinfo {author} {\bibfnamefont {F.~M.}\ \bibnamefont
  {Mayor}}, \bibinfo {author} {\bibfnamefont {S.}~\bibnamefont {Malik}},
  \bibinfo {author} {\bibfnamefont {A.~G.}\ \bibnamefont {Primo}}, \bibinfo
  {author} {\bibfnamefont {S.}~\bibnamefont {Gyger}}, \bibinfo {author}
  {\bibfnamefont {W.}~\bibnamefont {Jiang}}, \bibinfo {author} {\bibfnamefont
  {T.~P.~M.}\ \bibnamefont {Alegre}},\ and\ \bibinfo {author} {\bibfnamefont
  {A.~H.}\ \bibnamefont {Safavi-Naeini}},\ }\href
  {https://doi.org/10.1038/s41467-025-57948-7} {\bibfield  {journal} {\bibinfo
  {journal} {Nat. Commun.}\ }\textbf {\bibinfo {volume} {16}},\ \bibinfo
  {pages} {2576} (\bibinfo {year} {2025})}\BibitemShut {NoStop}%
\bibitem [{\citenamefont {Sonar}\ \emph {et~al.}(2025)\citenamefont {Sonar},
  \citenamefont {Hatipoglu}, \citenamefont {Meesala}, \citenamefont {Lake},
  \citenamefont {Ren},\ and\ \citenamefont
  {Painter}}]{sonar_high-efficiency_2025}%
  \BibitemOpen
  \bibfield  {author} {\bibinfo {author} {\bibfnamefont {S.}~\bibnamefont
  {Sonar}}, \bibinfo {author} {\bibfnamefont {U.}~\bibnamefont {Hatipoglu}},
  \bibinfo {author} {\bibfnamefont {S.}~\bibnamefont {Meesala}}, \bibinfo
  {author} {\bibfnamefont {D.~P.}\ \bibnamefont {Lake}}, \bibinfo {author}
  {\bibfnamefont {H.}~\bibnamefont {Ren}},\ and\ \bibinfo {author}
  {\bibfnamefont {O.}~\bibnamefont {Painter}},\ }\href
  {https://doi.org/10.1364/OPTICA.538557} {\bibfield  {journal} {\bibinfo
  {journal} {Optica}\ }\textbf {\bibinfo {volume} {12}},\ \bibinfo {pages} {99}
  (\bibinfo {year} {2025})}\BibitemShut {NoStop}%
\bibitem [{\citenamefont {Weiss}\ \emph {et~al.}(2021)\citenamefont {Weiss},
  \citenamefont {Gritsch}, \citenamefont {Merkel},\ and\ \citenamefont
  {Reiserer}}]{weiss_erbium_2021}%
  \BibitemOpen
  \bibfield  {author} {\bibinfo {author} {\bibfnamefont {L.}~\bibnamefont
  {Weiss}}, \bibinfo {author} {\bibfnamefont {A.}~\bibnamefont {Gritsch}},
  \bibinfo {author} {\bibfnamefont {B.}~\bibnamefont {Merkel}},\ and\ \bibinfo
  {author} {\bibfnamefont {A.}~\bibnamefont {Reiserer}},\ }\href
  {https://doi.org/10.1364/OPTICA.413330} {\bibfield  {journal} {\bibinfo
  {journal} {Optica}\ }\textbf {\bibinfo {volume} {8}},\ \bibinfo {pages} {40}
  (\bibinfo {year} {2021})}\BibitemShut {NoStop}%
\bibitem [{\citenamefont {Gritsch}\ \emph {et~al.}(2022)\citenamefont
  {Gritsch}, \citenamefont {Weiss}, \citenamefont {Fr\"uh}, \citenamefont
  {Rinner},\ and\ \citenamefont {Reiserer}}]{gritsch2022narrow}%
  \BibitemOpen
  \bibfield  {author} {\bibinfo {author} {\bibfnamefont {A.}~\bibnamefont
  {Gritsch}}, \bibinfo {author} {\bibfnamefont {L.}~\bibnamefont {Weiss}},
  \bibinfo {author} {\bibfnamefont {J.}~\bibnamefont {Fr\"uh}}, \bibinfo
  {author} {\bibfnamefont {S.}~\bibnamefont {Rinner}},\ and\ \bibinfo {author}
  {\bibfnamefont {A.}~\bibnamefont {Reiserer}},\ }\href
  {https://doi.org/10.1103/PhysRevX.12.041009} {\bibfield  {journal} {\bibinfo
  {journal} {Phys. Rev. X}\ }\textbf {\bibinfo {volume} {12}},\ \bibinfo
  {pages} {041009} (\bibinfo {year} {2022})}\BibitemShut {NoStop}%
\bibitem [{\citenamefont {Ourari}\ \emph {et~al.}(2023)\citenamefont {Ourari},
  \citenamefont {Dusanowski}, \citenamefont {Horvath}, \citenamefont {Uysal},
  \citenamefont {Phenicie}, \citenamefont {Stevenson}, \citenamefont {Raha},
  \citenamefont {Chen}, \citenamefont {Cava}, \citenamefont {de~Leon},\ and\
  \citenamefont {Thompson}}]{ourari_indistinguishable_2023}%
  \BibitemOpen
  \bibfield  {author} {\bibinfo {author} {\bibfnamefont {S.}~\bibnamefont
  {Ourari}}, \bibinfo {author} {\bibfnamefont {L.}~\bibnamefont {Dusanowski}},
  \bibinfo {author} {\bibfnamefont {S.~P.}\ \bibnamefont {Horvath}}, \bibinfo
  {author} {\bibfnamefont {M.~T.}\ \bibnamefont {Uysal}}, \bibinfo {author}
  {\bibfnamefont {C.~M.}\ \bibnamefont {Phenicie}}, \bibinfo {author}
  {\bibfnamefont {P.}~\bibnamefont {Stevenson}}, \bibinfo {author}
  {\bibfnamefont {M.}~\bibnamefont {Raha}}, \bibinfo {author} {\bibfnamefont
  {S.}~\bibnamefont {Chen}}, \bibinfo {author} {\bibfnamefont {R.~J.}\
  \bibnamefont {Cava}}, \bibinfo {author} {\bibfnamefont {N.~P.}\ \bibnamefont
  {de~Leon}},\ and\ \bibinfo {author} {\bibfnamefont {J.~D.}\ \bibnamefont
  {Thompson}},\ }\href {https://doi.org/10.1038/s41586-023-06281-4} {\bibfield
  {journal} {\bibinfo  {journal} {Nature}\ }\textbf {\bibinfo {volume} {620}},\
  \bibinfo {pages} {977} (\bibinfo {year} {2023})}\BibitemShut {NoStop}%
\bibitem [{\citenamefont {Higginbottom}\ \emph {et~al.}(2022)\citenamefont
  {Higginbottom}, \citenamefont {Kurkjian}, \citenamefont {Chartrand},
  \citenamefont {Kazemi}, \citenamefont {Brunelle}, \citenamefont {MacQuarrie},
  \citenamefont {Klein}, \citenamefont {Lee-Hone}, \citenamefont {Stacho},
  \citenamefont {Ruether}, \citenamefont {Bowness}, \citenamefont {Bergeron},
  \citenamefont {DeAbreu}, \citenamefont {Harrigan}, \citenamefont
  {Kanaganayagam}, \citenamefont {Marsden}, \citenamefont {Richards},
  \citenamefont {Stott}, \citenamefont {Roorda}, \citenamefont {Morse},
  \citenamefont {Thewalt},\ and\ \citenamefont
  {Simmons}}]{higginbottom_optical_2022}%
  \BibitemOpen
  \bibfield  {author} {\bibinfo {author} {\bibfnamefont {D.~B.}\ \bibnamefont
  {Higginbottom}}, \bibinfo {author} {\bibfnamefont {A.~T.~K.}\ \bibnamefont
  {Kurkjian}}, \bibinfo {author} {\bibfnamefont {C.}~\bibnamefont {Chartrand}},
  \bibinfo {author} {\bibfnamefont {M.}~\bibnamefont {Kazemi}}, \bibinfo
  {author} {\bibfnamefont {N.~A.}\ \bibnamefont {Brunelle}}, \bibinfo {author}
  {\bibfnamefont {E.~R.}\ \bibnamefont {MacQuarrie}}, \bibinfo {author}
  {\bibfnamefont {J.~R.}\ \bibnamefont {Klein}}, \bibinfo {author}
  {\bibfnamefont {N.~R.}\ \bibnamefont {Lee-Hone}}, \bibinfo {author}
  {\bibfnamefont {J.}~\bibnamefont {Stacho}}, \bibinfo {author} {\bibfnamefont
  {M.}~\bibnamefont {Ruether}}, \bibinfo {author} {\bibfnamefont
  {C.}~\bibnamefont {Bowness}}, \bibinfo {author} {\bibfnamefont
  {L.}~\bibnamefont {Bergeron}}, \bibinfo {author} {\bibfnamefont
  {A.}~\bibnamefont {DeAbreu}}, \bibinfo {author} {\bibfnamefont {S.~R.}\
  \bibnamefont {Harrigan}}, \bibinfo {author} {\bibfnamefont {J.}~\bibnamefont
  {Kanaganayagam}}, \bibinfo {author} {\bibfnamefont {D.~W.}\ \bibnamefont
  {Marsden}}, \bibinfo {author} {\bibfnamefont {T.~S.}\ \bibnamefont
  {Richards}}, \bibinfo {author} {\bibfnamefont {L.~A.}\ \bibnamefont {Stott}},
  \bibinfo {author} {\bibfnamefont {S.}~\bibnamefont {Roorda}}, \bibinfo
  {author} {\bibfnamefont {K.~J.}\ \bibnamefont {Morse}}, \bibinfo {author}
  {\bibfnamefont {M.~L.~W.}\ \bibnamefont {Thewalt}},\ and\ \bibinfo {author}
  {\bibfnamefont {S.}~\bibnamefont {Simmons}},\ }\href
  {https://doi.org/10.1038/s41586-022-04821-y} {\bibfield  {journal} {\bibinfo
  {journal} {Nature}\ }\textbf {\bibinfo {volume} {607}},\ \bibinfo {pages}
  {266} (\bibinfo {year} {2022})}\BibitemShut {NoStop}%
\bibitem [{\citenamefont {Komza}\ \emph {et~al.}(2024)\citenamefont {Komza},
  \citenamefont {Samutpraphoot}, \citenamefont {Odeh}, \citenamefont {Tang},
  \citenamefont {Mathew}, \citenamefont {Chang}, \citenamefont {Song},
  \citenamefont {Kim}, \citenamefont {Xiong}, \citenamefont {Hautier},\ and\
  \citenamefont {Sipahigil}}]{komza2024indistinguishable}%
  \BibitemOpen
  \bibfield  {author} {\bibinfo {author} {\bibfnamefont {L.}~\bibnamefont
  {Komza}}, \bibinfo {author} {\bibfnamefont {P.}~\bibnamefont
  {Samutpraphoot}}, \bibinfo {author} {\bibfnamefont {M.}~\bibnamefont {Odeh}},
  \bibinfo {author} {\bibfnamefont {Y.-L.}\ \bibnamefont {Tang}}, \bibinfo
  {author} {\bibfnamefont {M.}~\bibnamefont {Mathew}}, \bibinfo {author}
  {\bibfnamefont {J.}~\bibnamefont {Chang}}, \bibinfo {author} {\bibfnamefont
  {H.}~\bibnamefont {Song}}, \bibinfo {author} {\bibfnamefont {M.-K.}\
  \bibnamefont {Kim}}, \bibinfo {author} {\bibfnamefont {Y.}~\bibnamefont
  {Xiong}}, \bibinfo {author} {\bibfnamefont {G.}~\bibnamefont {Hautier}},\
  and\ \bibinfo {author} {\bibfnamefont {A.}~\bibnamefont {Sipahigil}},\ }\href
  {https://doi.org/10.1038/s41467-024-51265-1} {\bibfield  {journal} {\bibinfo
  {journal} {Nat. Commun.}\ }\textbf {\bibinfo {volume} {15}},\ \bibinfo
  {pages} {6920} (\bibinfo {year} {2024})}\BibitemShut {NoStop}%
\bibitem [{\citenamefont {Zhang}\ \emph {et~al.}(2023)\citenamefont {Zhang},
  \citenamefont {Zhang}, \citenamefont {Wei}, \citenamefont {Li}, \citenamefont
  {Liao}, \citenamefont {Li}, \citenamefont {Deng}, \citenamefont {Wang},
  \citenamefont {Song}, \citenamefont {You}, \citenamefont {Jing},
  \citenamefont {Chen}, \citenamefont {Guo},\ and\ \citenamefont
  {Zhou}}]{zhang2023}%
  \BibitemOpen
  \bibfield  {author} {\bibinfo {author} {\bibfnamefont {X.}~\bibnamefont
  {Zhang}}, \bibinfo {author} {\bibfnamefont {B.}~\bibnamefont {Zhang}},
  \bibinfo {author} {\bibfnamefont {S.}~\bibnamefont {Wei}}, \bibinfo {author}
  {\bibfnamefont {H.}~\bibnamefont {Li}}, \bibinfo {author} {\bibfnamefont
  {J.}~\bibnamefont {Liao}}, \bibinfo {author} {\bibfnamefont {C.}~\bibnamefont
  {Li}}, \bibinfo {author} {\bibfnamefont {G.}~\bibnamefont {Deng}}, \bibinfo
  {author} {\bibfnamefont {Y.}~\bibnamefont {Wang}}, \bibinfo {author}
  {\bibfnamefont {H.}~\bibnamefont {Song}}, \bibinfo {author} {\bibfnamefont
  {L.}~\bibnamefont {You}}, \bibinfo {author} {\bibfnamefont {B.}~\bibnamefont
  {Jing}}, \bibinfo {author} {\bibfnamefont {F.}~\bibnamefont {Chen}}, \bibinfo
  {author} {\bibfnamefont {G.}~\bibnamefont {Guo}},\ and\ \bibinfo {author}
  {\bibfnamefont {Q.}~\bibnamefont {Zhou}},\ }\href
  {https://doi.org/10.1126/sciadv.adf4587} {\bibfield  {journal} {\bibinfo
  {journal} {Sci. Adv.}\ }\textbf {\bibinfo {volume} {9}},\ \bibinfo {pages}
  {eadf4587} (\bibinfo {year} {2023})}\BibitemShut {NoStop}%
\bibitem [{\citenamefont {Rančić}\ \emph {et~al.}(2018)\citenamefont
  {Rančić}, \citenamefont {Hedges}, \citenamefont {Ahlefeldt},\ and\
  \citenamefont {Sellars}}]{rancic_coherence_2018}%
  \BibitemOpen
  \bibfield  {author} {\bibinfo {author} {\bibfnamefont {M.}~\bibnamefont
  {Rančić}}, \bibinfo {author} {\bibfnamefont {M.~P.}\ \bibnamefont
  {Hedges}}, \bibinfo {author} {\bibfnamefont {R.~L.}\ \bibnamefont
  {Ahlefeldt}},\ and\ \bibinfo {author} {\bibfnamefont {M.~J.}\ \bibnamefont
  {Sellars}},\ }\href {https://doi.org/10.1038/nphys4254} {\bibfield  {journal}
  {\bibinfo  {journal} {Nat. Phys.}\ }\textbf {\bibinfo {volume} {14}},\
  \bibinfo {pages} {50} (\bibinfo {year} {2018})}\BibitemShut {NoStop}%
\bibitem [{\citenamefont {Kristensen}\ \emph {et~al.}(2024)\citenamefont
  {Kristensen}, \citenamefont {Kralj}, \citenamefont {Langman},\ and\
  \citenamefont {Schliesser}}]{kristensen_2024}%
  \BibitemOpen
  \bibfield  {author} {\bibinfo {author} {\bibfnamefont {M.~B.}\ \bibnamefont
  {Kristensen}}, \bibinfo {author} {\bibfnamefont {N.}~\bibnamefont {Kralj}},
  \bibinfo {author} {\bibfnamefont {E.~C.}\ \bibnamefont {Langman}},\ and\
  \bibinfo {author} {\bibfnamefont {A.}~\bibnamefont {Schliesser}},\ }\href
  {https://doi.org/10.1103/PhysRevLett.132.100802} {\bibfield  {journal}
  {\bibinfo  {journal} {Phys. Rev. Lett.}\ }\textbf {\bibinfo {volume} {132}},\
  \bibinfo {pages} {100802} (\bibinfo {year} {2024})}\BibitemShut {NoStop}%
\bibitem [{\citenamefont {Johansson}\ \emph {et~al.}(2012)\citenamefont
  {Johansson}, \citenamefont {Nation},\ and\ \citenamefont
  {Nori}}]{johansson_qutip_2012}%
  \BibitemOpen
  \bibfield  {author} {\bibinfo {author} {\bibfnamefont {J.~R.}\ \bibnamefont
  {Johansson}}, \bibinfo {author} {\bibfnamefont {P.~D.}\ \bibnamefont
  {Nation}},\ and\ \bibinfo {author} {\bibfnamefont {F.}~\bibnamefont {Nori}},\
  }\href {https://doi.org/10.1016/j.cpc.2012.02.021} {\bibfield  {journal}
  {\bibinfo  {journal} {Comput. Phys. Commun.}\ }\textbf {\bibinfo {volume}
  {183}},\ \bibinfo {pages} {1760} (\bibinfo {year} {2012})}\BibitemShut
  {NoStop}%
\bibitem [{\citenamefont {Johansson}\ \emph {et~al.}(2013)\citenamefont
  {Johansson}, \citenamefont {Nation},\ and\ \citenamefont
  {Nori}}]{johansson_qutip_2013}%
  \BibitemOpen
  \bibfield  {author} {\bibinfo {author} {\bibfnamefont {J.~R.}\ \bibnamefont
  {Johansson}}, \bibinfo {author} {\bibfnamefont {P.~D.}\ \bibnamefont
  {Nation}},\ and\ \bibinfo {author} {\bibfnamefont {F.}~\bibnamefont {Nori}},\
  }\href {https://doi.org/https://doi.org/10.1016/j.cpc.2012.11.019} {\bibfield
   {journal} {\bibinfo  {journal} {Comput. Phys. Commun.}\ }\textbf {\bibinfo
  {volume} {184}},\ \bibinfo {pages} {1234} (\bibinfo {year}
  {2013})}\BibitemShut {NoStop}%
\bibitem [{\citenamefont {Gerry}\ and\ \citenamefont
  {Knight}(2023)}]{gerry2023introductory}%
  \BibitemOpen
  \bibfield  {author} {\bibinfo {author} {\bibfnamefont {C.~C.}\ \bibnamefont
  {Gerry}}\ and\ \bibinfo {author} {\bibfnamefont {P.~L.}\ \bibnamefont
  {Knight}},\ }\href@noop {} {\emph {\bibinfo {title} {Introductory quantum
  optics}}}\ (\bibinfo  {publisher} {Cambridge university press},\ \bibinfo
  {year} {2023})\BibitemShut {NoStop}%
\bibitem [{\citenamefont {Galland}\ \emph {et~al.}(2014)\citenamefont
  {Galland}, \citenamefont {Sangouard}, \citenamefont {Piro}, \citenamefont
  {Gisin},\ and\ \citenamefont {Kippenberg}}]{galland_heralded_2014}%
  \BibitemOpen
  \bibfield  {author} {\bibinfo {author} {\bibfnamefont {C.}~\bibnamefont
  {Galland}}, \bibinfo {author} {\bibfnamefont {N.}~\bibnamefont {Sangouard}},
  \bibinfo {author} {\bibfnamefont {N.}~\bibnamefont {Piro}}, \bibinfo {author}
  {\bibfnamefont {N.}~\bibnamefont {Gisin}},\ and\ \bibinfo {author}
  {\bibfnamefont {T.~J.}\ \bibnamefont {Kippenberg}},\ }\href
  {https://doi.org/10.1103/PhysRevLett.112.143602} {\bibfield  {journal}
  {\bibinfo  {journal} {Phys. Rev. Lett.}\ }\textbf {\bibinfo {volume} {112}},\
  \bibinfo {pages} {143602} (\bibinfo {year} {2014})}\BibitemShut {NoStop}%
\bibitem [{\citenamefont {Riedinger}(2018)}]{riedinger2018single}%
  \BibitemOpen
  \bibfield  {author} {\bibinfo {author} {\bibfnamefont {R.}~\bibnamefont
  {Riedinger}},\ }\emph {\bibinfo {title} {Single Phonon Quantum Optics}},\
  \href {https://doi.org/https://doi.org/10.25365/thesis.54532} {Ph.D.
  thesis},\ \bibinfo  {school} {University of Vienna} (\bibinfo {year}
  {2018})\BibitemShut {NoStop}%
\bibitem [{\citenamefont {Holewa}\ \emph {et~al.}(2024)\citenamefont {Holewa},
  \citenamefont {Vajner}, \citenamefont {Zieba-Ostoj}, \citenamefont {Wasiluk},
  \citenamefont {Gaál}, \citenamefont {Sakanas}, \citenamefont {Burakowski},
  \citenamefont {Mrowiński}, \citenamefont {Krajnik}, \citenamefont {Xiong},
  \citenamefont {Yvind}, \citenamefont {Gregersen}, \citenamefont {Musiał},
  \citenamefont {Huck}, \citenamefont {Heindel}, \citenamefont {Syperek},\ and\
  \citenamefont {Semenova}}]{holewa_high-throughput_2024}%
  \BibitemOpen
  \bibfield  {author} {\bibinfo {author} {\bibfnamefont {P.}~\bibnamefont
  {Holewa}}, \bibinfo {author} {\bibfnamefont {D.~A.}\ \bibnamefont {Vajner}},
  \bibinfo {author} {\bibfnamefont {E.}~\bibnamefont {Zieba-Ostoj}}, \bibinfo
  {author} {\bibfnamefont {M.}~\bibnamefont {Wasiluk}}, \bibinfo {author}
  {\bibfnamefont {B.}~\bibnamefont {Gaál}}, \bibinfo {author} {\bibfnamefont
  {A.}~\bibnamefont {Sakanas}}, \bibinfo {author} {\bibfnamefont
  {M.}~\bibnamefont {Burakowski}}, \bibinfo {author} {\bibfnamefont
  {P.}~\bibnamefont {Mrowiński}}, \bibinfo {author} {\bibfnamefont
  {B.}~\bibnamefont {Krajnik}}, \bibinfo {author} {\bibfnamefont
  {M.}~\bibnamefont {Xiong}}, \bibinfo {author} {\bibfnamefont
  {K.}~\bibnamefont {Yvind}}, \bibinfo {author} {\bibfnamefont
  {N.}~\bibnamefont {Gregersen}}, \bibinfo {author} {\bibfnamefont
  {A.}~\bibnamefont {Musiał}}, \bibinfo {author} {\bibfnamefont
  {A.}~\bibnamefont {Huck}}, \bibinfo {author} {\bibfnamefont {T.}~\bibnamefont
  {Heindel}}, \bibinfo {author} {\bibfnamefont {M.}~\bibnamefont {Syperek}},\
  and\ \bibinfo {author} {\bibfnamefont {E.}~\bibnamefont {Semenova}},\ }\href
  {https://doi.org/10.1038/s41467-024-47551-7} {\bibfield  {journal} {\bibinfo
  {journal} {Nat. Commun.}\ }\textbf {\bibinfo {volume} {15}},\ \bibinfo
  {pages} {3358} (\bibinfo {year} {2024})}\BibitemShut {NoStop}%
\bibitem [{\citenamefont {Chen}\ \emph {et~al.}(2024)\citenamefont {Chen},
  \citenamefont {Luo}, \citenamefont {Long}, \citenamefont {Shi}, \citenamefont
  {Shen},\ and\ \citenamefont {Liu}}]{chen_2024}%
  \BibitemOpen
  \bibfield  {author} {\bibinfo {author} {\bibfnamefont {R.}~\bibnamefont
  {Chen}}, \bibinfo {author} {\bibfnamefont {Y.-H.}\ \bibnamefont {Luo}},
  \bibinfo {author} {\bibfnamefont {J.}~\bibnamefont {Long}}, \bibinfo {author}
  {\bibfnamefont {B.}~\bibnamefont {Shi}}, \bibinfo {author} {\bibfnamefont
  {C.}~\bibnamefont {Shen}},\ and\ \bibinfo {author} {\bibfnamefont
  {J.}~\bibnamefont {Liu}},\ }\href
  {https://doi.org/10.1103/PhysRevLett.133.083803} {\bibfield  {journal}
  {\bibinfo  {journal} {Phys. Rev. Lett.}\ }\textbf {\bibinfo {volume} {133}},\
  \bibinfo {pages} {083803} (\bibinfo {year} {2024})}\BibitemShut {NoStop}%
\bibitem [{\citenamefont {Kolvik}\ \emph {et~al.}(2023)\citenamefont {Kolvik},
  \citenamefont {Burger}, \citenamefont {Frey},\ and\ \citenamefont
  {Van~Laer}}]{kolvik_clamped_2023}%
  \BibitemOpen
  \bibfield  {author} {\bibinfo {author} {\bibfnamefont {J.}~\bibnamefont
  {Kolvik}}, \bibinfo {author} {\bibfnamefont {P.}~\bibnamefont {Burger}},
  \bibinfo {author} {\bibfnamefont {J.}~\bibnamefont {Frey}},\ and\ \bibinfo
  {author} {\bibfnamefont {R.}~\bibnamefont {Van~Laer}},\ }\href
  {https://doi.org/10.1364/OPTICA.492143} {\bibfield  {journal} {\bibinfo
  {journal} {Optica}\ }\textbf {\bibinfo {volume} {10}},\ \bibinfo {pages}
  {913} (\bibinfo {year} {2023})}\BibitemShut {NoStop}%
\bibitem [{\citenamefont {Humphreys}\ \emph {et~al.}(2018)\citenamefont
  {Humphreys}, \citenamefont {Kalb}, \citenamefont {Morits}, \citenamefont
  {Schouten}, \citenamefont {Vermeulen}, \citenamefont {Twitchen},
  \citenamefont {Markham},\ and\ \citenamefont
  {Hanson}}]{humphreys_deterministic_2018}%
  \BibitemOpen
  \bibfield  {author} {\bibinfo {author} {\bibfnamefont {P.~C.}\ \bibnamefont
  {Humphreys}}, \bibinfo {author} {\bibfnamefont {N.}~\bibnamefont {Kalb}},
  \bibinfo {author} {\bibfnamefont {J.~P.~J.}\ \bibnamefont {Morits}}, \bibinfo
  {author} {\bibfnamefont {R.~N.}\ \bibnamefont {Schouten}}, \bibinfo {author}
  {\bibfnamefont {R.~F.~L.}\ \bibnamefont {Vermeulen}}, \bibinfo {author}
  {\bibfnamefont {D.~J.}\ \bibnamefont {Twitchen}}, \bibinfo {author}
  {\bibfnamefont {M.}~\bibnamefont {Markham}},\ and\ \bibinfo {author}
  {\bibfnamefont {R.}~\bibnamefont {Hanson}},\ }\href
  {https://doi.org/10.1038/s41586-018-0200-5} {\bibfield  {journal} {\bibinfo
  {journal} {Nature}\ }\textbf {\bibinfo {volume} {558}},\ \bibinfo {pages}
  {268} (\bibinfo {year} {2018})}\BibitemShut {NoStop}%
\bibitem [{\citenamefont {Yu}\ \emph {et~al.}(2020)\citenamefont {Yu},
  \citenamefont {Ma}, \citenamefont {Luo}, \citenamefont {Jing}, \citenamefont
  {Sun}, \citenamefont {Fang}, \citenamefont {Yang}, \citenamefont {Liu},
  \citenamefont {Zheng}, \citenamefont {Xie}, \citenamefont {Zhang},
  \citenamefont {You}, \citenamefont {Wang}, \citenamefont {Chen},
  \citenamefont {Zhang}, \citenamefont {Bao},\ and\ \citenamefont
  {Pan}}]{yu_entanglement_2020}%
  \BibitemOpen
  \bibfield  {author} {\bibinfo {author} {\bibfnamefont {Y.}~\bibnamefont
  {Yu}}, \bibinfo {author} {\bibfnamefont {F.}~\bibnamefont {Ma}}, \bibinfo
  {author} {\bibfnamefont {X.-Y.}\ \bibnamefont {Luo}}, \bibinfo {author}
  {\bibfnamefont {B.}~\bibnamefont {Jing}}, \bibinfo {author} {\bibfnamefont
  {P.-F.}\ \bibnamefont {Sun}}, \bibinfo {author} {\bibfnamefont {R.-Z.}\
  \bibnamefont {Fang}}, \bibinfo {author} {\bibfnamefont {C.-W.}\ \bibnamefont
  {Yang}}, \bibinfo {author} {\bibfnamefont {H.}~\bibnamefont {Liu}}, \bibinfo
  {author} {\bibfnamefont {M.-Y.}\ \bibnamefont {Zheng}}, \bibinfo {author}
  {\bibfnamefont {X.-P.}\ \bibnamefont {Xie}}, \bibinfo {author} {\bibfnamefont
  {W.-J.}\ \bibnamefont {Zhang}}, \bibinfo {author} {\bibfnamefont {L.-X.}\
  \bibnamefont {You}}, \bibinfo {author} {\bibfnamefont {Z.}~\bibnamefont
  {Wang}}, \bibinfo {author} {\bibfnamefont {T.-Y.}\ \bibnamefont {Chen}},
  \bibinfo {author} {\bibfnamefont {Q.}~\bibnamefont {Zhang}}, \bibinfo
  {author} {\bibfnamefont {X.-H.}\ \bibnamefont {Bao}},\ and\ \bibinfo {author}
  {\bibfnamefont {J.-W.}\ \bibnamefont {Pan}},\ }\href
  {https://doi.org/10.1038/s41586-020-1976-7} {\bibfield  {journal} {\bibinfo
  {journal} {Nature}\ }\textbf {\bibinfo {volume} {578}},\ \bibinfo {pages}
  {240} (\bibinfo {year} {2020})},\ \bibinfo {note} {publisher: Springer
  Science and Business Media LLC}\BibitemShut {NoStop}%
\bibitem [{\citenamefont {Benevides}\ \emph {et~al.}(2017)\citenamefont
  {Benevides}, \citenamefont {Santos}, \citenamefont {Luiz}, \citenamefont
  {Wiederhecker},\ and\ \citenamefont {Alegre}}]{benevides_ultrahigh_q_2017}%
  \BibitemOpen
  \bibfield  {author} {\bibinfo {author} {\bibfnamefont {R.}~\bibnamefont
  {Benevides}}, \bibinfo {author} {\bibfnamefont {F.~G.~S.}\ \bibnamefont
  {Santos}}, \bibinfo {author} {\bibfnamefont {G.~O.}\ \bibnamefont {Luiz}},
  \bibinfo {author} {\bibfnamefont {G.~S.}\ \bibnamefont {Wiederhecker}},\ and\
  \bibinfo {author} {\bibfnamefont {T.~P.~M.}\ \bibnamefont {Alegre}},\ }\href
  {https://doi.org/10.1038/s41598-017-02515-4} {\bibfield  {journal} {\bibinfo
  {journal} {Scientific Reports}\ }\textbf {\bibinfo {volume} {7}},\ \bibinfo
  {pages} {2491} (\bibinfo {year} {2017})}\BibitemShut {NoStop}%
\bibitem [{\citenamefont {Weaver}\ \emph {et~al.}(2025)\citenamefont {Weaver},
  \citenamefont {Arnold}, \citenamefont {Weaver}, \citenamefont {Gröblacher},\
  and\ \citenamefont {Stockill}}]{weaver_scalable_2025}%
  \BibitemOpen
  \bibfield  {author} {\bibinfo {author} {\bibfnamefont {M.~J.}\ \bibnamefont
  {Weaver}}, \bibinfo {author} {\bibfnamefont {G.}~\bibnamefont {Arnold}},
  \bibinfo {author} {\bibfnamefont {H.}~\bibnamefont {Weaver}}, \bibinfo
  {author} {\bibfnamefont {S.}~\bibnamefont {Gröblacher}},\ and\ \bibinfo
  {author} {\bibfnamefont {R.}~\bibnamefont {Stockill}},\ }\href@noop {}
  {\bibinfo {title} {Scalable {Quantum} {Computing} with {Optical} {Links}}}
  (\bibinfo {year} {2025}),\ \Eprint {https://arxiv.org/abs/2505.00542}
  {arXiv:2505.00542} \BibitemShut {NoStop}%
\bibitem [{\citenamefont {Vlassov}\ \emph {et~al.}(2022)\citenamefont
  {Vlassov}, \citenamefont {Pinto~Moura}, \citenamefont {Marinkovic},
  \citenamefont {Wallucks}, \citenamefont {Fiaschi},\ and\ \citenamefont
  {Groeblacher}}]{Vlassov_Filter_Cavity_Design_2022}%
  \BibitemOpen
  \bibfield  {author} {\bibinfo {author} {\bibfnamefont {M.}~\bibnamefont
  {Vlassov}}, \bibinfo {author} {\bibfnamefont {J.}~\bibnamefont
  {Pinto~Moura}}, \bibinfo {author} {\bibfnamefont {I.}~\bibnamefont
  {Marinkovic}}, \bibinfo {author} {\bibfnamefont {A.}~\bibnamefont
  {Wallucks}}, \bibinfo {author} {\bibfnamefont {N.}~\bibnamefont {Fiaschi}},\
  and\ \bibinfo {author} {\bibfnamefont {S.}~\bibnamefont {Groeblacher}},\
  }\href {https://doi.org/10.5281/zenodo.6827927} {\bibinfo {title} {Filter
  cavity design by groeblacherlab}} (\bibinfo {year} {2022})\BibitemShut
  {NoStop}%
\bibitem [{\citenamefont {Weis}\ \emph {et~al.}(2010)\citenamefont {Weis},
  \citenamefont {Rivière}, \citenamefont {Deléglise}, \citenamefont
  {Gavartin}, \citenamefont {Arcizet}, \citenamefont {Schliesser},\ and\
  \citenamefont {Kippenberg}}]{weis2010optomechanically}%
  \BibitemOpen
  \bibfield  {author} {\bibinfo {author} {\bibfnamefont {S.}~\bibnamefont
  {Weis}}, \bibinfo {author} {\bibfnamefont {R.}~\bibnamefont {Rivière}},
  \bibinfo {author} {\bibfnamefont {S.}~\bibnamefont {Deléglise}}, \bibinfo
  {author} {\bibfnamefont {E.}~\bibnamefont {Gavartin}}, \bibinfo {author}
  {\bibfnamefont {O.}~\bibnamefont {Arcizet}}, \bibinfo {author} {\bibfnamefont
  {A.}~\bibnamefont {Schliesser}},\ and\ \bibinfo {author} {\bibfnamefont
  {T.~J.}\ \bibnamefont {Kippenberg}},\ }\href
  {https://doi.org/10.1126/science.1195596} {\bibfield  {journal} {\bibinfo
  {journal} {Science}\ }\textbf {\bibinfo {volume} {330}},\ \bibinfo {pages}
  {1520} (\bibinfo {year} {2010})}\BibitemShut {NoStop}%
\bibitem [{\citenamefont {Barclay}\ \emph
  {et~al.}(2005{\natexlab{a}})\citenamefont {Barclay}, \citenamefont
  {Srinivasan},\ and\ \citenamefont {Painter}}]{barclay_nonlinear_2005}%
  \BibitemOpen
  \bibfield  {author} {\bibinfo {author} {\bibfnamefont {P.~E.}\ \bibnamefont
  {Barclay}}, \bibinfo {author} {\bibfnamefont {K.}~\bibnamefont
  {Srinivasan}},\ and\ \bibinfo {author} {\bibfnamefont {O.}~\bibnamefont
  {Painter}},\ }\href {https://doi.org/10.1364/OPEX.13.000801} {\bibfield
  {journal} {\bibinfo  {journal} {Opt. Express}\ }\textbf {\bibinfo {volume}
  {13}},\ \bibinfo {pages} {801} (\bibinfo {year}
  {2005}{\natexlab{a}})}\BibitemShut {NoStop}%
\bibitem [{\citenamefont {Chan}\ \emph {et~al.}(2011)\citenamefont {Chan},
  \citenamefont {Alegre}, \citenamefont {Safavi-Naeini}, \citenamefont {Hill},
  \citenamefont {Krause}, \citenamefont {Gröblacher}, \citenamefont
  {Aspelmeyer},\ and\ \citenamefont {Painter}}]{chan_laser_2011}%
  \BibitemOpen
  \bibfield  {author} {\bibinfo {author} {\bibfnamefont {J.}~\bibnamefont
  {Chan}}, \bibinfo {author} {\bibfnamefont {T.~P.~M.}\ \bibnamefont {Alegre}},
  \bibinfo {author} {\bibfnamefont {A.~H.}\ \bibnamefont {Safavi-Naeini}},
  \bibinfo {author} {\bibfnamefont {J.~T.}\ \bibnamefont {Hill}}, \bibinfo
  {author} {\bibfnamefont {A.}~\bibnamefont {Krause}}, \bibinfo {author}
  {\bibfnamefont {S.}~\bibnamefont {Gröblacher}}, \bibinfo {author}
  {\bibfnamefont {M.}~\bibnamefont {Aspelmeyer}},\ and\ \bibinfo {author}
  {\bibfnamefont {O.}~\bibnamefont {Painter}},\ }\href
  {https://doi.org/10.1038/nature10461} {\bibfield  {journal} {\bibinfo
  {journal} {Nature}\ }\textbf {\bibinfo {volume} {478}},\ \bibinfo {pages}
  {89} (\bibinfo {year} {2011})}\BibitemShut {NoStop}%
\bibitem [{\citenamefont {Barclay}\ \emph
  {et~al.}(2005{\natexlab{b}})\citenamefont {Barclay}, \citenamefont
  {Srinivasan},\ and\ \citenamefont {Painter}}]{barclay2005nonlinear}%
  \BibitemOpen
  \bibfield  {author} {\bibinfo {author} {\bibfnamefont {P.~E.}\ \bibnamefont
  {Barclay}}, \bibinfo {author} {\bibfnamefont {K.}~\bibnamefont
  {Srinivasan}},\ and\ \bibinfo {author} {\bibfnamefont {O.}~\bibnamefont
  {Painter}},\ }\href {https://doi.org/https://doi.org/10.1364/OPEX.13.000801}
  {\bibfield  {journal} {\bibinfo  {journal} {Opt. Express}\ }\textbf {\bibinfo
  {volume} {13}},\ \bibinfo {pages} {801} (\bibinfo {year}
  {2005}{\natexlab{b}})}\BibitemShut {NoStop}%
\bibitem [{\citenamefont {Meenehan}\ \emph
  {et~al.}(2014{\natexlab{a}})\citenamefont {Meenehan}, \citenamefont {Cohen},
  \citenamefont {Gröblacher}, \citenamefont {Hill}, \citenamefont
  {Safavi-Naeini}, \citenamefont {Aspelmeyer},\ and\ \citenamefont
  {Painter}}]{meenehan_silicon_2014}%
  \BibitemOpen
  \bibfield  {author} {\bibinfo {author} {\bibfnamefont {S.~M.}\ \bibnamefont
  {Meenehan}}, \bibinfo {author} {\bibfnamefont {J.~D.}\ \bibnamefont {Cohen}},
  \bibinfo {author} {\bibfnamefont {S.}~\bibnamefont {Gröblacher}}, \bibinfo
  {author} {\bibfnamefont {J.~T.}\ \bibnamefont {Hill}}, \bibinfo {author}
  {\bibfnamefont {A.~H.}\ \bibnamefont {Safavi-Naeini}}, \bibinfo {author}
  {\bibfnamefont {M.}~\bibnamefont {Aspelmeyer}},\ and\ \bibinfo {author}
  {\bibfnamefont {O.}~\bibnamefont {Painter}},\ }\href
  {https://doi.org/10.1103/PhysRevA.90.011803} {\bibfield  {journal} {\bibinfo
  {journal} {Phys. Rev. A}\ }\textbf {\bibinfo {volume} {90}},\ \bibinfo
  {pages} {011803} (\bibinfo {year} {2014}{\natexlab{a}})}\BibitemShut
  {NoStop}%
\bibitem [{\citenamefont {Kobayashi}\ \emph {et~al.}(1995)\citenamefont
  {Kobayashi}, \citenamefont {Yamashita}, \citenamefont {Mori}, \citenamefont
  {Nakato}, \citenamefont {Komeda},\ and\ \citenamefont
  {Nishioka}}]{kobayashi1995interface}%
  \BibitemOpen
  \bibfield  {author} {\bibinfo {author} {\bibfnamefont {H.}~\bibnamefont
  {Kobayashi}}, \bibinfo {author} {\bibfnamefont {Y.}~\bibnamefont
  {Yamashita}}, \bibinfo {author} {\bibfnamefont {T.}~\bibnamefont {Mori}},
  \bibinfo {author} {\bibfnamefont {Y.}~\bibnamefont {Nakato}}, \bibinfo
  {author} {\bibfnamefont {T.}~\bibnamefont {Komeda}},\ and\ \bibinfo {author}
  {\bibfnamefont {Y.}~\bibnamefont {Nishioka}},\ }\href
  {https://doi.org/10.1143/JJAP.34.959} {\bibfield  {journal} {\bibinfo
  {journal} {Jpn. J. Appl. Phys.}\ }\textbf {\bibinfo {volume} {34}},\ \bibinfo
  {pages} {959} (\bibinfo {year} {1995})}\BibitemShut {NoStop}%
\bibitem [{\citenamefont {Yamashita}\ \emph {et~al.}(1996)\citenamefont
  {Yamashita}, \citenamefont {Namba}, \citenamefont {Nakato}, \citenamefont
  {Nishioka},\ and\ \citenamefont {Kobayashi}}]{yamashita1996spectroscopic}%
  \BibitemOpen
  \bibfield  {author} {\bibinfo {author} {\bibfnamefont {Y.}~\bibnamefont
  {Yamashita}}, \bibinfo {author} {\bibfnamefont {K.}~\bibnamefont {Namba}},
  \bibinfo {author} {\bibfnamefont {Y.}~\bibnamefont {Nakato}}, \bibinfo
  {author} {\bibfnamefont {Y.}~\bibnamefont {Nishioka}},\ and\ \bibinfo
  {author} {\bibfnamefont {H.}~\bibnamefont {Kobayashi}},\ }\href
  {https://doi.org/https://doi.org/10.1063/1.361472} {\bibfield  {journal}
  {\bibinfo  {journal} {J. Appl. Phys.}\ }\textbf {\bibinfo {volume} {79}},\
  \bibinfo {pages} {7051} (\bibinfo {year} {1996})}\BibitemShut {NoStop}%
\bibitem [{\citenamefont {Sakurai}\ and\ \citenamefont
  {Sugano}(1981)}]{sakurai1981theory}%
  \BibitemOpen
  \bibfield  {author} {\bibinfo {author} {\bibfnamefont {T.}~\bibnamefont
  {Sakurai}}\ and\ \bibinfo {author} {\bibfnamefont {T.}~\bibnamefont
  {Sugano}},\ }\href {https://doi.org/https://doi.org/10.1063/1.329023}
  {\bibfield  {journal} {\bibinfo  {journal} {J. Appl. Phys.}\ }\textbf
  {\bibinfo {volume} {52}},\ \bibinfo {pages} {2889} (\bibinfo {year}
  {1981})}\BibitemShut {NoStop}%
\bibitem [{\citenamefont {Srivastava}(2022)}]{srivastava_physics_2022}%
  \BibitemOpen
  \bibfield  {author} {\bibinfo {author} {\bibfnamefont {G.~P.}\ \bibnamefont
  {Srivastava}},\ }\href@noop {} {\emph {\bibinfo {title} {The {Physics} of
  {Phonons}}}},\ \bibinfo {edition} {2nd}\ ed.\ (\bibinfo  {publisher} {CRC
  Press},\ \bibinfo {address} {Boca Raton},\ \bibinfo {year}
  {2022})\BibitemShut {NoStop}%
\bibitem [{\citenamefont {Meenehan}\ \emph
  {et~al.}(2014{\natexlab{b}})\citenamefont {Meenehan}, \citenamefont {Cohen},
  \citenamefont {Gr{\"o}blacher}, \citenamefont {Hill}, \citenamefont
  {Safavi-Naeini}, \citenamefont {Aspelmeyer},\ and\ \citenamefont
  {Painter}}]{meenehan2014silicon}%
  \BibitemOpen
  \bibfield  {author} {\bibinfo {author} {\bibfnamefont {S.~M.}\ \bibnamefont
  {Meenehan}}, \bibinfo {author} {\bibfnamefont {J.~D.}\ \bibnamefont {Cohen}},
  \bibinfo {author} {\bibfnamefont {S.}~\bibnamefont {Gr{\"o}blacher}},
  \bibinfo {author} {\bibfnamefont {J.~T.}\ \bibnamefont {Hill}}, \bibinfo
  {author} {\bibfnamefont {A.~H.}\ \bibnamefont {Safavi-Naeini}}, \bibinfo
  {author} {\bibfnamefont {M.}~\bibnamefont {Aspelmeyer}},\ and\ \bibinfo
  {author} {\bibfnamefont {O.}~\bibnamefont {Painter}},\ }\href
  {https://doi.org/https://doi.org/10.1103/PhysRevA.90.011803} {\bibfield
  {journal} {\bibinfo  {journal} {Phys. Rev. A}\ }\textbf {\bibinfo {volume}
  {90}},\ \bibinfo {pages} {011803} (\bibinfo {year}
  {2014}{\natexlab{b}})}\BibitemShut {NoStop}%
\bibitem [{\citenamefont {Guo}\ and\ \citenamefont
  {Gröblacher}(2022)}]{guo_integrated_2022}%
  \BibitemOpen
  \bibfield  {author} {\bibinfo {author} {\bibfnamefont {J.}~\bibnamefont
  {Guo}}\ and\ \bibinfo {author} {\bibfnamefont {S.}~\bibnamefont
  {Gröblacher}},\ }\href {https://doi.org/10.1038/s41377-022-00966-7}
  {\bibfield  {journal} {\bibinfo  {journal} {Light Sci. Appl.}\ }\textbf
  {\bibinfo {volume} {11}},\ \bibinfo {pages} {282} (\bibinfo {year}
  {2022})}\BibitemShut {NoStop}%
\bibitem [{\citenamefont {Stevens}\ \emph {et~al.}(2014)\citenamefont
  {Stevens}, \citenamefont {Glancy}, \citenamefont {Nam},\ and\ \citenamefont
  {Mirin}}]{stevens_third-order_2014}%
  \BibitemOpen
  \bibfield  {author} {\bibinfo {author} {\bibfnamefont {M.~J.}\ \bibnamefont
  {Stevens}}, \bibinfo {author} {\bibfnamefont {S.}~\bibnamefont {Glancy}},
  \bibinfo {author} {\bibfnamefont {S.~W.}\ \bibnamefont {Nam}},\ and\ \bibinfo
  {author} {\bibfnamefont {R.~P.}\ \bibnamefont {Mirin}},\ }\href
  {https://doi.org/10.1364/OE.22.003244} {\bibfield  {journal} {\bibinfo
  {journal} {Opt. Express}\ }\textbf {\bibinfo {volume} {22}},\ \bibinfo
  {pages} {3244} (\bibinfo {year} {2014})}\BibitemShut {NoStop}%
\bibitem [{\citenamefont {Davidovich}(1996)}]{davidovich_sub-poissonian_1996}%
  \BibitemOpen
  \bibfield  {author} {\bibinfo {author} {\bibfnamefont {L.}~\bibnamefont
  {Davidovich}},\ }\href {https://doi.org/10.1103/RevModPhys.68.127} {\bibfield
   {journal} {\bibinfo  {journal} {Rev. Mod. Phys.}\ }\textbf {\bibinfo
  {volume} {68}},\ \bibinfo {pages} {127} (\bibinfo {year} {1996})}\BibitemShut
  {NoStop}%
\bibitem [{\citenamefont {Fishman}\ \emph {et~al.}(2023)\citenamefont
  {Fishman}, \citenamefont {Patel}, \citenamefont {Hopper}, \citenamefont
  {Huang},\ and\ \citenamefont
  {Bassett}}]{fishman_photon-emission-correlation_2023}%
  \BibitemOpen
  \bibfield  {author} {\bibinfo {author} {\bibfnamefont {R.~E.}\ \bibnamefont
  {Fishman}}, \bibinfo {author} {\bibfnamefont {R.~N.}\ \bibnamefont {Patel}},
  \bibinfo {author} {\bibfnamefont {D.~A.}\ \bibnamefont {Hopper}}, \bibinfo
  {author} {\bibfnamefont {T.-Y.}\ \bibnamefont {Huang}},\ and\ \bibinfo
  {author} {\bibfnamefont {L.~C.}\ \bibnamefont {Bassett}},\ }\href
  {https://doi.org/10.1103/PRXQuantum.4.010202} {\bibfield  {journal} {\bibinfo
   {journal} {PRX Quantum}\ }\textbf {\bibinfo {volume} {4}},\ \bibinfo {pages}
  {010202} (\bibinfo {year} {2023})}\BibitemShut {NoStop}%
\bibitem [{\citenamefont {Børkje}\ \emph {et~al.}(2011)\citenamefont
  {Børkje}, \citenamefont {Nunnenkamp},\ and\ \citenamefont
  {Girvin}}]{borkje_proposal_2011}%
  \BibitemOpen
  \bibfield  {author} {\bibinfo {author} {\bibfnamefont {K.}~\bibnamefont
  {Børkje}}, \bibinfo {author} {\bibfnamefont {A.}~\bibnamefont
  {Nunnenkamp}},\ and\ \bibinfo {author} {\bibfnamefont {S.~M.}\ \bibnamefont
  {Girvin}},\ }\href {https://doi.org/10.1103/PhysRevLett.107.123601}
  {\bibfield  {journal} {\bibinfo  {journal} {Phys. Rev. Lett.}\ }\textbf
  {\bibinfo {volume} {107}},\ \bibinfo {pages} {123601} (\bibinfo {year}
  {2011})}\BibitemShut {NoStop}%
\bibitem [{\citenamefont {Gröblacher}\ \emph {et~al.}(2013)\citenamefont
  {Gröblacher}, \citenamefont {Hill}, \citenamefont {Safavi-Naeini},
  \citenamefont {Chan},\ and\ \citenamefont
  {Painter}}]{groblacher_highly_2013}%
  \BibitemOpen
  \bibfield  {author} {\bibinfo {author} {\bibfnamefont {S.}~\bibnamefont
  {Gröblacher}}, \bibinfo {author} {\bibfnamefont {J.~T.}\ \bibnamefont
  {Hill}}, \bibinfo {author} {\bibfnamefont {A.~H.}\ \bibnamefont
  {Safavi-Naeini}}, \bibinfo {author} {\bibfnamefont {J.}~\bibnamefont
  {Chan}},\ and\ \bibinfo {author} {\bibfnamefont {O.}~\bibnamefont
  {Painter}},\ }\href {https://doi.org/10.1063/1.4826924} {\bibfield  {journal}
  {\bibinfo  {journal} {Appl. Phys. Lett.}\ }\textbf {\bibinfo {volume}
  {103}},\ \bibinfo {pages} {181104} (\bibinfo {year} {2013})}\BibitemShut
  {NoStop}%
\end{thebibliography}
\end{document}